%% file: tol02_clean_manuscript.tex
\documentclass[a4paper,fleqn,usenatbib]{mnras}
\usepackage{newtxtext,newtxmath}
\usepackage[T1]{fontenc}
\usepackage{ae,aecompl}

\usepackage{graphicx}	% Including figure files
\usepackage{amsmath}	% Advanced maths commands
\usepackage{amssymb}	% Extra maths symbols
\usepackage{threeparttable} %permite poner notas al pie de tablas
\usepackage{subfigure} %permite usar subfigura 

\title[Optical/Near-IR study of Tol 02]{Optical/Near-IR spatially resolved study of the H II galaxy Tol 02.\thanks{Based on observations obtained at the Southern Astrophysical Research (SOAR) telescope, which is a joint project of the Minist\'{e}rio da Ci\^{e}ncia, Tecnologia, e Inova\c{c}\~{a}o (MCTI) da Rep\'{u}blica Federativa do Brasil, the U.S. National Optical Astronomy Observatory (NOAO), the University of North Carolina at Chapel Hill (UNC), and Michigan State University (MSU).}}

\author[A. Torres-Campos et al.]{
A. Torres-Campos$^{1}$\thanks{E-mail: tcampos@inaoep.mx (ATC)}, 
E. Terlevich$^{1}$,
D. Rosa-Gonz\'{a}lez$^{1}$,
R. Terlevich$^{1,3}$,
E. Telles$^{2}$,
\newauthor A. I. D\'{i}az$^{4}$
\\
$^{1}$Instituto Nacional de Astrof\'{i}sica \'{O}ptica y Electr\'{o}nica, L.E. Erro No. 1, Santa Mar\'{i}a Tonantzintla, Puebla, Mexico\\
$^{2}$Observatorio Nacional, Rua Jos\'{e} Cristino 77, 20921-400, Rio de Janeiro, Brasil\\
$^{3}$Institute of Astronomy, University of Cambridge, Madingley Road, Cambridge CB3 0HA, UK\\
$^{4}$Departamento de F\'{i}sica Te\'{o}rica, C-XI, Universidad Aut\'{o}noma de Madrid, Cantoblanco, E-28049 Madrid, Spain\\
}

\date{Accepted XXX. Received YYY; in original form ZZZ}

\pubyear{2016}

\begin{document}
\label{firstpage}
\pagerange{\pageref{firstpage}--\pageref{lastpage}}
\maketitle

% Abstract of the paper
\begin{abstract}

%--------------------------------ABSTRACT-----------------------------------------------------
%
The main goal of this study is to characterise the stellar populations in very low metallicity galaxies.
We have obtained  broad U, B, R, I, J, H, K, intermediate Str\"{o}mgren {\it y} and narrow ${\rm H\alpha}$ and ${\rm [OIII]}$ deep images of the Wolf-Rayet, Blue Compact Dwarf, H II galaxy Tol 02.
We have analysed the  low surface brightness component, the stellar cluster complexes and the H II regions.
The stellar populations in the galaxy have been characterised by comparing the observed broad band colours with those of single stellar population models.
The main results are consistent with Tol 02 being formed by a 1.5 Gyr old disk component at the centre of which a group of 8 massive (${\rm > 10^4 \ M_{\sun}}$) stellar cluster clumps is located.
Six of these clumps are 10 Myr old and their near infrared colours suggest that their light is dominated by Red Supergiant stars, the other two are young Wolf-Rayet cluster candidates of ages  3 and 5 Myr respectively.
12 H II regions in the star-forming region of the galaxy are also identified. 
These are immersed in a diffuse ${\rm H\alpha}$ and ${\rm [OIII]}$ emission that spreads towards the North and South covering the old stellar disk.
Our spatial-temporal analysis shows that star formation is more likely stochastic and simultaneous within short time scales. 
The mismatch between observations and models cannot be attributed alone to a mistreat of the RSG phase and still needs to be further investigated.
%

%--------------------------------ABSTRACT-----------------------------------------------------

%This is a simple template for authors to write new MNRAS papers.
%The abstract should briefly describe the aims, methods, and main results of the paper.
%It should be a single paragraph not more than 250 words (200 words for Letters).
%No references should appear in the abstract.
\end{abstract}

% Select between one and six entries from the list of approved keywords.
% Don't make up new ones.

%%%%%%%%%%%%%%%%%%%%%%%%%%%%%%%%%%%%%%%%%%%%%%%%%%

%%%%%%%%%%%%%%%%% BODY OF PAPER %%%%%%%%%%%%%%%%%%

%\section{Introduction}

%This is a simple template for authors to write new MNRAS papers.
%See \texttt{mnras\_sample.tex} for a more complex example, and \texttt{mnras\_guide.tex}
%for a full user guide.

%All papers should start with an Introduction section, which 
%sets the work in context, 
%cites relevant earlier studies in the field by \citet{Others2013},
%and describes the problem the authors aim to solve \citep[e.g.][]{Author2012}.

%=============================================================================================
%	1 INTRODUCTION
%=============================================================================================
\input{tol02_1v6c}

%=============================================================================================
%	2 OBSERVATIONS, DATA REDUCTION AND CALIBRATION
%=============================================================================================
\input{tol02_2v6c}

%=============================================================================================
%	3 RESULTS: PHOTOMETRIC RESULTS. 
%               - DESCRIPTION OF GALAXY MAPS, 
%               - LOW SURFACE BRIGHTNESS PROFILE
%               
%=============================================================================================
\input{tol02_3v6c}

%=============================================================================================
%	4 RESULTS: - IDENTIFICATION OF SCC AND H II REGIONS, HOST GALAXY CHARACTERIZATION
%=============================================================================================
\input{tol02_4v6c}

%=============================================================================================
%	5 RESULTS: SED FITTING TECHNIQUE, PHOTOMETRY VS MODELS: AGE, MASS, AV DISTRIBUTION
%=============================================================================================
\input{tol02_5v6c}

%=============================================================================================
%	5A DISCUSSION: COMPARISON OF THE CHARACTERIZED CLUSTERS/H II REGIONS WITH OTHERS INTH
%=============================================================================================
\input{tol02_5bv6c}

%=============================================================================================
%	5B CONCLUSIONS: 
%The last numbered section should briefly summarise what has been done, and describe
%the final conclusions which the authors draw from their work.
%=============================================================================================
\input{tol02_6v6c}

%=============================================================================================
%   8 ACKNOWLEDGEMENTS
%=============================================================================================
\input{tol02_7v6c}

%The Acknowledgements section is not numbered. Here you can thank helpful
%colleagues, acknowledge funding agencies, telescopes and facilities used etc.
%Try to keep it short.

%%%%%%%%%%%%%%%%%%%%%%%%%%%%%%%%%%%%%%%%%%%%%%%%%%

%%%%%%%%%%%%%%%%%%%% REFERENCES %%%%%%%%%%%%%%%%%%

% The best way to enter references is to use BibTeX:

\bibliographystyle{mnras}
%\bibliography{example} % if your bibtex file is called example.bib

\bibliography{biblio} % if your bibtex file is called example.bib

% Alternatively you could enter them by hand, like this:
% This method is tedious and prone to error if you have lots of references
%\begin{thebibliography}{99}
%\bibitem[\protect\citeauthoryear{Author}{2012}]{Author2012}
%Author A.~N., 2013, Journal of Improbable Astronomy, 1, 1
%\bibitem[\protect\citeauthoryear{Others}{2013}]{Others2013}
%Others S., 2012, Journal of Interesting Stuff, 17, 198
%\end{thebibliography}

%%%%%%%%%%%%%%%%%%%%%%%%%%%%%%%%%%%%%%%%%%%%%%%%%%

%%%%%%%%%%%%%%%%% APPENDICES %%%%%%%%%%%%%%%%%%%%%

\appendix

%\section{Some extra material}

\input{tol02_8}

If you want to present additional material which would interrupt the flow of the main paper,
it can be placed in an Appendix which appears after the list of references.

%%%%%%%%%%%%%%%%%%%%%%%%%%%%%%%%%%%%%%%%%%%%%%%%%%

% Don't change these lines
\bsp	% typesetting comment
\label{lastpage}
\end{document}

%% file: tol02_1v6c.tex
\section[]{INTRODUCTION}

%---------------------------------------------------------------------------------------------
% ESTABLISHING THE IMPORTANCE OF THE TOPIC FOR ALL ASTRONOMERS BY HIGHLIGHTING A GENERAL PROBLEM --> LOW METALLICITY SYSTEMS STAR CLUSTER PROPERTIES 
%---------------------------------------------------------------------------------------------

%
One of the greatest challenges in astronomy is the comprehension of the star formation process that took place inside the very first galaxies formed in the Universe. 
The key aspects to understand this process are its efficiency to form massive clusters ($> 10^4$ ${\rm M_{\sun}}$), stellar mass and stellar density of the newly formed star clusters \citep{Ad10a,Ma12,Br16,Gr16}.
%
%---------------------------------------------------------------------------------------------
% WHY ANALYSING H II GALAXIES? --> KEY INSTRUMENT OF STAR-FORMING PROCESS
%---------------------------------------------------------------------------------------------
An indirect manner to learn about these primordial star clusters is through the analysis of the star clusters formed in H II galaxies. 
%
%Although H II galaxies are not necessarily recently formed galaxies -as once thought-, they are low luminosity, metal poor (${\rm 1/50 Z_{\odot} - 1/3 Z_{\odot} }$) gas rich systems undergoing an intense burst of star formation, starbursts in H II galaxies resemble the star formation process in the first formed galaxies of the Universe.
%
Although H II galaxies are not necessarily recently formed galaxies, the fact that they are of low metallicity and high gas content implies that they represent relatively unevolved galaxies, probably the low z counterpart of the equally unevolved high z  galaxies.
This  is why a deep understanding of the star formation history of H II galaxies is essential to address the star-formation process in the first formed galaxies \citep{Te97,Te04,La11,Ma12}.

%---------------------------------------------------------------------------------------------
% HISTORY OF H II GALAXIES RESEARCH 
%---------------------------------------------------------------------------------------------
H II galaxies were first identified by -and named after- their optical spectrum, which is characterised by a weak stellar continuum dominated by strong emission lines and resembles the optical spectrum of H II regions \citep{Sa70, Te91}. 
Due to their spectroscopic features -high ${\rm EW(H\alpha)}$ and low metal content- it was believed that they were the youngest galaxies in the Universe producing their first generation of stars, until deep infrared imaging unveiled their old stellar population with low surface brightness \citep{Th83,Ca03,Mu09}.
Although they do have an older stellar component responsible for their non-zero metallicity, H II galaxies remain as the most unevolved galaxies in the near Universe.

Most H~II galaxies do not have an identified companion \citep{Te95} although in recent years, low surface brightness companions have been found around BCD, starburst, star forming dwarfs and even some objects classified as H~II galaxies \citep{Ri12,Le14,As17}.
The morphology of H II galaxies is varied -ranging from spiral to Blue Compact Dwarf (BCD)- and their star-forming regions are frequently composed by multiple knots  \citep{Te97,Me00,La11}.

%
%arose the idea that they were the youngest galaxies in the Universe producing their first generation of stars. 
%
%Later studies using deep optical and infrared imaging unveiled their low surface brightness old stellar population \citep{Th83,Ca03,Mu09}.  

%TOL 0957-278
%CROSS-IDENTIFICATIONS for ESO 435-IG 020 (Back to INDEX)
%Object Names 	        Type 	    Object Names 	Type
%ESO 435-IG 020	        GGroup	    ESO-LV 4350200 	        G
%ESO 095705-2753.5 	    GGroup	    6dF J0959212-280800 	G
%AM 0957-275 	        G 	        PGC 028863 	            G
%TOLOLO 00002 	        EmLS 	    NVSS J095921-280805 	RadioS
%TOLOLO 0957-278 	    EmLS 	 	 
%2MASX J09592122-2808000 IrS 
%---------------------------------------------------------------------------------------------
%       Tol 02 GENERAL DATA
%---------------------------------------------------------------------------------------------
Tol 02 (also known as Tol 0957-278, ESO 435-IG 020, AM 0957-275, PGC 028863, 2MASX J09592122-2808000, NVSS J095921-280805) was first identified as an H II galaxy due to its intense emission lines \citep{Sm76,Te91,Ke04}. 
It is an isolated galaxy located in the southern sky (R.A = 09\fh 59\fm 21.21\fs, Dec. = -28\fdg 08\farcm 00.3\farcs, J2000) at a redshift z=0.003 (Distance ${\rm \sim 11}$ Mpc) with an angular diameter of ${\rm \sim 50 }$ arcsec. 
A bright star-forming region with high intensity emission lines and low gaseous metallicity ${\rm12+log(O/H)\approx8.2}$ \citep{Va92,Ma94,Bu05,En08} lies at its centre. 
This galaxy has also been classified as a BCD galaxy because of its low B band magnitude ${\rm M_{B} = -16.26 \ mag}$ \citep{Ku88,Do99,Gi03} and as a Wolf-Rayet galaxy candidate due to the broad ${\rm He \, II \lambda4686}$ emission line observed in its integrated spectra \citep{Ku85,Sc99}.
%

%---------------------------------------------------------------------------------------------
%       PREVIOUS NOWLEDGE ABOUT THE GALAXY
%---------------------------------------------------------------------------------------------

%---------------------------------------------------------------------------------------------
%       MOTIVATION/OBJECTIVE OF THIS WORK
%---------------------------------------------------------------------------------------------
%
Previous research on Tol 02 has focused on the integrated spectrum of the galaxy \citep{Sm76,Ku85,Te91,Sc99,Ke04}. 
These have concluded that the galaxy is composed of multiple stellar populations, in which the dominance of extremely young -${\rm age < 3 }$ Myrs old- and Red Super Giant (RSG) stars are eminent, while the older stellar populations with ages greater than ${\rm 5 }$ Myr is modest \citep{Ra00,We04}.
Few photometric studies have been performed in this galaxy, and they have focused in the decomposition  of the galaxy into a disk -or host galaxy formed by the old stellar populations- and a star-forming region.
A previous characterisation of the galaxy H II regions was performed by \citet{Me00}, where they did not identify the stellar populations in the continuum images.
%

%---------------------------------------------------------------------------------------------
%       BRIEF EXPLANATION OF WHAT IS DONE IN THIS WORK
%---------------------------------------------------------------------------------------------
The characteristics of Tol 02 described before -isolated H II galaxy, BCD classification, projected size > 30 arcsec-, as well as the previous results found on the galaxy -extremely low brightness old stellar population, possibility of finding Wolf-Rayet and RSG stars- made us realise that this galaxy was an excellent target to analyse the star formation process.
%
%Nevertheless, a detailed study on the distribution and properties of this galaxy resolved stellar populations had not been made before this work. 
%But they haven't been performed on this galaxy.
The main goal of this study is to characterise the photometric properties, ages and masses of
 the stellar populations in the galaxy by identifying its structural components.

%ages and masses of the star cluster complexes (SCCs) in this galaxy splitting the galaxy into its components. 

In Section \ref{sec:obs}, we describe the observations and data reduction  procedures. 
In Section \ref{sec:phot}, we compare our results on the galaxy optical broad and narrow band integrated photometry with values reported in the literature. In Section \ref{sec:analysis}, we analyse the photometry of the galaxy low surface brightness component, the star-forming region, and the identified stellar cluster complexes and H II regions.
 In Section \ref{sec:fitbb}, we describe the procedure used to characterise the age, mass and extinction of the stellar populations. In Section \ref{sec:disc}, we discuss the photometric and modelled properties of the stellar populations derived and our conclusions on the stellar populations that dominate the galaxy photometry.

%---------------------------------------------------------------------------------------------
%       SECTIONS OF THE ARTICLE
%---------------------------------------------------------------------------------------------

%% file: tol02_2v6c.tex
\section[]{OBSERVATIONS, DATA REDUCTION AND CALIBRATION}
\label{sec:obs}

%The data were obtained at SOAR\footnote {Based on observations obtained at the Southern Astrophysical Research (SOAR) telescope, which is a joint project of the Minist\'{e}rio da Ci\^{e}ncia, Tecnologia, e Inova\c{c}\~{a}o (MCTI) da Rep\'{u}blica Federativa do Brasil, the U.S. National Optical Astronomy Observatory (NOAO), the University of North Carolina at Chapel Hill (UNC), and Michigan State University (MSU).}, the 4.1m telescope situated in Cerro Pach\'{o}n, Chile, in service mode. Images in optical and near infrared bands were taken and the instrumentation used is described below.

The data were obtained at SOAR, the 4.1m telescope situated in Cerro Pach\'{o}n, Chile, in service mode. Images in optical and near infrared bands were taken and the instrumentation used is described below.

% SOI
Optical U, B, R and I broad-, y Str\"{o}mgren intermediate- and ${\rm H\alpha}$ and ${\rm [OIII]\lambda5007}$ narrow-band images were obtained on February and May of 2010 using SOAR Optical Imager (SOI),
with a characteristic seeing of 0.8 arcsec. 
SOI is a mini-mosaic comprised of two CCDs of 2048x4096 pixels each and mounted with their long sides parallel and spaced 102 pixels apart. 
%Each CCD is read by two different amplifiers, resulting in 4 data frames that form the mosaic image. This image covers a total area on the sky of $5.3\times5.2$ arcmin$^2$ (with a central gap of 7.8 arcsec wide). 
%
Each CCD is read by two different amplifiers, resulting in a mosaic image formed by four data frames. 
%
%This image covers a total area on the sky of $5.3\times5.2$ arcmin$^2$ (with a central gap of 7.8 arcsec wide). 
%
The sky area covered by this image is $5.3\times5.2$ arcmin$^2$ (with a central gap of 7.8 arcsec wide) and a 0.1534 arcsec pixel$^{-1}$ plate scale in the 2x2 binned mode.
%
%The scale is 0.1534 arcsec pixel$^{-1}$ in the 2x2 binned mode. The characteristic seeing of the observing run was 0.8 arcsec.

% SPARTAN
%NIR broad-band J, H and K images were taken on March 2012 with the Spartan infrared camera. Spartan is a 4 HAWAII-2 detectors (2048x2048 pixels) mosaic. We used the wide field mode which has a plate scale of 0.07 arcsec pixel$^{-1}$ and covers a sky area of ${\rm 2.34\times2.32}$ arcmin$^2$ with a gap between detectors of 0.47 arcmin. The four detectors have different response.
%, number 3 being the best one. 
%
NIR broad-band J, H and K images were taken on March 2012 with Spartan, SOAR infrared camera.
Spartan is a mosaic formed by four HAWAII-2 detectors (2048x2048 ${\rm pixels^-2}$). 
The wide field mode was used, covering a sky area of ${\rm 2.34\times2.32}$ arcmin$^2$ with a gap between detectors of 0.47 arcmin, and a plate scale of 0.07 arcsec pixel$^{-1}$.
%
%We used the wide field mode which has a plate scale of 0.07 arcsec pixel$^{-1}$ and covers a sky area of ${\rm 2.34\times2.32}$ arcmin$^2$ with a gap between detectors of 0.47 arcmin. The four detectors have different response.
%, number 3 being the best one. 
%
The galaxy was centred on detector 3 (Tol 02 major diameter is only 0.8 arcmin),
since this is the one with the best response. 
The other detectors were not used.
%
%We centred Tol 02 on detector 3 (the galaxy major diameter is  0.8 arcmin) and did not use the other detectors image. The characteristic seeing of the observing run was 0.6 arcsec. 
%
The dithering pattern of the images followed a six points rectangular path. Each consecutive image was taken with a 5 arcseconds displacement in either right ascension or declination.

% FILTERS TRANSMISSION
The observing log is presented in Table \ref{tab:art_tol0957_tab_log}, where column 1 is the date of observation, column 2 the filter, column 3 the total exposure time in seconds (as a combination of several exposures), column 4 the mean airmass of the observations and column 5 the seeing of the image in arcseconds. 
The properties of the optical and NIR filters\footnote{http://www.ctio.noao.edu/soar/content/filters-available-soar} are given in Table {\ref{tab:tb_tol0957_filters}. Column 1 is the filter name, column 2 the central wavelength in \AA ngstroms, column 3 the FWHM in \AA ngstroms and column 4 the  relative transmission at the central wavelength.

%---------------------------------------------------------------------------------------------

% OBSERVATION LOG
\input{art_tol0957_tab_log_9v3}

% FILTERS TABLE
\input{tb_tol0957_filters}

%---------------------------------------------------------------------------------------------
\subsection{Optical data}
%---------------------------------------------------------------------------------------------

Standard image reduction was performed individually to each of the optical frames using IRAF\footnote{IRAF (Image Reduction and Analysis Facility) is distributed by the National Optical Astronomy Observatories, which are operated by the Association of Universities for Research in Astronomy (AURA), Inc., under cooperative agreement with the National Science Foundation.}. 
This includes bias subtraction, flat-fielding and cosmic-ray elimination by combining  multiple exposures. 
Astrometric solution was applied using the tasks {\it ccxymatch} and {\it ccsetwcs} from the {\it images} package. 
%
%Each galaxy image was formed by grouping the four reduced frames using  {\it wregister} and {\it imcombine} and then subtracting the sky value of this combined image, which was calculated as the median of the Gaussian that best fitted the image counts distribution. 
%
Each galaxy image was formed by grouping the four reduced frames using  {\it wregister} and {\it imcombine}. 
Once the complete image was obtained the median sky value was subtracted.
The sky value of this combined image was calculated as the median of the Gaussian that best fitted the image counts distribution. 
%
%Once the complete image was formed the median sky value was subtracted. 
%
We have degraded all images into the same spatial resolution given by the image with the worst  seeing (which turned out to be the ${\rm H\alpha}$ image), calculated using the field stars  full width at half maximum (FWHM). 
The  optical images final resolution is 1 arcsec.
%---------------------------------------------------------------------------------------------

%---------------------------------------------------------------------------------------------
\subsection{NIR data}
%---------------------------------------------------------------------------------------------

The NIR reduction steps are based on \citet{Lo11} and  \citet{Ma05}. 
We used IRAF  tasks to perform  the image processing. 
The process included correction for bad pixels, sky subtraction in order to achieve zero mean sky value, field stars subtraction, sky flat correction of pixel to pixel CCD response variation and astrometric alignment of the images. % in order to obtain colours.

\subsection{Flux calibration}
\label{sec:obs:fluxcal}

%To calibrate the optical images (both broad and narrow bands) we used two photometric standard stars from the southern spectrophotometric catalogue \citep{Ha92}. 
%
%Standard methods were aplied to calibrate the images, 

The images were calibrated using standard methods described in Appendix \ref{ap:a}.
Two spectrophotometric standard stars from \citet{Ha92} were used to calibrate the optical images (both broad and narrow bands).
The NIR images were calibrated using four faint NIR standard stars from \citet{Pe98}. 

We have compared the magnitudes of field stars identified in both SOAR and 2MASS images (Figure \ref{fig:comp2mass} and Table \ref{tb:soar_2mass}).
Unfortunately, most of the stars are dimmer than the 2MASS average limits, J = 15.8 mag, 
H = 15.1 mag and Ks = 14.3 mag \citep{Sk06}, although all of them were classified as having a 
SNR > 10 thus  good photometry\footnote{The 2MASS user's guide indicates that the point source photometric sensitivity 
varies from frame to frame, therefore the SNR = 10 level is usually achieved at fainter 
magnitudes than the ones imposed as the survey limits, as is the case for the image frames of Tol 02 
%(http:\/\/www.ipac.caltech.edu\/2mass\/releases\/allsky\/doc\/sec2_2.html)}, 
(http://www.ipac.caltech.edu/2mass/releases/allsky/doc/sec2\_2.html). }.
Table \ref{tb:soar_2mass} shows that our  J and H values are ${\rm \sim 0.2}$ magnitudes lower than the 
ones calculated by 2MASS while a much better agreement is found in the K band. A possible cause for this systematic difference could be  different filter response between the 2MASS and SOAR systems.

%
%This discrepancy is believed to arise from SOAR images being deeper than 2MASS.
%

% 2MASS COMPARISON
%---------------------------------------------------------------------------------------------
\begin{figure}
\centering
\resizebox{0.45\textwidth}{!}{\includegraphics{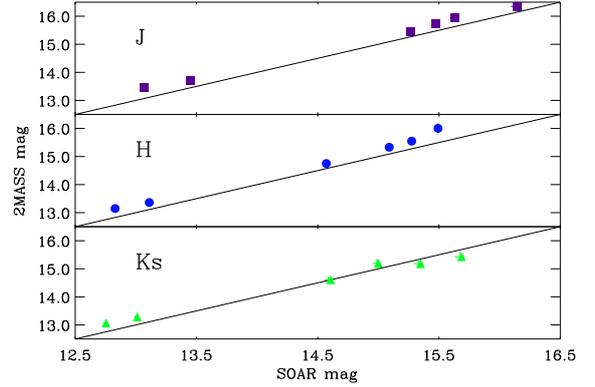}} 
\caption{ \small Near Infrared field stars magnitude; comparison between 2MASS and this work.}
\label{fig:comp2mass}
\end{figure}
%---------------------------------------------------------------------------------------------

% Table with the median values of the comparison stars
%---------------------------------------------------------------------------------------------
\input{tb.dif.soar.2mass}

\subsection{Emission line images}
\label{sec:obs:puremilin}

Emission line images were obtained after subtracting the continuum contribution to the narrow band filter images. 
The amount of continuum in the narrow band images was estimated by comparing the field stars flux in narrow, broad and intermediate band  images.
This procedure is based on the premise that the field stars flux in the narrow band images is none other than the continuum flux contribution to the image, since stars usually do not present emission lines.
Therefore, the field stars broad and intermediate band flux (${\rm Flux_{B/I} }$) was scaled by a constant $C$ to match the flux of the stars in the narrow band image ${\rm Flux_{NB}}$ (${\rm Flux_{NB} = C \dot Flux_{B/I} }$).
The scaled R and y Str\"{o}mgren images were used to estimate the continuum flux in the ${\rm H\alpha+[NII]+cont}$  and ${\rm [OIII]\lambda5007+cont}$ images respectively.

%\begin{equation}
%\mathrm{ Flux_{NB} = Flux_{line} +  C_{BB}Flux_{cont \, BB} } \\
%Flux_{NB} = CFlux_{B/I}  \\
%\end{equation}

%The total flux in the narrow band image is composed of the emission line plus the continuum flux: 

%\begin{equation}
% Flux_{NB} = Flux_{line} +  Flux_{cont \, NB}  \\
%\end{equation}

%\noindent The flux of the  broad and  intermediate band field stars was scaled by a constant $C_{BB}$ to match the flux of the stars in the narrow band image.

%\begin{equation}
%\mathrm{ Flux_{NB} = Flux_{line} +  C_{BB}Flux_{cont \, BB} } \\
%Flux_{NB} = Flux_{line} +  C_{BB}Flux_{cont \, BB}  \\
%\end{equation}

%

%To estimate the continuum flux in the ${\rm H\alpha+[NII]+cont}$  and in the ${\rm [OIII]\lambda5007+cont}$ images, we used the scaled  broad R band and  y Str\"{o}mgren images respectively. These scaled continua were subtracted from the corresponding narrow band image to obtain a pure emission line one.

%The  ${\rm H\alpha}$  line lies in the R band filter transmission curve. This is one of the brightest lines in the optical spectrum and the most important contribution of nebular emission inside the R image. For this reason we have subtracted the ${\rm H\alpha}$ pure emission flux image from the R band flux image before performing any analysis. All the results  regarding the R band  refer to this R band image after  the ${\rm H\alpha}$ emission line has been subtracted. 

The  ${\rm H\alpha}$  emission line is one of the brightest lines in the optical spectrum and the most important contribution of nebular emission inside the R band image. 
For this reason we have subtracted the pure emission ${\rm H\alpha}$ image from the R band  image before performing any analysis. 
All the results  regarding the R band  refer to this ${\rm H\alpha}$ subtracted R band image.

%% file: art_tol0957_tab_log_9v3.tex
%--------------------------------------------------------------------------------------------------
%	OBSERVING LOG TABLE
%       EXP.TIME, ZP. AIRMASS COMES FROM iraf_redIm/phot/fpc.txt TABLE
%--------------------------------------------------------------------------------------------------
\begin{table}
\centering
\caption{\label{tab:art_tol0957_tab_log} Log of observations.}
{\small
%\begin{threeparttable}
\begin{tabular}{c c c c c}
\hline
\hline
%\multicolumn{3}{|c|}{Imagen 63a65} & Imagen 60a62 & Imagen 57a59 \\
%     &      & \multicolumn{3}{c}{Exposure Time} &    \\
Date & Filter & Exp. Time & Airmass & Seeing\\
 &  & (s) &  & (\arcsec) \\
% &  & \footnotesize{(s)} &  \\
\hline
%(1)  & (2)  & (3)  & (4)\\
%\hline
%---------------------------------------------------------------------------------------------
11-Feb-2010  & U                         & 1200  & 1.08 & 0.8\\
11-Feb-2010  & B                         & 900   & 1.05 & 0.8\\
11-Feb-2010  & R                         & 900   & 1.03 & 0.6\\
11-Feb-2010  & I                         & 900   & 1.02 & 0.6\\
09-May-2010  & ${\rm H\alpha}$           & 1800  & 1.17 & 1.0\\ 
09-May-2010  & [OIII]${\rm \lambda5007}$ & 1800  & 1.10 & 1.0\\
11-Feb-2010  & y Str\"{o}mgren           & 600   & 1.01 & 0.7\\
3-Mar-2012   & J                         & 1700  & 1.01 & 0.9\\
3-Mar-2012   & H                         & 2500  & 1.08 & 0.7\\
3,4-Mar-2012 & K                         & 3570  & 1.07 & 0.7\\
%---------------------------------------------------------------------------------------------
\hline
\hline
\end{tabular}
%     {\scriptsize
%     \begin{description}  \addtolength{\itemsep}{-0.8\baselineskip} %tablenote
%       \item[] Notes: (1) Stars seem elongated, (2) h:m:s,(3) d:m:s, (4) Z, (5)  calibrated,
%       \item[]  (5) tasa de fotones capaces de ionizar el hidr\'ogeno, (6) densidad del gas de la regi\'on HII asociada al SCE y (7) radio fotom\'etrico en H$\alpha$.
       %\item[] (6) Densidad del gas de la regi\'on HII asociada al SCE y (7) radio fotom\'etrico en H$\alpha$.
       %\item[3] Densidad estelar del SCE.
       %\item[4] Masa del SCE \citep{Me05}.
%     \end{description} %tablenotes
%     } %scritpstize
%  \end{threeparttable}
} %small
\end{table}
%--------------------------------------------------------------------------------------------------

%% file: tb_tol0957_filters.tex
%--------------------------------------------------------------------------------------------------
%	FILTERS TABLE
%--------------------------------------------------------------------------------------------------
\begin{table}
\centering
\caption{\label{tab:tb_tol0957_filters} Filter characteristics. The central wavelength ($\lambda_{c}$) and FWHM are presented in units of angstroms (\AA) for optical filters and microns (${\rm\mu m}$) for Near Infrared filters (J, H and K). The relative transmission at the filter central wavelength is given in percentage (\%).}
{\small
\begin{threeparttable}
\begin{tabular}{c c c c c}
\hline
\hline
%\multicolumn{3}{|c|}{Imagen 63a65} & Imagen 60a62 & Imagen 57a59 \\
%
Filter & $\lambda_{c}$ & FWHM & ${\rm \tau ( \lambda_{c} ) }$ \\
%     &      & (s)       & (s)    & (s)    & (arcsec) \\
 %\hline
%(1)  & (2)  & (3)       & (4)  \\
\hline
U (SOI)         & 3624  &  784 &  60.68     \\
B (SOI)         & 4326  & 1269 &  73.76     \\
R (SOI)         & 6289  & 1922 &  76.97     \\
I (SOI)         & 8665  & 3914 &  95.28     \\
H$\alpha$ (CTIO 660075-4) & 6600 & 67  & 84.77 \\
${\rm [OIII]}$  (CTIO 5019)            & 5027 & 50  & 79.48 \\
y Str\"{o}mgren	& 5478  & 244 & 70.83 \\
 & & & \\
J  & 1.25       & 0.16  & 84.77 \\
H  & 1.64       & 0.29  & 95.41 \\
K  & 2.20       & 0.34  & 87.27 \\
\hline
\hline
\end{tabular}
%     {\scriptsize
%     \begin{description}  \addtolength{\itemsep}{-0.3\baselineskip} %tablenote
%       \item[] Notes: The filters central wavelength ($\lambda_{c}$) and FWHM are presented in units of angstroms (\AA) for optical filters and microns (${\rm\mu m}$) for Near Infrared filters.
       %\item[] (6) Densidad del gas de la regi\'on H II asociada al SCE y (7) radio fotom\'etrico en H$\alpha$.
       %\item[3] Densidad estelar del SCE.
       %\item[4] Masa del SCE \citep{Me05}.
%     \end{description} %tablenotes
%     } %scritpstize
  \end{threeparttable}
} %small
\end{table}
%--------------------------------------------------------------------------------------------------

%% file: tb.dif.soar.2mass.tex
%------------------------------------------------------------------------------------------------
\begin{table}
\centering
\caption{\label{tb:soar_2mass} {Difference SOAR-2MASS for field stars photometry. }}
{\small
\begin{threeparttable}
\begin{tabular}{ c c c}
\hline
\hline
{\footnotesize Filter} & {\footnotesize Mean} & {\footnotesize Median}\\
{\footnotesize } & {\footnotesize (mag)} & {\footnotesize (mag)}  \\
\hline                          
J & $-0.27$ & $-0.25$  \\ 
H & $-0.29$ & $-0.25$  \\ 
K & $-0.06$ & $-0.01$ \\ 
\hline
\hline
\end{tabular}
\end{threeparttable}
} % END SMALL
\end{table}

%% file: tol02_3v6c.tex
\section[]{PHOTOMETRY}
\label{sec:phot}

\subsection{Broad band}
\label{subsec:phot_bb}

% Tol 02 RGB IMAGE 
%---------------------------------------------------------------------------------------------
\begin{figure*}
\centering
\resizebox{0.9\textwidth}{!}{\includegraphics{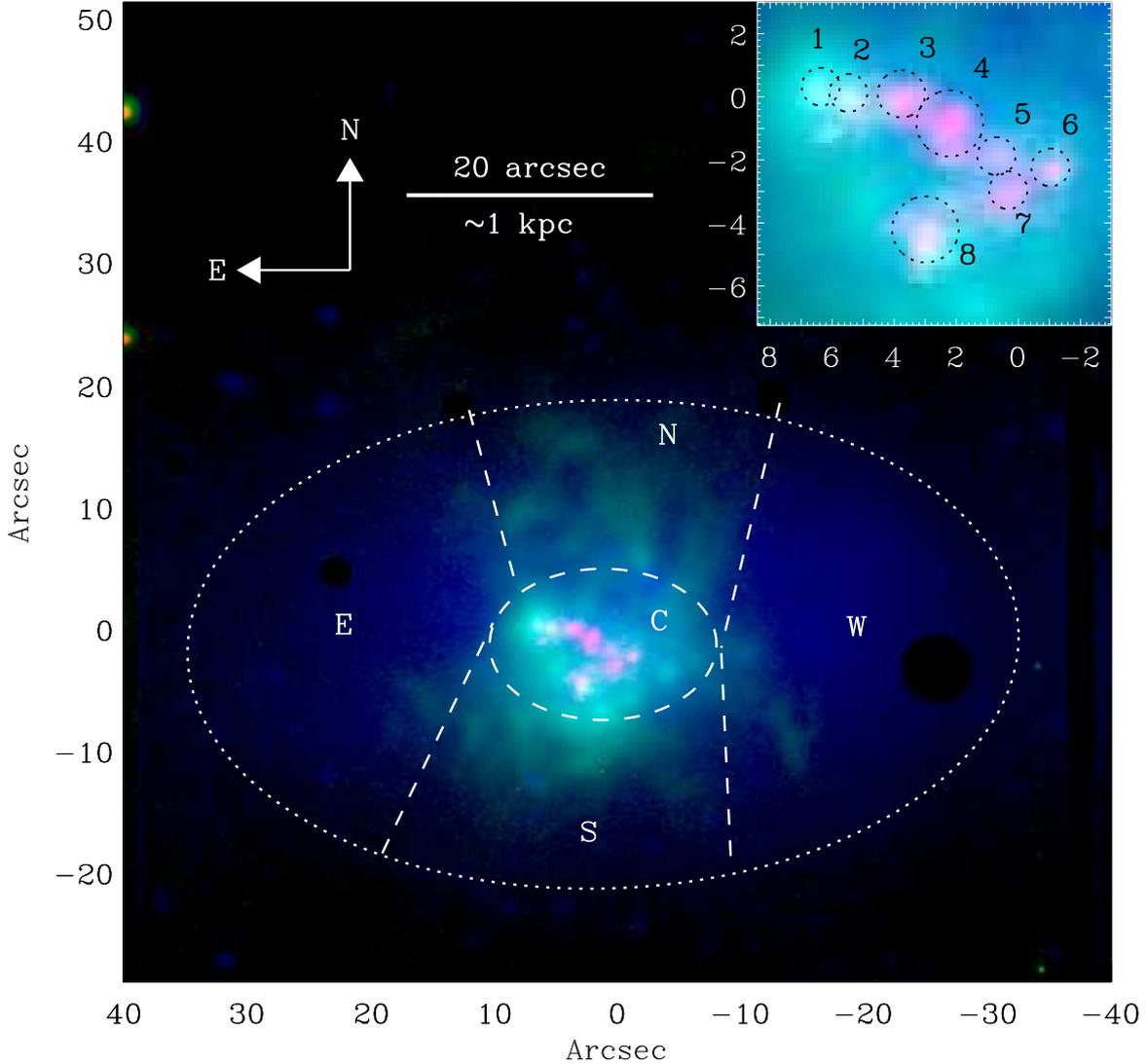}} 
\caption{ \small Optical-NIR composite image of Tol 02. J filter in red, B filter in blue and ${\rm H\alpha}$ emission in green. At the top-right corner a zoom of the galaxy centre exhibits eight stellar knots bright in the J image. The ellipse that best fits the B 25 mag ${\rm arcsec^2}$ isophote and five zones into which the galaxy has been divided for the analysis (C-Centre, N-North, W-West, S-South and E-East) are plotted; see text.}
\label{fig:tol0957rgb}
\end{figure*}
%---------------------------------------------------------------------------------------------

A composite optical-NIR image is shown in Figure \ref{fig:tol0957rgb}. It is formed by the NIR J band image in red, optical B band image in blue and ${\rm H \alpha+[NII]}$ emission image in green.
The figure shows in blue the B band surface brightness decreasing from the centre outwards, forming an elongated body.
In this figure, the  25 B  magnitude isophote is  fitted by an ellipse with a major axis ${\rm A=68}$ arcsec and a minor axis B = 40 arcsec ($A\sim3.6$ kpc and $B\sim2$ kpc at the adopted distance). 
The centre of this ellipse will be referred to as the (photometric) centre of the galaxy (R.A. 09h59m21.00s, Dec. -28d07m59.40s).
% 

%---------------------------------------------------------------------------------------------
% GALAXY INTEGRATED OPTICAL COLOURS PHOTOMETRY COMPARISON WITH THE LITERATURE
%---------------------------------------------------------------------------------------------
The optical magnitudes (U, B, R and I) of the galaxy were obtained by integrating the light collected inside the B 25 magnitude  isophote.
%
%---------------------------------------------------------------------------------------------
%       CORRECTS FOR GALACTIC EXTINCTION
%---------------------------------------------------------------------------------------------
The absolute magnitudes and luminosities of both broad and narrow band filters (assuming a distance\footnote{Calculated assuming a uniform Hubble flow with ${\rm H_0 = 73.4 \, km \,  s^{-1} \, Mpc^{-1}}$ and 826 ${\rm km \, s^{-1} } $ \citep{Mo00} for the recession velocity of the object. } of 11 Mpc) were corrected for galactic extinction using \citet{Ca89} equations and a colour excess  of E(B-V) = 0.094 mag obtained from the NASA/IPAC Extragalactic Database (NED). No internal extinction correction was applied. 
Our results -with and without galactic extinction correction- are shown in Table \ref{tab:BBlitcomp} together with those from \citet{Do99,Me00,Gi03}, hereinafter DO99, ME00 and GI03, respectively.
The table shows  a spread of ${\rm \Delta \sim 1}$ mag,  our B and R  values being the highest ones. 
Our B band magnitude (corrected from galactic extinction) is ${\rm \approx 0.13}$ mag higher than the B band magnitudes in DO99 and GI03.
This discrepancy is comparable with the ${\rm +0.36\pm0.34}$ magnitude difference found between GI03 and DO99 for the 23 galaxies they have in common.
Our R band value differs 0.02 and -0.28 magnitudes from GI00 and DO99, respectively. 
The difference between our and DO99 values can be accounted for by the different methods used   to calculate the integrated magnitude of the galaxy.
DO99 used a modelled light profile from the galaxy.  
Both GI03 and us obtained the integrated magnitudes of the galaxy  by using aperture photometry instead.
Therefore, we attribute the discrepancy  to the use of different methods to remove field stars and nearby galaxies.

The U band magnitude for this galaxy has   been published in ME00.
Our U and B extinction corrected  values are 0.52 mag greater.
This difference is primarily due to the different extinction correction performed,
since  ME00   considered internal extinction, using the Balmer decrement reported by \citet{Va92} and \citet{Wh58} extinction law.
%la fotometría de ME00 fue corregida por extinción interna, estimada utilizando e decremento de balmer y la ley de Wh58 
%
Our extinction corrected colours are in agreement with those from DO99, GI03 (${\rm B-R=0.9}$ mag) and ME00 (${\rm U-B=0.6 }$ mag). 
We did not find in the literature an I band integrated magnitude value for this galaxy.

% HANDMADE TABLE!!! the values for Tol 02 are obtained from the file
% Documents/inaoe/d_tesis/iraf_redIm/im_thesis/galaxies/g.mag.obs.txtF
%---------------------------------------------------------------------------------------------
\input{tb_tol0957_BBlitcomp}

% HANDMADE TABLE USING tb_photopt4, tb_photscc (p_imthesis.pro),
% pix.rad.zones.txt, pos.rad.hii.txt, pos.rad.scc.txt (p_art_plots.pro)
\input{tb_photopt3zones_mix}

%By looking at U-B map we decided to divide the galaxy into five zones of study: North, West, South, East and Centre. 
The galaxy was divided into five zones: North, West, South, East and Centre.
The limits were imposed by the U-B colour map morphology of the galaxy and are indicated in Figure \ref{fig:tol0957rgb}. 
% TABLE WITH INTEGRATED, ZONE AVERAGED, ZONE INTEGRATED AND SCC PHOTOMETRIC RESULTS
% THIS TABLE IS GENERATED AT P_IMTHESIS. p_imthesis.pro tb_photopt2
%---------------------------------------------------------------------------------------------
The B absolute magnitude and the optical colours of the integrated galaxy and its five zones are presented in Table  \ref{tab:photopt3_zone}. 
Column 1 is the zone's name, column 2 is the B band absolute magnitude and columns 3 to 5 are the U-B, B-R and R-I colours, respectively. The uncertainties of the parameters presented in the table are lower than ${\rm \Delta = \pm 0.005 \ mag}$ except for  U-B with a  ${\rm \Delta = \pm 0.01 \ mag}$ uncertainty for all  zones.
The Centre zone is delimited by the  22 B mag ${\rm arcsec^{-2}}$ isophote, has an elliptical shape with a major axis of ${\rm \sim 20 }$ arcsecs (${A \sim 1 }$ kpc) and a blue U-B = -0.7 mag colour. %encloses a collection of eight bright knots .
The North and South zones also show blue colours with ${\rm U-B \sim -0.5}$ mag and ${\rm U-B \sim -0.7}$ mag, respectively. 
Together, these three zones form a blue strip that separates the East and West zones, which have redder colours of ${\rm U-B \approx -0.2 \ mag}$.
% 
%---------------------------------------------------------------------------------------------
% B-R COLOUR MAP 
%---------------------------------------------------------------------------------------------
The B-R map %(Figure \ref{fig:phot_mapBR}) 
shows a homogeneous red ${\rm B-R \approx 1.2 \ mag}$ disk-like structure formed by all the zones but the Centre. 
A blue (${\rm B-R \approx 1 \ mag}$) circular structure with a $\sim 10$ arcsecs radius is observed at the centre of the galaxy, exceeding the  Centre zone.

Only discrete sources can be identified in the infrared data due to the insufficient depth. They will be discussed in Section  \ref{subsec:phot_bbphot_scc}.

%=============================================================================================
%       COMPARISON WITH MENDEZ & ESTEBAN
%=============================================================================================
\subsection{Narrow band}
\label{subsec:phot_nbphot_res}

\begin{figure}
\centering
\resizebox{0.45\textwidth}{!}{\includegraphics{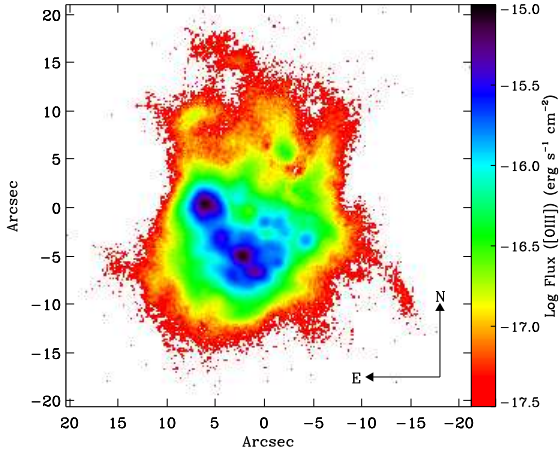}} 
\caption{ \small ${\rm [OIII]\lambda5007}$ emission line map. }
\label{fig:mapOIII}
\end{figure}

%---------------------------------------------------------------------------------------------
%        [OIII] MAP 
%---------------------------------------------------------------------------------------------
The present star-forming region  is delimited by the ${\rm H\alpha}$ emission (Figure \ref{fig:tol0957rgb}) which is located at the Centre zone of the galaxy and extends towards the North and South zones. 
Figure \ref{fig:mapOIII} shows the ${\rm [OIII]}$ emission map, which shows a similar spatial distribution
 as the ${\rm H\alpha}$ emission,
albeit on a different flux scale. 
Structures as faint as ${\rm \approx3.0 \times 10^{-18} \, erg \, s^{-1} \, cm^{-2}}$ (with more than ${\rm 3 \sigma}$ confidence) are identified in both ${\rm [OIII]\lambda 5007}$ and ${\rm H\alpha}$ images.

% [OIII]/Ha  HISTOGRAM (p_imthesis--mesh_maps)
%---------------------------------------------------------------------------------------------

% LOCATION OF THE NARROW BAND EMISSION
%---------------------------------------------------------------------------------------------
%
The integrated flux, luminosity  (corrected by galactic extinction and the contribution of the [NII]  lines and equivalent width of both ${\rm H\alpha}$ and ${\rm [OIII]\lambda5007}$  of the  star-forming region (SF) are given in Table \ref{tab:tb_hiilit}.
The [NII] contribution to the ${\rm H\alpha+[NII]}$ flux was removed using  the values obtained from \citet{Ke04} long slit spectrophotometry ${\rm [NII]\lambda6548/H\alpha = 0.009}$ and ${\rm [NII]\lambda6584/H\alpha = 0.045}$  and assuming that these values are constant across the starburst.
The equivalent widths were calculated by using the nebular and continuum light within the aperture.

% HANDMADE TABLE!!! the values for Tol 02 are obtained from the file
% Documents/inaoe/d_tesis/iraf_redIm/im_thesis/galaxies/g.lumNB.log.txt,g.NBratio.txt
% AND TABLE tb_hiinb.tex
%---------------------------------------------------------------------------------------------
%\input{tb_hiinb2_mix}
\input{tb_hiilit}

${\rm H\alpha}$ integrated flux on Tol 02 had been previously analysed in ME00 and GI03. As in
GI03 we use the R band image to determine the continuum  contribution to the ${\rm H\alpha}$ image, while a narrow band filter adjacent  to the ${\rm H\alpha}$ filter central wavelength is used in ME00. 
The three studies show a similar ${\rm H\alpha}$ luminosity; ours is 4\% and 10\% higher than those of  ME00 and GI03, respectively\footnote{The luminosity has been  corrected  to  the same (11 Mpc) adopted distance.}. 
On the other hand, the ${\rm EW(H\alpha)}$ from ME00 is $\approx 2$ times higher than those in GI03 and this work.
This difference is believed to arise from the sensitivity of the ${\rm EW(H\alpha)}$ to the continuum subtraction.
%
%---------------------------------------------------------------------------------------------

%% file: tb_tol0957_BBlitcomp.tex
%--------------------------------------------------------------------------------------------------
%	FILTERS TABLE
%--------------------------------------------------------------------------------------------------
\begin{table}
\centering
\caption{\label{tab:BBlitcomp} Optical broad band integrated magnitude comparison with the literature.}
{\small
%\begin{threeparttable}
\begin{tabular}{c c c c c}
\hline
\hline
%\multicolumn{3}{|c|}{Imagen 63a65} & Imagen 60a62 & Imagen 57a59 \\
%
U & B & R & I & Reference \\
%     &      & (s)       & (s)    & (s)    & (arcsec) \\
\hline
14.04   & 14.59 & 13.57 & 13.20 & This work (Observed) \\
13.59   & 14.20 & 13.32 & 13.05 & This work (E(B-V) corrected) \\
%        & 15.30 & 13.90 & &  \citet{Ku88} \\
        & 14.08 & 13.04 &       & DO99 \\
13.07   & 13.68 &       &       & ME00  \\
        & 14.06 & 13.34 &       & GI03  \\

\hline
\hline
\end{tabular}
%     {\scriptsize
%     \begin{description}  \addtolength{\itemsep}{-0.8\baselineskip} %tablenote
%       \item[] Notes: (1) Stars seem elongated, (2) h:m:s,(3) d:m:s, (4) Z, (5)  calibrated,
%       \item[]  (5) tasa de fotones capaces de ionizar el hidr\'ogeno, (6) densidad del gas de la regi\'on H II asociada al SCE y (7) radio fotom\'etrico en H$\alpha$.
       %\item[] (6) Densidad del gas de la regi\'on H II asociada al SCE y (7) radio fotom\'etrico en H$\alpha$.
       %\item[3] Densidad estelar del SCE.
       %\item[4] Masa del SCE \citep{Me05}.
%     \end{description} %tablenotes
%     } %scritpstize
%  \end{threeparttable}
} %small
\end{table}
%--------------------------------------------------------------------------------------------------

%% file: tb_photopt3zones_mix.tex
%------------------------------------------------------------------------------------------------
\begin{table}
\centering
\caption{\label{tab:photopt3_zone} B absolute magnitude and optical colours. ${\rm R_{c} }$ is the radius (in parsecs) of a circle that has the same are as the zone. }
{\small
\begin{threeparttable}
\begin{tabular}{ c c c c c c}
\hline
\hline
{\footnotesize ID} & {\footnotesize ${\rm R_{c} }$ } & {\footnotesize ${\rm M_B}$} & {\footnotesize ${\rm U-B}$} & {\footnotesize ${\rm B-R}$} & {\footnotesize ${\rm R-I}$}  \\
% & {\footnotesize (mag)} & {\footnotesize (mag)} & {\footnotesize (mag)} & {\footnotesize (mag)}  \\
%
\hline
Galaxy & 1250 &$-16.03$ & $-0.61$ & $0.88$ & $0.27$ \\ 
North  & 515 &$-13.37$ & $-0.53$ & $1.07$ & $0.31$ \\ 
West   & 674 &$-13.38$ & $-0.21$ & $1.19$ & $0.44$ \\
South  & 531 &$-13.35$ & $-0.71$ & $1.01$ & $0.24$ \\  
East   & 635 &$-13.24$ & $-0.28$ & $1.17$ & $0.38$ \\
Centre & 395 &$-15.54$ & $-0.69$ & $0.72$ & $0.18$ \\ 
\hline
\hline
\end{tabular}
%     {\scriptsize
%     \begin{description}  \addtolength{\itemsep}{-0.8\baselineskip} %tablenote
%       \item[]  Notes: The units are: R.A. and Declination coordinates respecto to galaxy centre in [arcsec].
%       \item[] 
%     \end{description} %tablenotes
%     } %scritpstize
\end{threeparttable}
} % END SMALL
\end{table}

%% file: tb_hiilit.tex
%------------------------------------------------------------------------------------------------
 
%------------------------------------------------------------------------------------------------
\begin{table}
\centering
\caption{\label{tab:tb_hiilit} Star-forming region ${\rm H\alpha}$ and ${\rm [OIII]\lambda5007}$ fluxes, luminosities and derived properties comparison with the literature. The units for the flux, luminosity and equivalent width are [${\rm erg \, s^{-1} \, cm^{-2}}$], [${\rm erg \, s^{-1}}$], and [\AA] respectively. }
{\small
\begin{threeparttable}
\begin{tabular}{ c c c c}
\hline
\hline
{\footnotesize ${\rm \log F_{H\alpha}}$} & {\footnotesize ${\rm \log L_{H\alpha}}$} & {\footnotesize ${\rm EW(H\alpha)}$} & {\footnotesize Reference}\\
%{\footnotesize (${\rm erg \, s^{-1} \, cm^{-2}}$)} & {\footnotesize (${\rm erg \, s^{-1}}$) } & {\footnotesize (${\rm erg \, s^{-1}}$) } & {\footnotesize (\AA)}  \\
%
\hline
 ${-11.92}$ & ${40.29}$ & ${143\pm109}$ & This work \\
 & ${40.27}$ & ${290\pm5}$ & ME00 \\
 ${-11.92}$ & ${40.24}$ & ${127}$ & GI03 \\

\hline
{\footnotesize ${\rm \log F_{[OIII]}}$} & {\footnotesize ${\rm \log L_{[OIII]}}$} & {\footnotesize ${\rm EW([OIII])}$} & {\footnotesize Reference} \\
%{\footnotesize (${\rm erg \, s^{-1} \, cm^{-2}}$)} & {\footnotesize (${\rm erg \, s^{-1}}$) } & {\footnotesize (${\rm erg \, s^{-1}}$) } & {\footnotesize (\AA)}  \\
 ${-11.97}$ &  ${40.28}$  & ${\rm 64\pm11}$ & {\footnotesize This work}\\
\hline
\hline
\end{tabular}
%     {\scriptsize
%     \begin{description}  \addtolength{\itemsep}{-0.8\baselineskip} %tablenote
%       \item[] Notes: The units are: R.A. and Declination coordinates respecto to galaxy centre in [arcsec], Flux [${\rm erg \, s^{-1} \, cm^{-2}}$], Luminosity [${\rm erg \, s^{-1}}$], Equivalent width [${\rm \AA}$].
%       \item[] 
%     \end{description} %tablenotes
%     } %scritpstize
\end{threeparttable}
} % END SMALL
\end{table}

%% file: tol02_4v6c.tex
%%%%%%%%%%%%%%%%%%%%%%%%%%%%%%%%%%%%%%%%%%%%%%%%%%%%%%%%%%%%%%%%%%%%%%%%%%%%%%%%%%%%%%%%%%%%%%
%   
%%%%%%%%%%%%%%%%%%%%%%%%%%%%%%%%%%%%%%%%%%%%%%%%%%%%%%%%%%%%%%%%%%%%%%%%%%%%%%%%%%%%%%%%%%%%%%
\section[]{ANALYSIS}
\label{sec:analysis}

%*********************************************************************************************
% BROAD BAND:
% SURFACE BRIGHTENS PROFILES AND SEPARATION BETWEEN STAR-FORMING REGION AND OLD COMPONENT
%*********************************************************************************************
\subsection{Extended Low Surface Brightness Component}
\label{sec:host}

%  Describe the galaxy low surface brightness component using surface brightness profiles
%---------------------------------------------------------------------------------------------
Most blue compact dwarf galaxies consist of two main stellar structures (\citet{Ku00,Pa96,Te97} and references therein): a young star-forming region and an older low surface brightness component (LSBC) which pervades the complete body of the galaxy. This LSBC can only be appreciated outside the area covered by the ${\rm H\alpha}$ emission, and its light dominates the galaxy surface brightness profiles (SBP) faint levels ($\mu > 24 $ mag arcsec${^{-2}}$).
In the previous sections we described the photometric properties of the star-forming regions and of the individual star clusters.  Here we use the optical broad band images to analyse the LSBC.

%=============================================================================================
\subsubsection{Surface Brightness Profiles}
\label{subsec:host_met}

% Methods used to calculate the surface brightness profiles of  a galaxy
%---------------------------------------------------------------------------------------------
There are several methods to obtain  SBP \citep{Be99,Pa96,Ca01,Mi13}.  
Most of them represent the galaxy flux through a spherically-symmetric distribution $\mu(R_{eq})$, where $ R_{eq}$ is the photometric radius corresponding to the surface brightness level $\mu$. 
Apart from being simple in performance and visualisation, a spherically-symmetric distribution allows analysis of the colour profiles change as a function of distance to the centre.

% Method I used to derive the galaxy SBP (elliptical apertures)
%---------------------------------------------------------------------------------------------
Therefore we  use concentric ellipses with fixed axis ratio to determine the  surface brightness level ${\mu(R_{eq})}$ at an equivalent radius $R_{eq} = \sqrt{ab}$, where $a$ and $b$ are the ellipse semi-axes.
The equivalent radius ranges from ${R_{eq}=0}$ to $R_{eq}=30$ arcsec with fixed steps of $\Delta R_{eq} = 1.0$ arcsec.
The concentric ellipses centre and axis ratio ($a/b = 1.7$) was set by the ellipse that best fitted the 25 mag ${\rm arcsec^{-2}}$ isophote in the B band image. 
The SBPs were computed using an increment $\Delta R_{eq} = 1$ arcsec. 

\begin{figure}
\centering
\resizebox{0.45\textwidth}{!}{\includegraphics{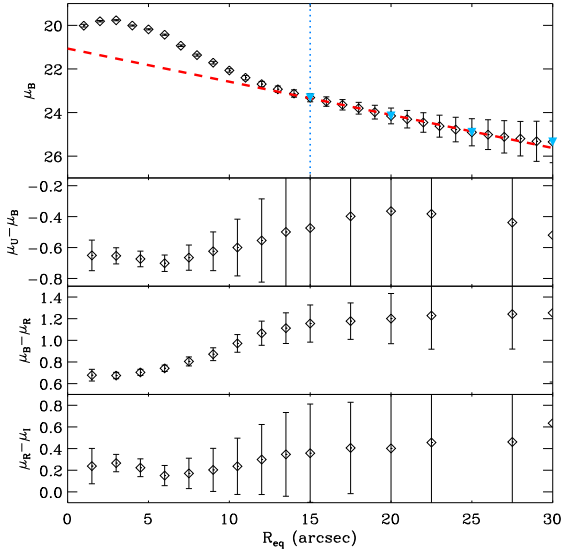}} 
\caption{ \small Surface brightness profiles (top to bottom: B band, U-B, B-R and R-I). The blue triangles indicate the position of the ellipses overplotted on the B band map, the vertical blue dotted line is the minimum ${R_{eq}}$ used for the fitting of the disc model and the red dashed line is the best fitted disc model profile.}
\label{fig:sbp_shape}
\end{figure}

% Description of the galaxy surface brightness profile and colours 
%---------------------------------------------------------------------------------------------
The  B SBP at the top panel of Figure \ref{fig:sbp_shape} shows a clear exponential decay at equivalent radii greater than 15 arcsec. 
The U, R and I SBPs (not shown) also describe an exponential decay, similar in shape to the one in  B  but shifted in magnitude. 
% Even if the exponential disc should be better sampled in the I band, those images present fringing problem at low intensities,  so we decided not to use them for the characterisation of the LSBC disk.
%
 Even if the exponential disc should be better sampled in the I band, the low intensity values uncertainties are considerably increased by  a fringing problem.

The U-B, B-R and R-I colour profiles are also shown in  Figure \ref{fig:sbp_shape}, 
uncertainties were determined as: $\sigma_{F1-F2} = \sqrt{\sigma_{F1}^2 + \sigma_{F2}^2}$. 
The colours (albeit with large errors) are redder from the centre outwards, as previously noticed by \citet{Ku88} and DO99.  
We cannot determine whether the U-B colour profile really becomes bluer from the $R_{eq}=20$ arcsec outwards or if that is just an artefact of the large uncertainties present.

% B BAND SURFACE BRIGHTNESS PROFILES
%---------------------------------------------------------------------------------------------

%=============================================================================================
\subsubsection{LSBC exponential disc model}
\label{subsec:host_disc}

% Why we used a disc to fit the SBP of the galaxy
%--------------------------------------------------------------------------------------------- 
As already mentioned, the exponential decay observed in the SBP has been found previously by \citet{Ku88} and DO99. We fitted the observed SBP for each band with an exponential disc model, $\mu(R) = \mu(0) + 1.086\left(\frac{R}{R_d} \right)$, assuming that $R = R_{eq}$. 
%
%The fitting was performed separately for each of the broad band SBP. 
To avoid contamination from the nebular emission due to recent star formation, the model was fitted to the profiles in the radial range where the ${\rm H\alpha}$ contamination is negligible (${R_{eq} = 15}$ to ${R_{eq} = 30}$ arcsec).

% Exponential disc fitting results and comparison with the literature
%---------------------------------------------------------------------------------------------
Table \ref{tb:newI_sbpfit} shows the parameters for each fitting; column 1 is the filter name, column 2 is the surface brightness magnitude in units of ${\rm mag \ arcsec^{-2}}$ at ${R = 0}$ arcsec, column 3 is the scale length in parsecs and column 4 the I band surface brightness magnitude in ${\rm mag \ arcsec^{-2}}$ at $R = 0$ arcsec scaled to match the SBP of the other filters at $R = 25$ arcsec.
The exponential disc model that best fit the B SBP is shown in Figure \ref{fig:sbp_shape} as a red-dashed line; 
the blue dotted vertical line indicates the innermost equivalent radius at which the fitting was performed (${R_{eq} = 15}$ arcsec).  

\input{tb.newI.sbpfit}

% Galaxy modeled disc colours using different apertures
%---------------------------------------------------------------------------------------------
%
The modelled disc parameters, individually determined for each band, were used to create images of the LSBC in all  the optical bands.
We also used the I band fitted disc model (scaled to match the SBP of the other filters at the radius $R = 25$ arcsec) to create images of the disc population as represented by the I band  dominant stellar population (${\rm LSBC_{I}}$).
In both cases we used two elliptical apertures to measure the photometry in the modelled LSBC and ${\rm LSBC_{I}}$, as well as in the data.
The first elliptical aperture encloses the star-forming area using an equivalent radius of $ R_{eq} = 15$ arcsec. 
The second represents  the galaxy 25 B  mag  isophote with $R_{eq} = 26$ arcsec. 

% INTEGRATED OPTICAL VALUES FOR DIFFERENT APERTURES IN THE GALAXY
% handmade table using values from phot_tol0957_278/tb.new.colorsbp
\input{tb.new3.colorsbp}

With the calculated photometry we have analysed the disc stellar population (LSBC and ${\rm LSBC_I}$) contribution to the star-forming area and overall galaxy light emission.
We  found that although the star-forming area aperture accounts for 99\% of the optical light observed in the galaxy, the disc stellar population light contribution is not negligible. 
%
%The LSBC disc stellar population emits 54\% of the total light observed inside the star-forming area aperture, .
%
Half of the light observed inside the star-forming area aperture is emitted by the disc stellar population \mbox{(LSBC = 54\%,} \mbox{${\rm LSBC_I = 45 }$\% ).}
%
%The LSBC accounts for 43\% of the U band flux and up to 69\% of the whole galaxy I band flux.
%
The contribution of the disc population to the total U band flux of the whole galaxy is  higher than 30\%, independently of the model used \mbox{(LSBC = 43\%} and \mbox{${\rm LSBC_I = 33}$\%)}, and 68\% for the I band.
%
%This important flux contribution from the disc stellar population \mbox{(LSBC} and \mbox{${\rm LSBC_I}$)} translates into a reddening of the colours in the star-formation zone \mbox{(SF} and \mbox{${\rm SF_I}$, respecively)}. 

This important flux contribution from the disc stellar population affects both the colours of the galaxy integrated photometry and of the galaxy star-forming area. 
The colours of the two apertures: star-forming area and galaxy are presented in Table \ref{tab:colorsbp}, where column 2 is the absolute B magnitude and columns 3 to 5 are the U-B, B-R and R-I colours respectively. 
%
%The table presents the colours of the star-forming area, observed (${\rm SF_{Obs}}$) and
% minus the LSBC light contribution.
For the galaxy aperture we tabulate the observed colours of the galaxy as well as those for the modelled disc populations \mbox{(LSBC} and \mbox{${\rm LSBC_I }$).}
As well, for  the  star-forming region we tabulate the observed values (${\rm SF_{obs}}$) and the 
values after subtracting the disc stellar population contribution \mbox{(SF} and \mbox{${\rm SF_I}$)}.
The net effect of the disc population is a reddening of the star-forming area colours by ${\rm \Delta_{ U-B}\approx0.15}$, ${\rm \Delta_{B-R}\approx 0.32}$ and ${\rm \Delta_{R-I} \approx 0.14}$ magnitudes, more pronounced at the U-B colour for the LSBC than for the ${\rm LSBC_I}$ model, while for R-I is the other way round. 
We find that by identifying the light contribution of the disc population to the galaxy not only  we can characterise  the old stellar component, but it also helps to constrain the colours of the stellar populations in the star-forming region.
%

%*********************************************************************************************
% NARROW BAND:
% STAR FORMING REGION PROPERTIES
%*********************************************************************************************

%=============================================================================================
\subsection{Star-forming region}

To describe the nebular properties of the galaxy in two dimensions, the narrow band images were divided  in square cells of 1.5 arcseconds side and the light within each cell was integrated. 
Only those cells with value above ${\rm 3\sigma}$ were considered for the photometry.
The cells size was  chosen to be greater than the seeing (${\rm \sim 1}$ \, arcsec).  
Given the physical size of the cells (${\rm \approx 80}$ pc at the adopted distance of the galaxy), each cell might contain one \mbox{H II} region, although most \mbox{H II} regions are larger in size (\mbox{H II} regions sizes can reach up to a few hundred parsecs).

% EQUIVALENT WIDTHS
%---------------------------------------------------------------------------------------------
%With the narrow band pure emission line and continuum images (Section \ref{sec:obs:puremilin}) ${\rm H\alpha}$ and [OIII] equivalent width maps were created. 

The ${\rm H\alpha}$ and [OIII] equivalent width (EW) maps were created by using the narrow
 band emission line and continuum images (Section \ref{sec:obs:puremilin}).
Both maps show a linear shape structure with high equivalent width values
 (${\rm EW > 200}$  \AA) that corresponds to the position of the bright nebular emission peaks observed in Figure \ref{fig:mapOIII}.
%

% Ha/Hb MESH MAP BALMER DECREMENT
%---------------------------------------------------------------------------------------------

% Ha/Hb MESH MAP (p_imthesis--mesh_maps)
\begin{figure}
\centering
\resizebox{0.45\textwidth}{!}{\includegraphics{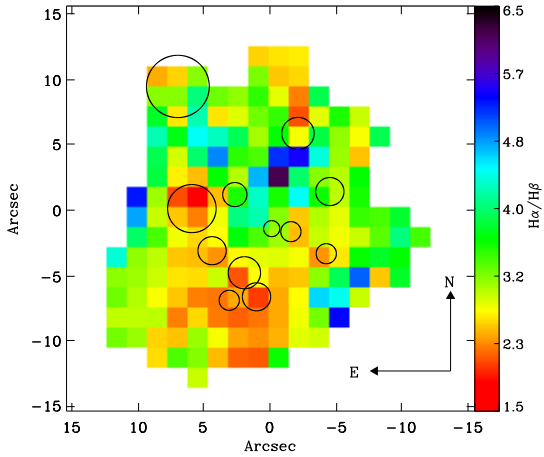}}  
\caption{ \small ${\rm H\alpha / H\beta}$ Balmer decrement map, low values indicate low nebular extinction. The 
nebular emission peaks observed in Figure \ref{fig:mapOIII} are indicated.} %It was built using \citet{La07} ${\rm H\beta}$ and our  ${\rm H\alpha}$ line image.}
\label{fig:mesh_hahb}
\end{figure}

%We used an  ${\rm H\beta}$ emission line image of Tol 02 from  \citet{La07} to create the ${\rm H \alpha / H \beta}$ Balmer decrement map displayed Figure \ref{fig:mesh_hahb}. 

The dust extinction distribution in the star-forming region of the galaxy is described by the ${\rm H\alpha / H\beta}$ map (Figure \ref{fig:mesh_hahb}), 
which was created using an ${\rm H\beta}$ emission line image from \citet{La07}.
The map suggests that the cells with the lowest extinction values are located on top and to the South of the chain of \mbox{H II} regions and the cells with the highest extinction values are grouped North-West.
The ${\rm  H\alpha / H\beta}$ theoretical value of an extinction free ISM is 2.86 \citep{Os06}, and it increases exponentially with the amount of dust. 
The ${\rm H\alpha / H\beta}$ map encompass values from 1.5 to 6.5.  Although the ${\rm H\alpha / H\beta}$ calculated value is lower than 2.86 for ${\sim 40}$\% of the cells, this percentage reduces to 16\% when the uncertainties are taken into consideration (${\rm H\alpha / H\beta + \sigma_{H \alpha / H\beta} < 2.86}$).
Still, the ${ \rm H\alpha / H\beta}$ extremely low values might be a product of a calibration mismatch between images, originated by a shift in the absolute photometry due to differences in atmospheric conditions, calibration standard stars, image continuum subtraction, etc.
If there was indeed a mismatch in the calibration between the different programmes when H$\alpha$ and H$\beta$ were obtained, such as ${\rm H\alpha/H\beta < 2.8}$ it would affect all the cells. However, when combined with our ${\rm 3\sigma}$ cutoff such a mismatch would produce an underestimate of the flux of ${\rm H\beta}$  (the weakest line) in the fainter regions which in turn would increase the Balmer decrement.

% [OIII]/Ha  HISTOGRAM (p_imthesis--mesh_maps)
%---------------------------------------------------------------------------------------------
\begin{figure}
\centering
\resizebox{0.45\textwidth}{!}{\includegraphics{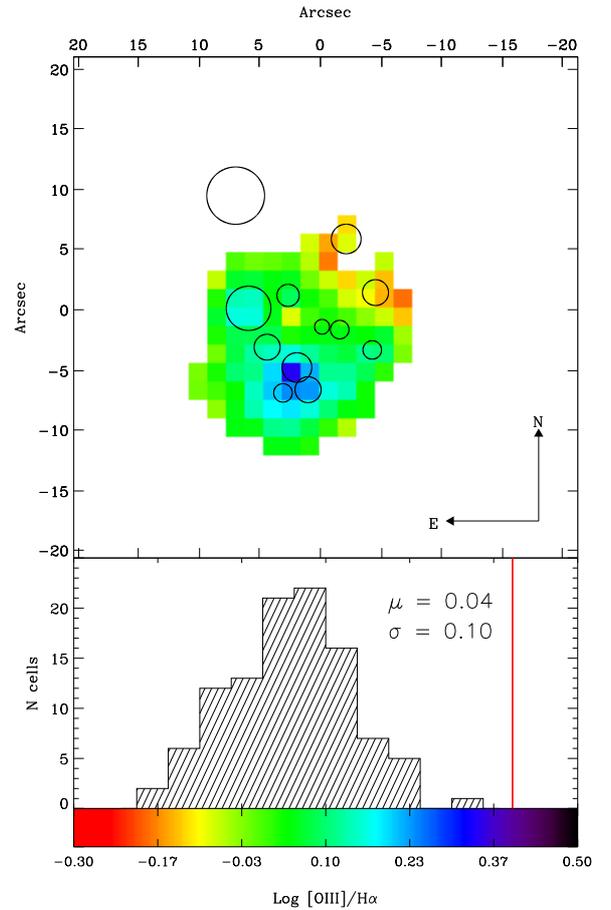}} 
\caption{ \small  ${\rm \log \, [OIII]/H\alpha}$ distribution across the galaxy with the position of nebular emission peaks observed in Figure \ref{fig:mapOIII} overplotted.
 The histogram of the distribution, as well
 as the mean (${\rm \mu}$) and standard deviation (${\rm \sigma}$) values are also presented.
 The red line at ${\rm \log \, [OIII] / H\alpha = 0.4}$ represents \citet{Ke01} star-forming regions diagnostic  threshold  assuming ${\rm A_{V} = 0}$ and ${\rm  \log \, [NII]\lambda 6584/H\alpha} = -1.3$ (see text).}
\label{fig:mesh_histOIIIHa}
\end{figure}

% ionisATION HARDNESS MAPS
%---------------------------------------------------------------------------------------------
The ionisation hardness map, obtained through the ${\rm \log \, [OIII]}$ and ${\rm H \beta}$ 
 images ratio, identify the area where the ionising photons are concentrated.
 These two emission lines are close enough in wavelength for the ratio to be insensitive to
 extinction.  
The ${\rm \log \, [OIII]/H \alpha}$ ratio can be assumed as an extinction dependent analogous
 of the  ${\rm \log \, [OIII]/H \beta}$ ratio.
Both ratios map display the same morphology, 
 indicating that the extinction is not severe and  quite uniform.
 The ${\rm \log \, [OIII]/H \alpha}$ map is shown in Figure \ref{fig:mesh_histOIIIHa}. 
A low excitation area is observed towards the NW, suggesting that the mean energy of the 
 ionising photons produced by the stellar sources in this area is lower than in the rest of 
 the map.
In emission line maps, this area is attributed to what seems to be diffuse emission. 
The highest values of ionisation hardness are observed at the position of the southernmost
 nebular peak.

The distribution of  ${\rm \log \, [OIII]/H \alpha}$ is presented at the bottom of 
 Figure \ref{fig:mesh_histOIIIHa}.
We have used the intensity of the emission lines from the star-forming region \citep[from][]{Te91}, 
 to calculate the nebular photoionisation threshold line \citep{Ke01}.
 This was calculated assuming that the interstellar medium in the galaxy is extinction-free 
 (${\rm H\alpha = 2.86 \, H \beta}$) and that its [NII] to ${\rm H \alpha}$ ratio is uniform, 
 with an estimated value of ${\rm \log \, [NII] \lambda 6584/ H \alpha = -1.3}$.
We notice that all the ${\rm \log \, [OIII]/H \alpha}$ value of the cells lie to the left 
 of the threshold line value, with a mean value of ${\rm \log \, [OIII]/H \alpha = 0.04}$, 
thus indicating that the interstellar medium is indeed being photoionised by  young massive stars.

%*********************************************************************************************
% DISCRETE COMPONENTS
%*********************************************************************************************
%\subsection{ Discrete components }
%\label{subsec:disccrete}

%*********************************************************************************************
% BROAD BAND:
% STELLAR CLUSTER COMPLEXES
%*********************************************************************************************
\subsubsection{ Star Cluster Complexes }
\label{subsec:phot_bbphot_scc}

% Maps with all three bands
%---------------------------------------------------------------------------------------------
\begin{figure}
\centering
\begin{tabular}{c}
% TOL 0957-278
\subfigure[Optical B]{
\resizebox{0.5\textwidth}{!}{\label{fig:scchii_position_B} \includegraphics{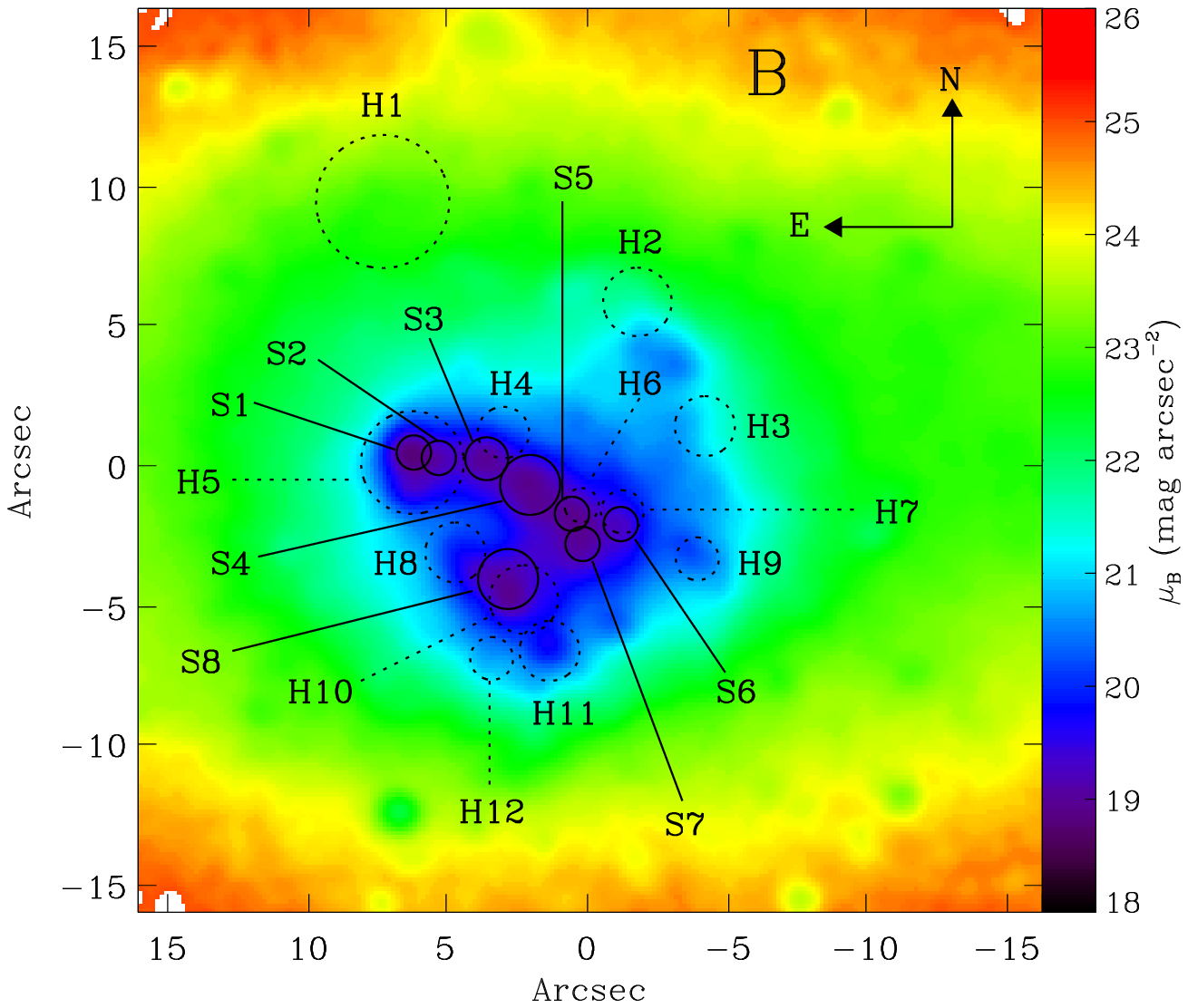}} } \\
\subfigure[NIR J]{
\resizebox{0.5\textwidth}{!}{\label{fig:scchii_position_J}\includegraphics{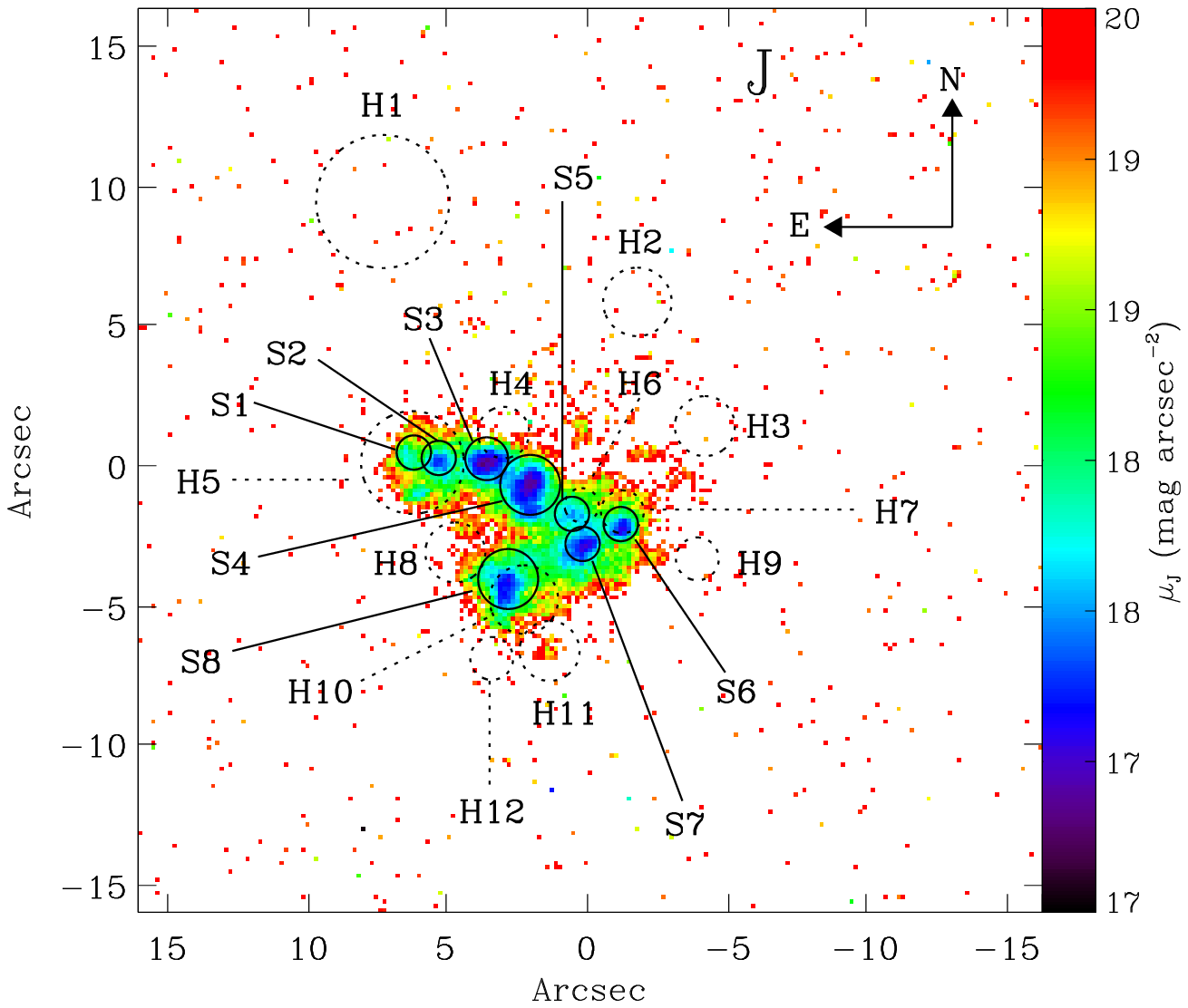}}} \\
\subfigure[${\rm H\alpha+[NII]}$ emission]{
\resizebox{0.5\textwidth}{!}{\label{fig:scchii_position_Ha}\includegraphics{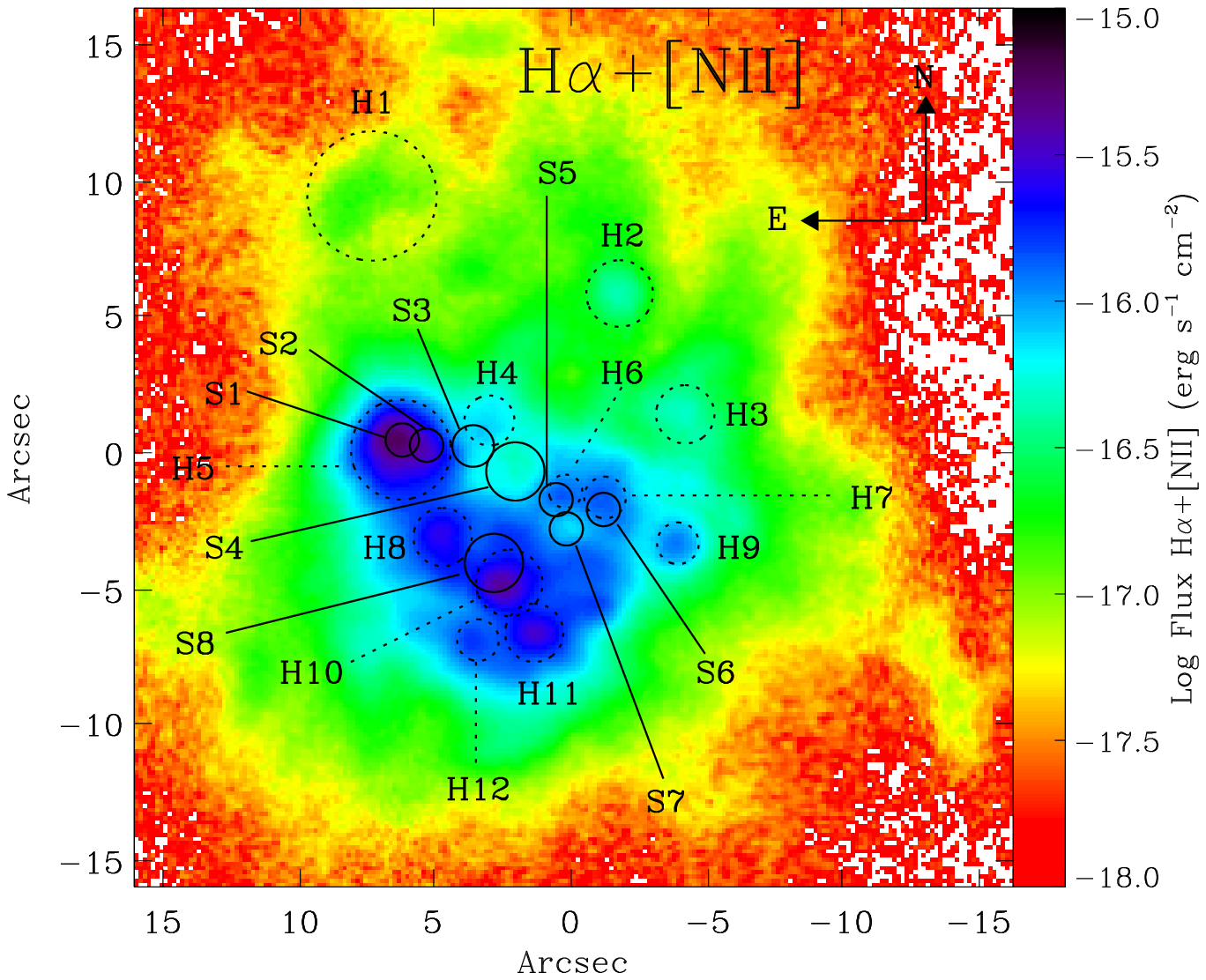}}} \\
\end{tabular}
%\caption[] {\color{red} {\bf {\small Position of the stellar cluster complexes (S) and H II regions (H) in the optical B,  NIR J, and  ${\rm  H\alpha}$+[NII] and ${\rm [OIII]\lambda5007}$  maps.}  }}
\caption[] {\small Position of the stellar cluster complexes (S) and H II regions (H) in the optical B,  NIR J, and  ${\rm  H\alpha}$+[NII]  maps.}
\label{fig:scchii_position}
\end{figure}

% 8 IDENTIFIED STAR CLUSTERS CLUMPS (J MAP WITH CLUSTERS POSITION)
%---------------------------------------------------------------------------------------------
A relevant feature seen in  Figure \ref{fig:tol0957rgb} is the eight bright knots at the Centre zone of the galaxy in both NIR and  optical images, which can be clearly identified by eye (see the insert of Figure \ref{fig:tol0957rgb}).
These stellar knots are aligned in a ``>'' shape structure. 
Although this chain of knots has been previously identified in optical bands \citep{Do99}, this is the first time that their photometry is analysed.
All of the knots are outlined by circular shapes with radii in the range of 30 to 60 pc.
Since typical star cluster sizes vary from 1 to 20 pc, each of these knots might contain more than one star cluster.
Therefore, these knots are interpreted as star cluster complexes (SCCs).
For a given SCC, the aperture size depends on the  seeing, the curve of growth of its light profile and its proximity to the nearest neighbour.
As a first approximation, the SCC centre is set as the position of the local maximum 
 in the B band surface brightness map. 
The  curve of growth rarely decreases due
to the bright background and proximity to  neighbours.
Therefore, we 
analyse the  surface brightness isocontours to delimit the
knots using a uniform increment of ${\rm \Delta = 0.1 \ mag}$. The isocontours
 enclosing each knot centre are fitted by  
circles. The adopted radius for each object is that of the
circle  fitted to the outermost isocontour without overlapping with another object, unless the overlapping was needed
in order to reach a radius of at least the FWHM size.
Figure \ref{fig:scchii_position} presents the SCCs apertures over the optical B, NIR J and nebular ${\rm H\alpha+[NII]}$ maps.

To calculate their photometry, the local background emission which was determined as the median value of the light within a concentric ring around the aperture \citep{Ad10a,La11,Al15} was subtracted from 
the light within the aperture. 
To avoid overestimating the level of the local background due to the neighbouring SCC, only pixels that are not shared with  other SCC regions were used to estimate the local background emission.
The SCCs position, aperture radius and optical and NIR photometry are presented in Table \ref{tab:photopt_scc}, where  column 1 is the SCC ID number, columns 2 and 3 are the right ascension and declination 
distance from the centre  in arcsecs respectively, column 4 is the radius in parsecs of a circle that has the same area as the region, column 5 and 6 are the B and J band absolute magnitudes and columns 7 to 11 are the U-B, B-R, R-I, I-J and J-H colours, respectively. The K band photometry is not reported because for all of the SCCs we could not determine accurately the local background level, since up to ${\rm \sim 40}$\% of the pixels inside the background aperture annulus had flux values below ${\rm 3 \sigma}$ level.
% 

%In it column 1 is the B band absolute magnitude, column 2 to 4 are the U-B, B-R and R-I colours, column 5 is the J band absolute magnitude and columns 6 and 7 are the J-H and H-K colours. 

%\ojo\ los labels de Tabla 4 debe decir: Integrated ... y de Tabla 5: SCC ...

% HANDMADE TABLE BUILT USING tb_photscc CREATED AT P_IMTHESIS
%---------------------------------------------------------------------------------------------
\input{tb_scc_phot}

SCCs \#1 and \#2 are the bluest and are located at the eastern tip of the chain of SCCs.
These SCCs have a radius of ${R=32}$ pc and overlap each other.
It is worth noticing that even though SCCs \#1 and \#2 have similar optical colours, SCC \#1 has different NIR colours than SCC \#2 by ${\rm \Delta(I-J)=-0.6}$ mag and ${\rm \Delta(J-H)=-0.8 }$ mag. These differences in colour will be essential when analysing dominant stellar population of the SCCs, as  discussed  in Section \ref{fitbb_scc}.

Another interesting SCC is \#4, which is located in the centre of the upper line of knots. 
It is the brightest knot in both optical and NIR bands (${\rm M_{B} = -11.6 \ mag \ and \ M_{J} = -13.2 \ mag}$) 
and its colours are similar to the mean values of the knots.
The U-B colour of the complexes seems to become redder towards the west, although no clear colour gradient is observed. 
As for the other colours (B-R, R-I, I-J and J-H), the SCCs do not show any tendency or pattern at all.
The SCCs optical broad band photometry uncertainties are typically 5\%,
the highest value is 12\% and corresponds to SCC \# 6.
The NIR bands uncertainties are higher, around 14\% for the magnitudes and up to 18\% for 
the NIR colour of SCC \#5.

%*********************************************************************************************
% NARROW BAND:
% H II REGIONS
%*********************************************************************************************
\subsubsection{ \mbox{H II} regions }
\label{subsec:phot_bbphot_hii}

% 12 IDENTIFIED \mbox{H II} REGIONS PROPERTIES
%---------------------------------------------------------------------------------------------
Identifying and setting boundaries for \mbox{H II} regions is not an easy task \citep[see, e.g.][]{Al15}. %(see, e.g. \citet{Al15}) 
${\rm H\alpha}$ images show \mbox{H II} regions as bright spheroids enclosed by a diffuse and irregular extended emission whose boundaries might overlap with those of a neighbouring \mbox{H II} region.
Even sophisticated software packages designed to find local maxima and minima in images, such as SExtractor \citep{Be96} and FOCAS \citep{Va82}, are likely to fail 
since they tend to confuse \mbox{H II} regions with diffuse emission and detect elongated objects 
which might be thinner than the image resolution.
%with one length smaller than the image seeing resolution limit. \ojo\ esto ultimo no se entiende \ojo\

 On the contrary, the identification and border delimitation of \mbox{H II} regions through careful eye inspection has been a widely used method and has proven to be effective albeit time consuming. 
Thus, to identify and delimit the \mbox{H II} regions in the ${\rm H\alpha}$ image of this galaxy,  we have chosen this method. % over the ones mentioned before. 
We were able to identify twelve \mbox{H II} regions. 
Following the same procedure as  for the  SCCs, the objects were identified  by eye and the position of the maximum flux in each knot was set as first approximation of the object centre. The curve of growth  and the flux increment for each object  was calculated. 
If at some radius the flux increment began decreasing, then
the radii at the minimum flux value was set as the maximum  radii for the object.
Otherwise the maximum radius was set free.
The isocontours of the ${\rm H\alpha}$ images were drawn using uniform flux increments ${\rm \Delta = 10^{-18} \ erg \ s^{-1} \ cm^{-2} }$ \AA${\rm ^{-1}}$  and  a circle was fitted to the isophotes until it reached the maximum radius value set by the curve of growth or when the isophote starts engulfing other object.
%Just like for the identification of the SCCs, the objects were enclosed inside circular apertures with a radius dependent on the growth curve of the object flux and its proximity to a companion.  
%
All \mbox{H II} regions in the galaxy are located at the Centre and North zones. 
The circular apertures used to calculate the photometry were shown with dotted lines in Figure \ref{fig:scchii_position}. 
The five brightest \mbox{H II} regions (5, 8, 10, 11 and 12) correspond to the emission peaks observed in the ${\rm H\alpha}$ and ${\rm [OIII]}$ maps.
They are aligned from Centre to South and show a shift from the position of the 
SCCs (dotted lines). Only the \mbox{H II} region \#5 engulfs two SCCs (${\rm \#1}$ and ${\rm \#2}$).
%These regions are delimited by dotted lines in Figure \ref{fig:logOIIIhii},
%where it is observed that only the \mbox{H II} region \#5 engulfs two SCCs (${\rm \#1}$ and ${\rm \#2}$).

% \mbox{H II} REGIONS PHOTOMETRY
%---------------------------------------------------------------------------------------------
The photometry of the \mbox{H II} regions was calculated with the same method as for the SCCs. 
%The position, radius (${\rm R_{H II}}$) and narrow band photometry results of the \mbox{H II} 
%regions are reported in Table \ref{tab:tb_hiinb2_mix}.
%
Table \ref{tab:tb_hiinb2_mix} presents the narrow band photometry results of the \mbox{H II} 
 regions. Column 1 is the H II region ID, columns 2 and 3 are the right ascension and declination 
 distance from the centre in arcsecs respectively, column 4 is the H II region radius in parsecs, 
 columns 5 and 6 are the observed ${\rm H\alpha}$ and ${\rm [OIII]}$ flux in ${\rm  ergs \ s^{-1} cm^{-2}}$,
 columns 7 and 8 are the extinction corrected luminosity in ${\rm  ergs \ s^{-1}}$ and columns
 9 and 10 are the equivalent widths in \AA.
For 6 H II regions we could also measure the optical broad band photometry,
 which is presented in Table \ref{tab:tb_hii_color} where column 1 is the H II region ID, 
 column 2 the B band absolute magnitude and columns 3 to 5 are the U-B, B-R and R-I colours,
 respectively.

% HANDMADE TABLE!!! the values for Tol 02 are obtained from the file
% Documents/inaoe/d_tesis/iraf_redIm/im_thesis/galaxies/g.lumNB.log.txt,g.NBratio.txt
% AND TABLE tb_hiinb.tex
%---------------------------------------------------------------------------------------------
\input{tb_hiinb2_mix}

% H II regions optical photometry values
\input{mnras.tb.hii.color}

Important parameters of the \mbox{H II} regions derived from the narrow band luminosities are 
the ${\rm \log \,  [OIII] / H\alpha}$, the star formation rate (SFR), the equivalent number 
 of O7V ionising stars (${\rm N_{\star} O7 V}$) and the Str\"{o}mgren radius (${\rm R_{Str}}$). 
These are described in the following paragraphs and their values are presented in Table 
 \ref{tab:hiideriv}, where column 1 is the ID of the \mbox{H II} 
 region, column 2 is the ${\rm [OIII]\lambda 5007/H\alpha }$ ratio in logarithm, column 3 is the 
SFR per surface area (${\rm \Sigma_{SFR} }$) in units of 
 ${\rm M_{\sun} \, yr^{-1} \, kpc^{-2}}$, column 4 is the number of main sequence ionising 
 stars and column 5 is the Str\"{o}mgren radius in parsecs.
%
%Another important parameter derived from the narrow band photometry is the star formation rate (SFR), an indicator of the amount of gas transformed into stars at a time interval.
The SFR is an indicator of the amount of gas transformed into stars at a time interval.
We derived it using Calzetti's rendition \citep{Ca10} ${\rm SFR = 8.3\times10^{-42} L_{H\alpha}}$ , 
which assumes a Salpeter IMF with stellar mass range ${\rm 0.1 \, to \, 100 \, M_{\sun}}$. 
Since \mbox{H II} regions are short-lived objects, it is assumed that they formed 
 instantaneously, and the specific SFR per unit area 
 (${\rm \Sigma_{SFR}  = SFR / [\pi \, R_{H II}^2] }$) can be interpreted as an indicator 
 of the surface density of massive stars.
% 
% density of massive stars. 
%the SFR is divided by the \mbox{H II} region projected area  
%(${\rm \Sigma_{SFR}  = SFR / (\pi \, R_{H II}^2) }$) it can be an indicator of the surface 
% density of massive stars. 
%This was divided by the aperture area in ${\rm kpc^{2}}$ to determine the ${\rm \Sigma_{SFR}}$.
%
The number\footnote{This is a lower limit since we are assuming no escaping photons.} of massive stars needed to ionise the observed \mbox{H II} emission 
(assuming no escaping photons) can be described by the ${\rm H\alpha}$ luminosity as 
${\rm N_{\star}O7V = Q(H^0) \times 10^{-48}}$, 
where ${\rm Q(H^0) = 7.31 \times 10^{11} L_{H \alpha} \ s^{-1}}$ is the ionising photons rate.

Assuming a homogeneous ISM, the Str\"{o}mgren sphere \citep{Os06} defines the volume in which all of the 
ionising photons emitted by the ionising stellar source are absorbed and re-emitted by the ISM and where the hydrogen is fully ionised. 
The Str\"{o}mgren radius of the identified H II regions is given by
${\rm R_{Str} = \left[ \left( \frac{3}{4\pi} \frac{Q(H^0)}{n_{e}^2 \alpha_{\beta}} \right)  \right]^{1/3} \ cm} $, 
 where ${\rm \alpha_{\beta} = 2.59 \times 10^{-13} \ cm^3 \ s^{-1}}$ is the H I recombination
 coefficient and ${\rm n_{e} = 100 \ cm^{-3}}$ the electron density of the nebula calculated 
 through the sulphur lines ratio (${\rm [SII] \lambda 6717 / \lambda 6731 }$) from \citet{Ke04}.

% [OIII]/Ha COMPARISON WITH KERIGH
%---------------------------------------------------------------------------------------------
Tol 02 \mbox{H II} regions are all brighter in ${\rm [OIII]\lambda5007}$ than in ${\rm H\alpha}$, 
 have a luminosity of ${\rm L_{H\alpha} > 10^{37} \, erg \, s^{-1}}$, 
 ${\rm EW(H\alpha) > 100}$ \AA \ and U-B colours bluer than ${\rm -0.7}$ mag.
The ${\rm \log \, [OIII]\lambda 5007/H\alpha }$ values of the H II regions range from 0 to 0.5, 
 typical values for \mbox{H II} galaxies \citep[e.g.][]{Ke04} %(e.g.~\citet{Ke04}). 
These authors also obtained the ${\rm \log \, [OIII]\lambda 5007/H\alpha }$ values 
for the regions \#5 and \#11 as 0.05 and 0.04, respectively.
These values are lower than the ones we found, but are similar to the median value of the 
cells (${\rm \log \, [OIII]\lambda 5007/H\alpha }$ = 0.04, Figure \ref{fig:mesh_histOIIIHa}).
It is worth noticing that the ${\rm \log \, [OIII]\lambda 5007/H\alpha }$ values of the 
 \mbox{H II} regions is greater than the values from the cells of the mesh.
The \mbox{H II} regions with the highest values (\#10, \#11 and \#13) are located south
 and coincide with the highest peak of the  ${\rm \log \, [OIII]\lambda 5007/H\alpha }$ map.
The four brightest \mbox{H II} regions (aligned from the centre to the southwest) are the
 ones with the highest ${\rm \Sigma_{SFR}}$ values, and the brightest \mbox{H II} 
 region (\#5) is the one with the highest value of all (${\rm \log \Sigma_{SFR} = -3.8 \ M_{\sun} \, yr^{-1} \, kpc^{-2}}$), 
 two orders of magnitude higher than the median (${ \rm \log \Sigma_{SFR} = -5.2 \ M_{\sun} \, yr^{-1} \, kpc^{-2}}$). 
As well, \mbox{H II} region \#5 can be associated to $\sim 100$ massive ionising stars,
 followed by \mbox{H II} region \#10 with 23 stars.
Only these four \mbox{H II} regions are associated to more than 10 massive stars, while the rest could be photoionised by  less than four.
These regions are also those for which the difference between the Str\"{o}mgren and ${\rm  H\alpha}$ radii is the smallest, albeit the ${\rm R_{Str} / R_{H\alpha} < 7}$ for all of them.

% tabla creada en p_art_plots.pro

\input{tab.hiideriv2}

% COMPARISON WITH MENDEZ & ESTEBAN
%---------------------------------------------------------------------------------------------
%\subsubsection{\citet{Me00} \mbox{H II} regions}
%\label{subsec:phot6_nbphot_res}

%The comparison with ME00 is less direct. 
%
%They provide the luminosity of twelve \mbox{H II} regions in the galaxy. 
%
Eight of the \mbox{H II} regions identified in this work were previously studied by ME00 using ${\rm H\alpha}$, U, B and V images taken with the 2.6 m Nordic Optical Telescope (NOT) in La Palma. 
To identify and analyse the \mbox{H II} regions they used FOCAS, which is a set of programs primarily designed to identify small faint objects and segregate them in stars, galaxies or noise. 
FOCAS identifies local maxima in images and encloses them using irregular areas built after analysing the image isocontours.
Unfortunately, FOCAS is no longer supported by the NOAO-IRAF community, which makes it impossible for the authors to install the package for analysing our images. 
Therefore, in order to achieve a comparison as similar as possible to ME00, we have recalculated the luminosity values of our identified \mbox{H II} regions by using irregular apertures delineated by the image isocontours. Since the area covered by the \mbox{H II} regions is not shown in ME00, the isocontours used were chosen so that the total luminosity of the \mbox{H II} region resembled the value provided by ME00. 
The ${\rm H\alpha}$ luminosity was subtracted from local background emission and corrected for intrinsic extinction\footnote{The intrinsic extinction was estimated using the same approach as ME00 applying \citet{Wh58} extinction law and \citet{Va92} ${\rm H\alpha / H\beta}$ value}, assuming the same distance to the galaxy as ME00 for comparison.
With this approach, our ${\rm H\alpha }$ luminosities are 30\% higher than those  by ME00, 
which is not surprising since our images are deeper than theirs.
The luminosity difference can also be severely affected by variations in the local background emission subtracted (due to a difference in the \mbox{H II} region shape).

%% file: tb.newI.sbpfit.tex
%------------------------------------------------------------------------------------------------
\begin{table}
\centering
\caption{\label{tb:newI_sbpfit} Parameter of the fitted exponential disk model.}
{\small
\begin{threeparttable}
\begin{tabular}{ c c c c}
\hline
\hline
{\footnotesize Filter} & {\footnotesize ${\rm \mu(0)}$} & {\footnotesize ${\rm R_{d}}$}  & {\footnotesize ${\rm \mu _{I}(0)}$} \\
{\footnotesize } & {\footnotesize ${\rm mag \, \, arcsec^2}$} & {\footnotesize (pc)} & {\footnotesize ${\rm mag \, \, arcsec^2}$} \\
\hline                          % effective radius
U & $20.66$ & $98$  & $21.24$ \\  % $82$ & 
B & $21.06$ & $97$  & $21.67$ \\ % $87$ & 
R & $20.02$ & $103$ & $20.43$ \\ % $103$ & 
I & $19.88$ & $114$ & $19.88$ \\ % $114$ &
\hline
\hline
\end{tabular}
\end{threeparttable}
} % END SMALL
\end{table}

%% file: tb.new3.colorsbp.tex
%------------------------------------------------------------------------------------------------
\begin{table*}
\centering
\caption{\label{tab:colorsbp} Absolute B band magnitude and colours on each different aperture.}
{\small
\begin{threeparttable}
\begin{tabular}{ c c c c c c}
\hline
\hline
{\footnotesize Aperture} & {\footnotesize Object} & {\footnotesize ${\rm M_B}$} & {\footnotesize ${\rm U-B}$} & {\footnotesize ${\rm B-R}$} & {\footnotesize ${\rm R-I}$} \\
\hline
Star-forming       &   ${\rm SF_{Obs} }$    & $-16.07 \pm 0.03$ & $-0.62 \pm 0.07$ & $0.89 \pm 0.03$ & $0.29 \pm 0.10$ \\ 
Star-forming       &   SF                   & $-15.29 \pm 0.05$ & $-0.79 \pm 0.12$ & $0.55 \pm 0.07$ & $0.20 \pm 0.31$ \\ 
Star-forming       &   ${\rm SF_I}$         & $-15.54 \pm 0.04$ & $-0.72 \pm 0.10$ & $0.60 \pm 0.06$ & $-0.10 \pm 0.30$ \\ 
25 B mag           &   LSBC                 & $-15.36 \pm 0.00$ & $-0.42 \pm 0.00$ & $1.15 \pm 0.00$ & $0.34 \pm 0.00$ \\ 
25 B mag           &   ${\rm LSBC_I}$       & $-15.06 \pm 0.00$ & $-0.43 \pm 0.00$ & $1.24 \pm 0.00$ & $0.55 \pm 0.00$ \\ 
25 B mag           &   Galaxy               & $-16.08 \pm 0.03$ & $-0.61 \pm 0.07$ & $0.90 \pm 0.03$ & $0.30 \pm 0.10$ \\ 
\hline
\hline
\end{tabular}
\end{threeparttable}
} % END SMALL
\end{table*}

%% file: tb_scc_phot.tex
%------------------------------------------------------------------------------------------------
\begin{table*}
\centering
\caption{\label{tab:photopt_scc} SCCs Broad band optical absolute magnitudes and colours. The SCCs displacement respect to the galaxy centre is given in arcseconds and the SCCs aperture redius is given in parsecs.}
{\small
\begin{threeparttable}
\begin{tabular}{ c c c c c c c c c c c}
\hline
\hline
{\footnotesize SCC}& {\footnotesize ${\rm \Delta R.A.}$} & {\footnotesize ${\rm \Delta Dec.}$} & {\footnotesize ${\rm R}$}  & {\footnotesize ${\rm M_B}$} &  {\footnotesize ${\rm M_J}$} & {\footnotesize ${\rm U-B}$} & {\footnotesize ${\rm B-R}$} & {\footnotesize ${\rm R-I}$} & {\footnotesize ${\rm I-J}$} & {\footnotesize ${\rm J-H}$} \\
%
% {\footnotesize SCC}& {\footnotesize (\arcsec)} & {\footnotesize (\arcsec)} & {\footnotesize (pc)} & {\footnotesize (mag)} & {\footnotesize (mag)} & {\footnotesize (mag)} & {\footnotesize (mag)} \\
%
\hline
1 & -6.37 &  0.32 & 33 & $-10.82\pm0.05$ & $-11.20\pm0.16$ & $-0.83\pm0.06$ & $0.45\pm0.06$ & $-0.20\pm0.08$  & $0.13\pm0.17$ & $-0.12\pm0.23$ \\ 
2 & -5.48 &  0.13 & 33 & $-9.97\pm0.07$  & $-11.25\pm0.16$ & $-0.83\pm0.10$ & $0.59\pm0.10$ & $-0.04\pm0.10$  & $0.74\pm0.17$ & $0.73\pm0.18$ \\  
3 & -3.76 &  0.10 & 41 & $-10.68\pm0.05$ & $-12.55\pm0.08$ & $-0.73\pm0.07$ & $0.56\pm0.07$ & $0.40\pm0.07$   & $0.90\pm0.09$ & $0.63\pm0.10$  \\
4 & -2.21 & -0.84 & 58 & $-11.55\pm0.04$ & $-13.22\pm0.06$ & $-0.67\pm0.06$ & $0.62\pm0.05$ & $0.36\pm0.04$   & $0.69\pm0.07$ & $0.67\pm0.08$ \\ 
5 & -0.71 & -1.88 & 33 & $-9.81\pm0.07$  & $-11.16\pm0.16$ & $-0.79\pm0.10$ & $0.32\pm0.11$ & $0.13\pm0.13$   & $0.90\pm0.19$ & $0.48\pm0.21$ \\
6 &  1.04 & -2.24 & 33 & $-9.82\pm0.12$  & $-11.73\pm0.12$ & $-0.66\pm0.18$ & $0.65\pm0.15$ & $0.32\pm0.13$   & $0.95\pm0.15$ & $0.70\pm0.15$ \\
7 & -0.33 & -2.96 & 33 & $-9.79\pm0.08$  & $-11.60\pm0.12$ & $-0.74\pm0.11$ & $0.51\pm0.11$ & $0.48\pm0.09$   & $0.81\pm0.14$ & $0.86\pm0.14$ \\
8 & -2.99 & -4.21 & 58 & $-11.26\pm0.04$ & $-12.80\pm0.08$ & $-0.67\pm0.05$ & $0.64\pm0.05$ & $0.27\pm0.05$   & $0.63\pm0.08$ & $0.56\pm0.09$ \\  
\hline
\hline
\end{tabular}
%     {\footnotesize
%     \begin{description}  \addtolength{\itemsep}{-0.8\baselineskip} %tablenote
%       \item[]  Notes: The units are: R.A. and Declination coordinates respecto to galaxy centre in [arcsec].
%       \item[] 
%     \end{description} %tablenotes
%     } %scritpstize
\end{threeparttable}
} % END SMALL
\end{table*}

%% file: tb_hiinb2_mix.tex
%------------------------------------------------------------------------------------------------
 
%------------------------------------------------------------------------------------------------
\begin{table*}
\centering
\caption{\label{tab:tb_hiinb2_mix} H II regions ${\rm H\alpha}$ and ${\rm [OIII]\lambda5007}$ fluxes, luminosities and derived properties. The H II regions position are in arcseconds and relative to the galaxy centre, and the radius (R) in parsecs. The units for the flux, luminosity and equivalent width are [${\rm erg \, s^{-1} \, cm^{-2}}$], [${\rm erg \, s^{-1}}$], and [\AA] respectively.}
{\small
\begin{threeparttable}
\begin{tabular}{ c c c c c c c c c c}
\hline
\hline
{\footnotesize H II} & {\footnotesize ${\rm \Delta R.A.}$} & {\footnotesize ${\rm \Delta Dec.}$} & {\footnotesize ${\rm R_{H II}}$} & {\footnotesize ${\rm \log F_{H\alpha}}$} & {\footnotesize ${\rm \log F_{[OIII]}}$} & {\footnotesize ${\rm \log L_{H\alpha}}$} & {\footnotesize ${\rm \log L_{[OIII]}}$} & {\footnotesize ${\rm EW(H\alpha)}$} & {\footnotesize ${\rm EW([OIII])}$}  \\
% & {\footnotesize (\arcsec)} & {\footnotesize (\arcsec)} & {\footnotesize (pc)} & {\footnotesize (${\rm erg \, s^{-1} \, cm^{-2}}$)} & {\footnotesize (${\rm erg \, s^{-1} \, cm^{-2}}$)} & {\footnotesize (${\rm erg \, s^{-1}}$) } & {\footnotesize (${\rm erg \, s^{-1}}$) } & {\footnotesize (\AA)} & {\footnotesize (\AA)}  \\
%
%{ } & {\footnotesize (arcsec)} & {\footnotesize (arcsec)} & {\footnotesize (arcsec)} & {\footnotesize (${\rm erg \, s^{-1} \, cm^{-2}}$) } & {\footnotesize (${\rm erg \, s^{-1} \, cm^{-2}}$) } & {\footnotesize (${\rm erg \, s^{-1} }$) } & {\footnotesize (${\rm erg \, s^{-1} }$) } &  {\footnotesize (\AA) } & {\footnotesize (\AA) }  \\
%
\hline
%SB &       &       &      & ${-11.92^{+0.00}_{-0.00}}$ & ${-11.97^{+0.00}_{-0.00}}$ & ${40.29^{+0.00}_{-0.00}}$ & ${40.28^{+0.00}_{-0.00}}$  & ${142.81\pm108.87}$  & ${\rm 63.83\pm11.17}$\\
 1 & -7.48 &  9.31 & 128 & ${-14.63^{+0.08}_{-0.09}}$ & ${-14.60^{+0.09}_{-0.12}}$ & ${37.59^{+0.08}_{-0.09}}$ & ${37.66^{+0.09}_{-0.12}}$  & ${305\pm144}$        & ${250\pm146}$   \\ 
 2 &  1.63 &  5.72 & 66 & ${-14.61^{+0.08}_{-0.09}}$ & ${-14.60^{+0.09}_{-0.12}}$ & ${37.61^{+0.08}_{-0.09}}$ & ${37.66^{+0.09}_{-0.12}}$  & ${958\pm922}$        & ${1913\pm5297}$ \\ 
 3 &  4.05 &  1.27 & 58 & ${-14.89^{+0.10}_{-0.13}}$ & -                          & ${37.33^{+0.10}_{-0.13}}$ & -                         & -            & -               \\ 
 4 & -3.16 &  1.04 & 50 & ${-14.74^{+0.10}_{-0.13}}$ & ${-14.53^{+0.09}_{-0.11}}$ & ${37.48^{+0.10}_{-0.13}}$ & ${37.72^{+ 0.09}_{-0.11}}$ & ${107\pm52}$         & ${84\pm39}$     \\ 
 5 & -6.42 & -0.04 & 99 & ${-13.09^{+0.01}_{-0.02}}$ & ${-12.94^{+0.02}_{-0.02}}$ & ${39.13^{+0.01}_{-0.02}}$ & ${39.31^{+0.02}_{-0.02}}$ & ${295\pm24}$         & ${263\pm24}$    \\ 
 6 & -0.36 & -1.57 & 33 & ${-14.77^{+0.09}_{-0.11}}$ & ${-14.57^{+0.09}_{-0.11}}$ & ${37.45^{+0.09}_{-0.11}}$ & ${37.68^{+0.09}_{-0.11}}$ & -            & -               \\ 
 7 &  1.10 & -1.80 & 41 & ${-14.56^{+0.07}_{-0.09}}$ & ${-14.49^{+0.08}_{-0.10}}$ & ${37.66^{+0.07}_{-0.09}}$ & ${37.76^{+0.08}_{-0.10}}$ & ${96\pm28}$          & ${67\pm22}$     \\ 
 8 & -4.89 & -3.26 & 57 & ${-14.01^{+0.04}_{-0.05}}$ & ${-13.81^{+0.04}_{-0.05}}$ & ${38.21^{+0.04}_{-0.05}}$ & ${38.44^{+0.04}_{-0.05}}$ & ${844\pm417}$        & ${1548\pm1488}$ \\ 
 9 &  3.78 & -3.49 & 41 & ${-14.49^{+0.07}_{-0.08}}$ & ${-14.33^{+0.07}_{-0.08}}$ & ${37.73^{+0.07}_{-0.08}}$ & ${37.93^{+0.07}_{-0.08}}$ & ${314\pm127}$        & ${292\pm133}$   \\ 
10 & -2.43 & -4.95 & 66 & ${-13.71^{+0.03}_{-0.03}}$ & ${-13.28^{+0.02}_{-0.02}}$ & ${38.51^{+0.03}_{-0.03}}$ & ${38.97^{+0.02}_{-0.02}}$ & ${472\pm102}$        & ${919\pm245}$   \\ 
11 & -1.51 & -6.79 & 57 & ${-13.90^{+0.04}_{-0.04}}$ & ${-13.63^{+0.03}_{-0.04}}$ & ${38.32^{+0.04}_{-0.04}}$ & ${38.63^{+0.03}_{-0.04}}$ & ${293\pm59}$         & ${377\pm85}$    \\ 
12 & -3.58 & -7.05 & 41 & ${-14.49^{+0.07}_{-0.09}}$ & ${-14.26^{+0.07}_{-0.09}}$ & ${37.73^{+0.07}_{-0.09}}$ & ${37.99^{+0.07}_{-0.09}}$ & ${405\pm196}$        & ${598\pm384}$   \\ 
\hline
\hline
\end{tabular}
%     {\scriptsize
%     \begin{description}  \addtolength{\itemsep}{-0.8\baselineskip} %tablenote
%       \item[] Notes: The units are: R.A. and Declination coordinates respecto to galaxy centre in [arcsec], Flux [${\rm erg \, s^{-1} \, cm^{-2}}$], Luminosity [${\rm erg \, s^{-1}}$], Equivalent width [${\rm \AA}$].
%       \item[] 
%     \end{description} %tablenotes
%     } %scritpstize
\end{threeparttable}
} % END SMALL
\end{table*}

%% file: mnras.tb.hii.color.tex
%------------------------------------------------------------------------------------------------
 
%------------------------------------------------------------------------------------------------
\begin{table*}
\centering
\caption{\label{tab:tb_hii_color} Optical absolute magntiude and colours of the identified H II regions.}
{\small
\begin{threeparttable}
\begin{tabular}{ c c c c c c}
\hline
\hline
{\footnotesize H II} & {\footnotesize ${\rm M_{B}}$} & {\footnotesize U-B} & {\footnotesize B-R} & {\footnotesize R-I} \\
%{\footnotesize H II} & {\footnotesize (mag)} & {\footnotesize (mag)} & {\footnotesize (mag)} & {\footnotesize (mag)} \\
\hline
    1 & ${\rm -8.24 \pm 0.13}$  & ${\rm -0.81 \pm 0.16}$ & ${\rm 1.17 \pm 0.17}$ & ${\rm 0.61 \pm 0.14}$ \\ 
    5 & ${\rm -12.70 \pm 0.03}$ & ${\rm -0.84 \pm 0.05}$ & ${\rm 0.60 \pm 0.05}$ & ${\rm -0.02 \pm 0.05}$ \\ 
    7 & ${\rm -10.06 \pm 0.14}$ & ${\rm -0.73 \pm 0.20}$ & ${\rm 0.79 \pm 0.17}$ & ${\rm 0.33 \pm 0.14}$ \\ 
    9 & ${\rm -9.22 \pm 0.09}$  & ${\rm -0.77 \pm 0.12}$ & ${\rm 0.53 \pm 0.13}$ & ${\rm -0.22 \pm 0.18}$ \\ 
   10 & ${\rm -10.76 \pm 0.10}$ & ${\rm -0.71 \pm 0.14}$ & ${\rm 0.49 \pm 0.15}$ & ${\rm 0.11 \pm 0.17}$ \\ 
   11 & ${\rm -10.70 \pm 0.05}$ & ${\rm -0.78 \pm 0.07}$ & ${\rm 0.59 \pm 0.07}$ & ${\rm -0.07 \pm 0.08}$ \\ 
\hline
\hline
\end{tabular}
%     {\scriptsize
%     \begin{description}  \addtolength{\itemsep}{-0.8\baselineskip} %tablenote
%       \item[] Notes: The units are: R.A. and Declination coordinates respecto to galaxy centre in [arcsec], Flux [${\rm erg \, s^{-1} \, cm^{-2}}$], Luminosity [${\rm erg \, s^{-1}}$], Equivalent width [${\rm \AA}$].
%       \item[] 
%     \end{description} %tablenotes
%     } %scritpstize
\end{threeparttable}
} % END SMALL
\end{table*}

%% file: tab.hiideriv2.tex
%------------------------------------------------------------------------------------------------
\begin{table}
\centering
\caption{\label{tab:hiideriv} Narrow band images derived quantities.
 The ${\rm \log \Sigma_{SFR}}$ is in units of [${\rm M_{\sun} \, yr^{-1} \, kpc^{-2} }$],
 the Str\"{o}mgren radius is in [pc].  }

{\small
\begin{threeparttable}
\begin{tabular}{ c c c c c }
\hline
\hline
{\footnotesize H II} & {\footnotesize ${\rm \log \, [OIII]/H_{\alpha}}$} & {\footnotesize ${\rm \log \Sigma_{SFR}}$} & {\footnotesize ${\rm N_{\star} O7 V}$} & {\footnotesize ${\rm R_{Str}}$} \\
%{\footnotesize H II}  &   & {\footnotesize (${\rm M_{\sun} \, yr^{-1} \, kpc^{-2} }$)} &  & {\footnotesize (pc)}  \\
\hline
   1 & ${0.07^{0.12}_{-0.16}}$ &  ${-2.20^{0.08}_{-0.09}}$ & $2.83$  & $4.46$  \\ 
   2 & ${0.05^{0.11}_{-0.16}}$ & ${-1.61^{0.08}_{-0.09}}$ & $2.97$  & $4.53$  \\ 
   3 &   -                     & ${-1.77^{0.10}_{-0.13}}$ & $1.58$  & $3.67$   \\ 
   4 & ${0.24^{0.13}_{-0.18}}$ & ${-1.50^{0.10}_{-0.13}}$ & $2.22$  & $4.11$  \\ 
   5 & ${0.18^{0.02}_{-0.02}}$ & ${-0.44^{0.01}_{-0.02}}$ & $97.88$ & $14.53$ \\ 
   6 & ${0.23^{0.12}_{-0.17}}$ & ${-1.16^{0.09}_{-0.11}}$ & $2.05$  & $4.00$  \\ 
   7 & ${0.11^{0.11}_{-0.14}}$ & ${-1.14^{0.07}_{-0.09}}$ & $3.33$  & $4.71$  \\ 
   8 & ${0.23^{0.06}_{-0.07}}$ & ${-0.88^{0.04}_{-0.05}}$ & $11.87$ & $7.19$  \\ 
   9 & ${0.19^{0.09}_{-0.12}}$ & ${-1.07^{0.07}_{-0.08}}$ & $3.95$  & $4.98$  \\ 
  10 & ${0.46^{0.04}_{-0.04}}$ & ${-0.71^{0.03}_{-0.03}}$ & $23.79$ & $9.07$  \\ 
  11 & ${0.31^{0.05}_{-0.05}}$ & ${-0.77^{0.04}_{-0.04}}$ & $15.11$ & $7.79$  \\ 
  12 & ${0.27^{0.10}_{-0.13}}$ & ${-1.07^{0.07}_{-0.09}}$ & $3.90$  & $4.96$  \\ 
\hline
\hline
\end{tabular}
\end{threeparttable}
} % END SMALL
\end{table}

%% file: tol02_5v6c.tex
\section[]{STELLAR POPULATIONS CHARACTERISATION}
\label{sec:fitbb}

%=============================================================================================
%       SSP MODELS AND RODRIGUEZ-MERINO ET AL X^2 FITTING METHOD
%=============================================================================================

\subsection[]{Spectral Energy Distribution Fitting}
\label{fitbb_sedfit}

The age, mass and extinction values of a specific stellar population can be inferred by comparing its photometric colours with those of synthetic stellar populations \citep{An03,Ad10a,Ro11,Ma12}. 
PopStar single stellar populations (SSP) models \citep{Mo09} cover an age range from 0.1 Myr to 15 Gyr, metallicities from Z=0.0001 to Z=0.05 and take into consideration the nebular continuum contribution to the stellar spectral energy distribution (SED).  
The wide age range covered by PopStar models allows to analyse both young and old stellar populations contained in BCD galaxies while the low metallicity models  are ideal to study \mbox{H II} galaxies. 
Therefore, we use a Z=0.004 
PopStar  model to characterise the stellar populations in Tol 02 since it is the one that best suits the galaxy Z=0.0035 value\footnote{Assuming ${\rm 12+\log\left(\frac{O}{H}\right) = \log \left(\frac{Z}{0.0126} \right) + 8.69}$ where ${\rm \frac{O}{H} = 1.37 \times 10^{-4}}$ \citep{Va92}.}.
We followed \citet{Ro11}  ${\rm \chi^2}$ SED fitting method to compare the observed colours with those of the SSP models.
The method is described briefly as:

\begin{enumerate}

\item{\it Build a set of reddened SSP SEDs}. The magnitudes and zero points at a given epoch were calculated by convolving SOAR filters transmission curves with the SED of both the modelled SSP and the star Vega  \citep{Pi98}.
The magnitudes were calculated at the 106 irregularly sampled PopStar time points that span  the age range from $t=0.1$ Myr to $t=13$ Gyr. 
These were then reddened using an extinction vector ranging from $A_{V}=0$ to $A_{V}=1.5$ mag at ${\rm \Delta A_{V}=0.02}$ magnitudes steps.

\item{\it Obtain SSP uniformly sampled time}. The temporal evolution (in logarithmic scale) of the reddened magnitudes was linearly interpolated using a $\Delta \log t=0.005$ Myr time step, to keep all 106 PopStar original points.
The final time vector consists of 1041 uniformly distributed time steps in the range $t=0.1$ Myr to $t=13$ Gyr.

\item {\it Defining the merit functions.} The colours of each object were compared to those of the modelled SSPs using two merit functions. 
The first one is defined as:

%\begin{equation}
%\chi^2 = \sum \left( \frac{ colour_{i}^{mod} - colour_{i}^{obs} }{\sigma_{i}^{obs}} \right)^2  \label{eqn:chi2}
%\end{equation}

\begin{equation}
\chi^2 = \sum_{i=1}^{N} \left( \frac{ Colour \left( i \right)_{Mod} - Colour \left( i \right) _{Obs}  }{  \sigma \left( i \right)_{Obs}}  \right)^2  \label{eqn:chi2}
\end{equation}

\noindent where N is the number of different colours to analyse, $Colour_{Mod}$ and $Colour_{Obs}$ are the modelled and observed colours respectively, and $\sigma_{Obs}$ is the object photometric colour uncertainty. The colours used in this first merit function were built using the R band as reference: U-R, B-R, R-I, R-J and R-H. The second merit function relies on the U-B colour potential to differentiate the ages of stellar populations younger than 500 Myr (the U-B colour of a SSP spans 1.3 magnitudes from 0.1 to 500 Myr) and is defined as: 

\begin{equation} 
\beta^2 = \chi^2 \left[ (U-B)_{Mod} - (U-B)_{Obs} \right]^2 
\end{equation}

The use of this ${\rm \beta}$ function strengthen the unicity of the solution. 
We selected the 50 models associated with the lowest ${\chi^2}$ values, and of these kept only those that are within the 10\% of the minimum value of ${\beta}$.

%\item {\it Simulated datasets}. We used Monte Carlo simulations to create 5000 mock sets associated with the  observed bands. 
%
\item {\it Simulated data sets}. We used Monte Carlo simulations to create a set of 5000 fake observations at each of the seven bands. 
The flux of the fake observations at each of the bands was randomly chosen from a Gaussian distribution with a mean value equal to the  observed flux and a sigma value equal to its uncertainty. 
We used the merit functions to determine the capability of the models to reproduce the colours of the fake observations. 
For each of the 5000 fake observations we kept the models associated to the lowest ${\chi^2}$ and ${\beta}$ values, and combined them with the model that best fit the object photometry.
This gave as a result a collection of various thousands of models related to the photometric values of the object under study.

\item {\it Age and extinction determination}. A Gaussian function was fitted to the 
 distribution of ages (in logarithmic scale) and extinction values that best reproduced
 the object photometric SED.
The mean value of the fitted Gaussians are adopted as the $\log \rm{Age} $ and $A_{V}$
 values of the object representative stellar population, 
 and the Gaussian standard deviation is taken as an uncertainty measure of the adopted
 ${\rm \log Age }$ and ${\rm A_{V}}$ values.

\item {\it Age and extinction uncertainties}. The uncertainties of the adopted age and
 extinction values only take into consideration the uncertainties from the  photometry.
To consider the uncertainty of the fitted age due to the discrete time steps we have
added to the age uncertainty the difference between the adopted age and its older and
younger closest PopStar model.
We have also analysed the change that the fitted age and extinction values would suffer if the 
 luminosity of the models is related to an uncertainty caused by: 1) the luminosity 
 variation among contiguous time steps or 2) model intrinsic uncertainties.
To compute this change in the adopted age and extinction values, we estimated again 
 the properties of the stellar populations that best fit the photometric SED from the 
 objects, this time considering the uncertainties of the models. 
Therefore Equation  \ref{eqn:chi2} becomes:

\begin{equation}
\chi^2 = \sum_{i=1}^{N}  \frac{ \left[ Colour \left( i \right)_{Mod} - Colour \left( i \right) _{Obs} \right]^2 }{  \sigma ^2 \left( i \right)_{Mod}  + \sigma ^2 \left( i \right)_{Obs}}  \label{eqn:chi2b}  
\end{equation}

\noindent where the uncertainty of each of the colours of the models ($\sigma_{Mod}$) is derived from the estimated uncertainty of the model luminosity ($L$) at a given epoch, which is defined as: 1) $ \sigma_{L}(t) = \left[ L(t+1) + L(t-1) - 2 L(t) \right] / 2 $ when accounting for the luminosity change in time or 2) $ \sigma_{L}(t) = \pm 0.05 \times L(t) $ when assuming that the modelled luminosity has a 95\% accuracy.
Finally, the uncertainty of the  adopted age (and extinction) is set as the average difference between the adopted value (error-free model) and the values obtained when an error was ascribed to the model.
\item {\it Mass}. The mass of the dominant stellar population is estimated by scaling the object R band luminosity to that of a ${\rm 10^6 \, M_{\sun}}$ model SSP with the adopted age and extinction. 
The uncertainties were estimated as the highest and lowest possible values obtained if a band other  than R had been used instead.

%\begin{equation}
%\chi^2 = \sum  \frac{ \left( colour_{i}^{mod} - colour_{i}^{obs} \right)^2 }{\sigma_{i}^{mod}^2 + \sigma_{i}^{obs}^2}   
%\end{equation}

\end{enumerate}

%=============================================================================================
%       disk AND STAR-FORMATION POPULATIONS
%=============================================================================================

\subsection{Stellar populations}

\subsubsection{LSBC and star-forming region}
\label{fitbb_SBLSBC}

%The LSBC is expected to be formed by multiple old and dust-free stellar populations.  
Although the LSBC is expected to be formed by multiple old and dust-free stellar populations,  
we have estimated the age of its dominant population using the LSBC aperture colours (Table \ref{tab:colorsbp} in Section \ref{subsec:host_disc}) assuming it does not suffer from extinction. 
%
%Once an age is adopted, the R luminosity is used to calculate a photometric mass. 
%
The results obtained after applying the method described above are indicative  of a dominant population in the LSBC that is ${\rm 1.5}$ Gyr old and ${\rm \approx 4\times10^{8}}$ ${\rm M_{\sun}}$ in  mass. 
The stellar population that best resembles the colours of the star-forming region (after subtracting the LSBC emission) is 5 Myr old, 
which is consistent with the ${\rm \approx 4}$ Myr found by ME00 who compared the ${\rm EW(H\alpha)}$ with that of SSP models.
The estimated stellar mass of the complete star-forming region is ${\rm \approx16 \times 10^{6} }$ ${\rm M_{\sun}}$, approximately 20 times less massive than the LSBC. 
%
%This population is still at the epoch of the Wolf-Rayet stars, 
%
%in agreement with the detection of Wolf-Rayet spectral features \citep {Sc99,Ku85}. 
%
%Our 5 Myr fitted age is consistent with the ${\rm \approx 4}$ Myr found by ME00, who compared the ${\rm EW(H\alpha)}$ with that of SSP models.

%---------------------------------------------------------------------------------------------

We have compared our results with those obtained by \citet{Ra00} and \citet{We04},
who characterise the stellar populations in the galaxy using age-dependent spectral features.
%
%
%Turning this comparison into a delicate matter due to the difference in methodology used. 
% 
%Still, it can provide a coarse idea on the galaxy dominant stellar populations ages.
%
%Also, it should not be overlooked that the photometric mass values are age dependent.
%
%As for that, we prefere to use them together with the age they are associated to.
%
%As explained previously, in this study we compared the LSBC colours with a broad set of SSP with ages ranging from 0.1 Myr to 10 Gyr. 
%
\citet{Ra00} characterise the intermediate and old stellar populations using spectral features of stellar populations at 100 Myr, 500 Myr and 10 Gyr.
They concluded that  a 10 Gyr old stellar population is not necessary in order to reproduce the spectra.
\citet{We04} also determined that a 5 Gyr old stellar population (the oldest dominant stellar population in the galaxy) is 100 times more massive than the young population,  estimated to be 10 Myr old.
Then, our estimation that the galaxy LSBC old dominant stellar population is 1.5 Gyr old, and 20 times more massive than the complete star-forming area seems to be in agreement with the results  obtained by \citet{Ra00} and \citet{We04}.

%=============================================================================================
%       GALAXY ZONES
%=============================================================================================
\subsubsection{Galaxy zones}
\label{fitbb_SBLSBC}

The dominant stellar population in each zone was characterised using their photometric colours (Section \ref{subsec:phot_bb}),
 with and without subtracting the LSBC contribution for the Centre zone.
Table \ref{tab:art.sspfit} shows the results, column 1 is the aperture identification name, 
 column 2 the dominant population age in logarithm, column 3 the stellar extinction and 
 column 4 the stellar population estimated mass.
The 1.5 Gyr age of the LSBC disk matches the ages of the SSPs that best fit the colours
 of the East and West zones ($\rm \sim2$ Gyr).
The  estimated extinction for the East and West zones (${\rm A_{V} = 0.0 \ mag}$) supports
 our previous assumption that the LSBC disk is extinction free.
The dominant stellar populations both in the North and South zones are attributed a high extinction value of  ${A_{V} \approx 1.5}$ mag. They are both young, being the population in the North 10 Myr old, and the one in the south only 5 Myr old. 
The difference in age causes the 0.2 mag difference between their U-B colours.
%The dominating stellar populations in the Centre and South zones are estimated to be
% ${\rm \sim 10}$ Myr old with a high extinction value of ${A_{V} \approx 1.5}$ mag. 
%
%These younger populations are responsible for the extremely blue ${\rm U-B\sim -0.8 \ mag}$ 
% colour observed in these zones. 
%
The colours of the Centre zone were best reproduced by a 5 Myr old SSP and an extinction
 ${\rm A_{V}=1.5 \ mag}$; after correcting   for the LSBC light contribution, however, 
 both age and extinction are reduced to half those values 
 (2.5 Myr and ${\rm A_{V}=0.8}$ mag). 
%

% TABLA HECHA A MANO COMBINANDO TABLAS CONSTRUIDAS EN p_phot11_mc_zones.pro p_phot11_mc_zones_elena.pro
%   art.zones.m.2sp.prop.15av y art.zones.m.2sp.prop.fd.15av
% (BUILT IN p_phot11_mc.pro--> tb_obs_disk.pro

% EN ESTA TABLA SE AÑADIO LA INCERTIDUMBRE DE LA DIFERENCIA DE LAS EDADES ALEDAÑAS A LA AJUSTADA
\input{art.sspfit2}

%=============================================================================================
%       STAR CLUSTER COMPLEXES AGE, MASS AND EXTINCTION
%=============================================================================================

\subsubsection{Star cluster complexes}
\label{fitbb_scc}

\begin{figure*}
\centering
\begin{tabular}{c c}
\subfigure{
\resizebox{0.3\textwidth}{!}{\label{fig:sccfit28} \includegraphics{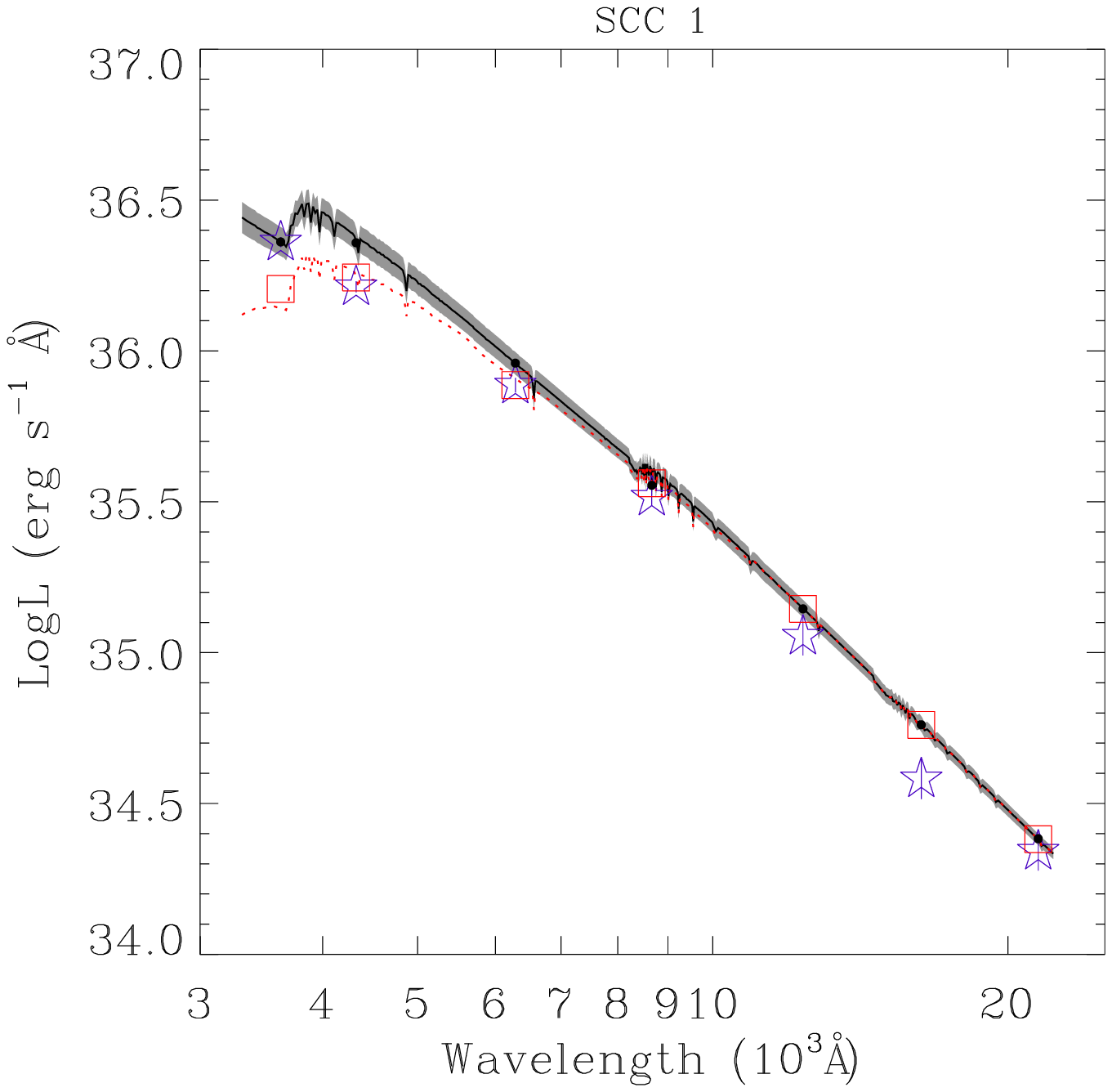}}} &
\subfigure{
\resizebox{0.3\textwidth}{!}{\label{fig:sccfit29} \includegraphics{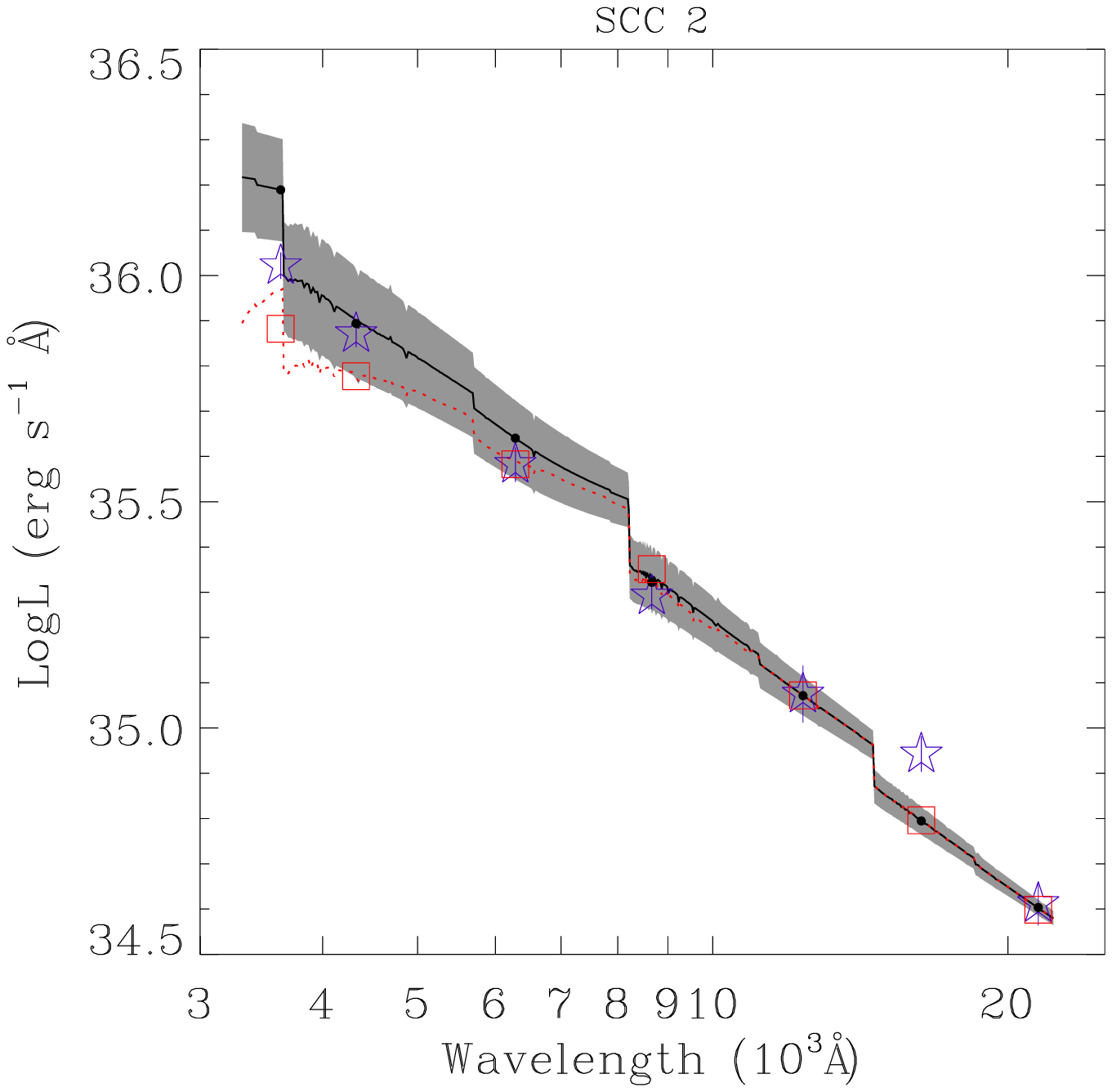}}} \\
\subfigure{
\resizebox{0.3\textwidth}{!}{\label{fig:sccfit30} \includegraphics{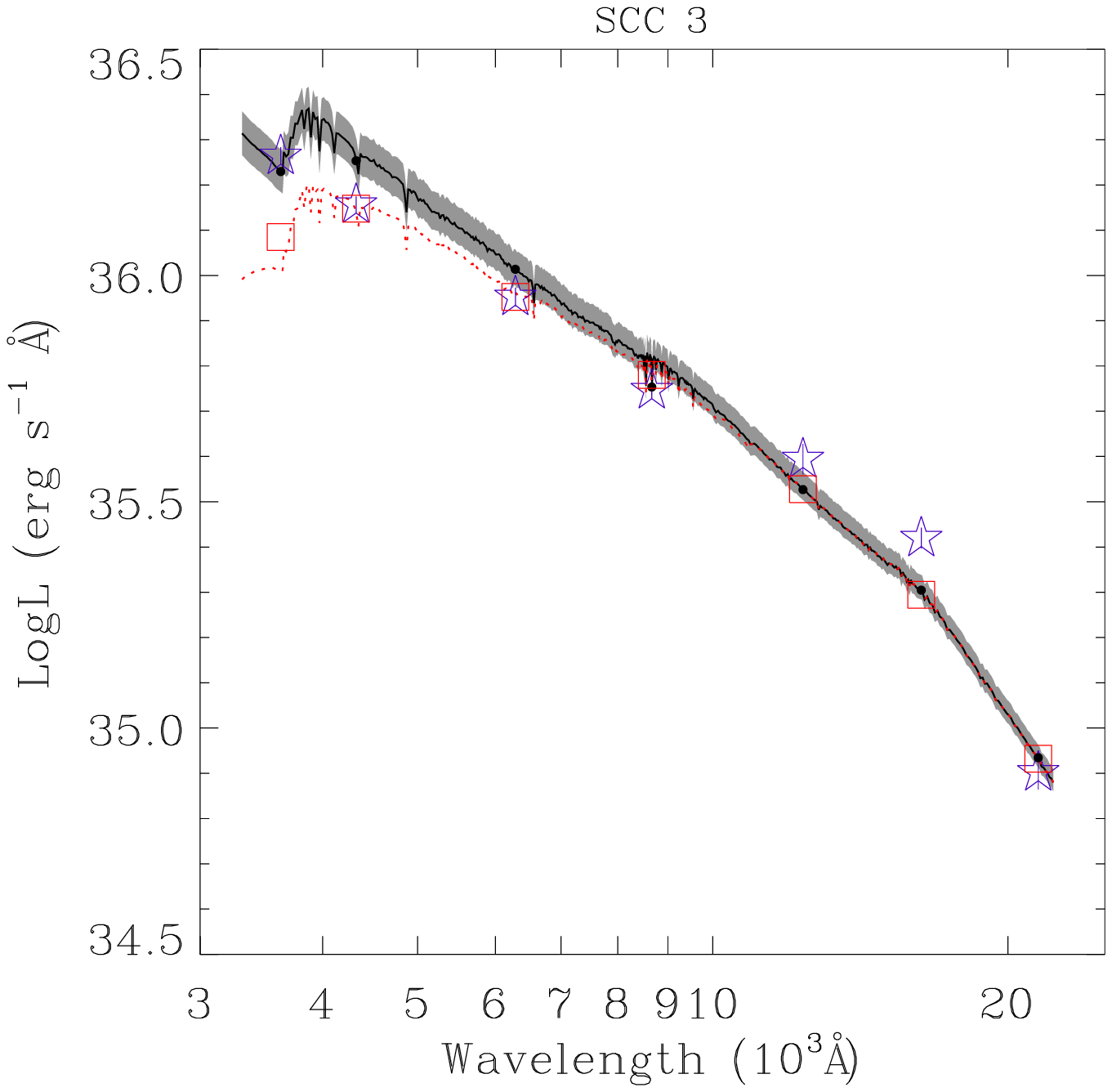}}} &
\subfigure{
\resizebox{0.3\textwidth}{!}{\label{fig:sccfit32} \includegraphics{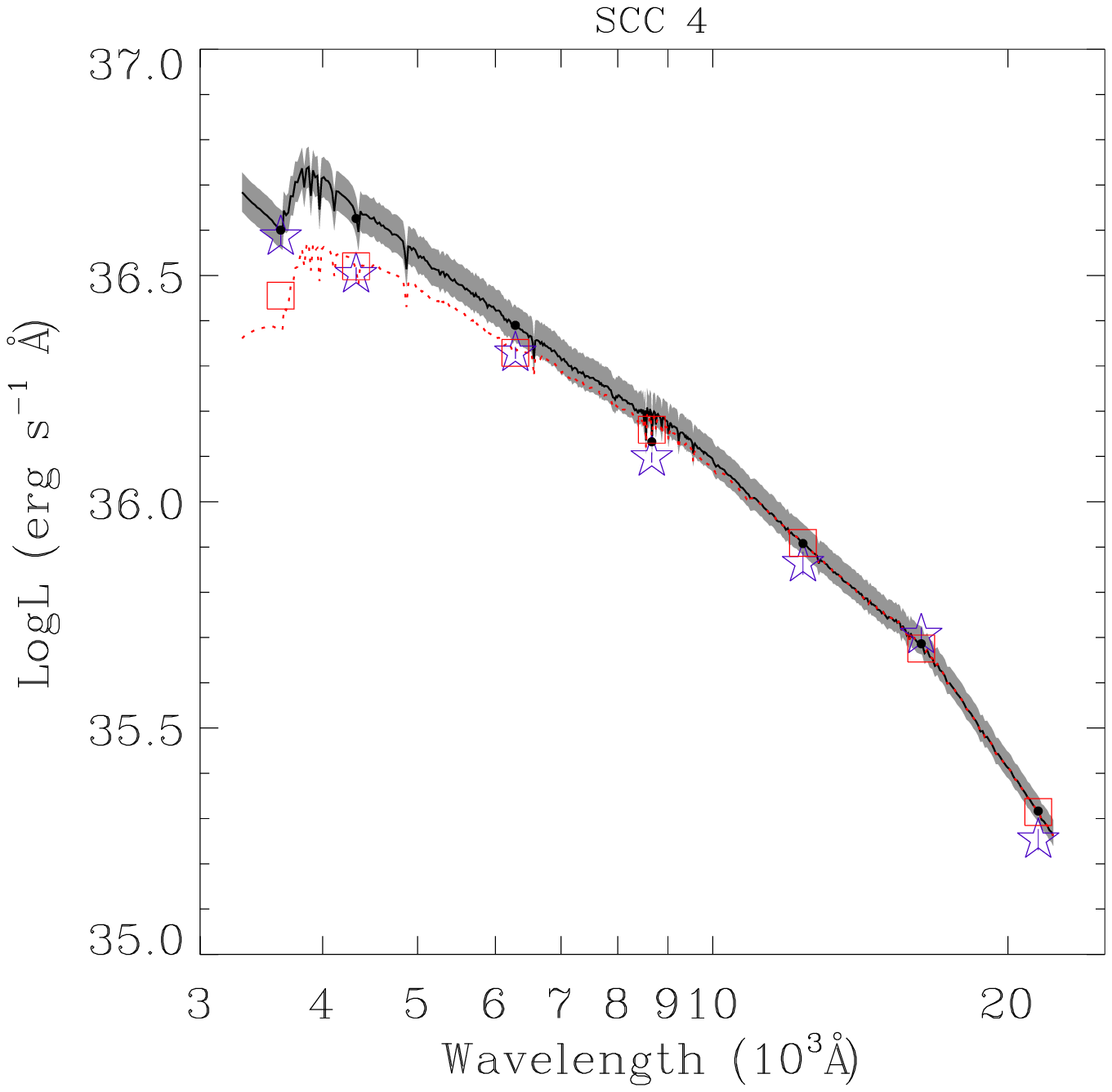}}} \\
\subfigure{
\resizebox{0.3\textwidth}{!}{\label{fig:sccfit33} \includegraphics{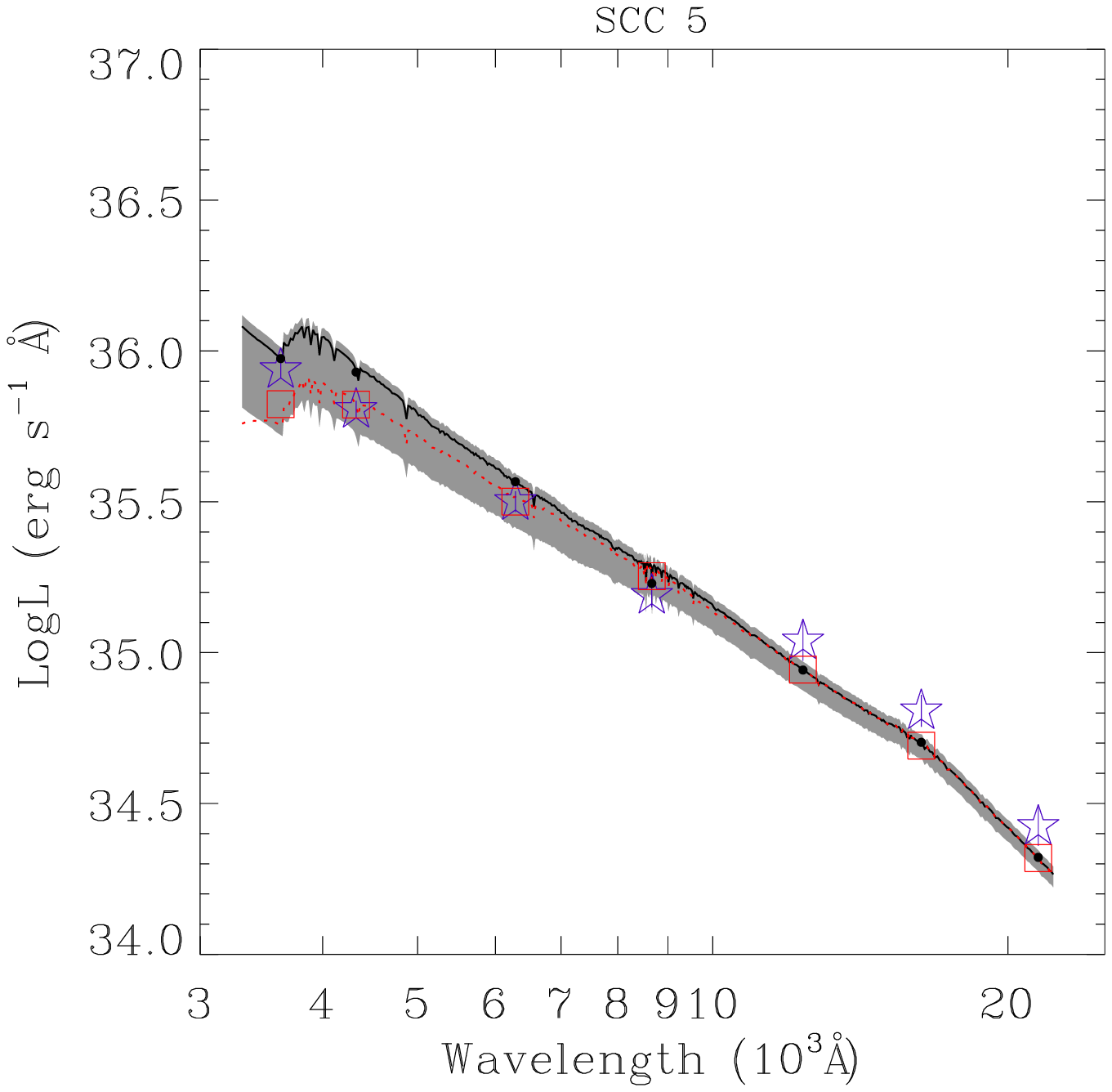}}} &
\subfigure{
\resizebox{0.3\textwidth}{!}{\label{fig:sccfit34} \includegraphics{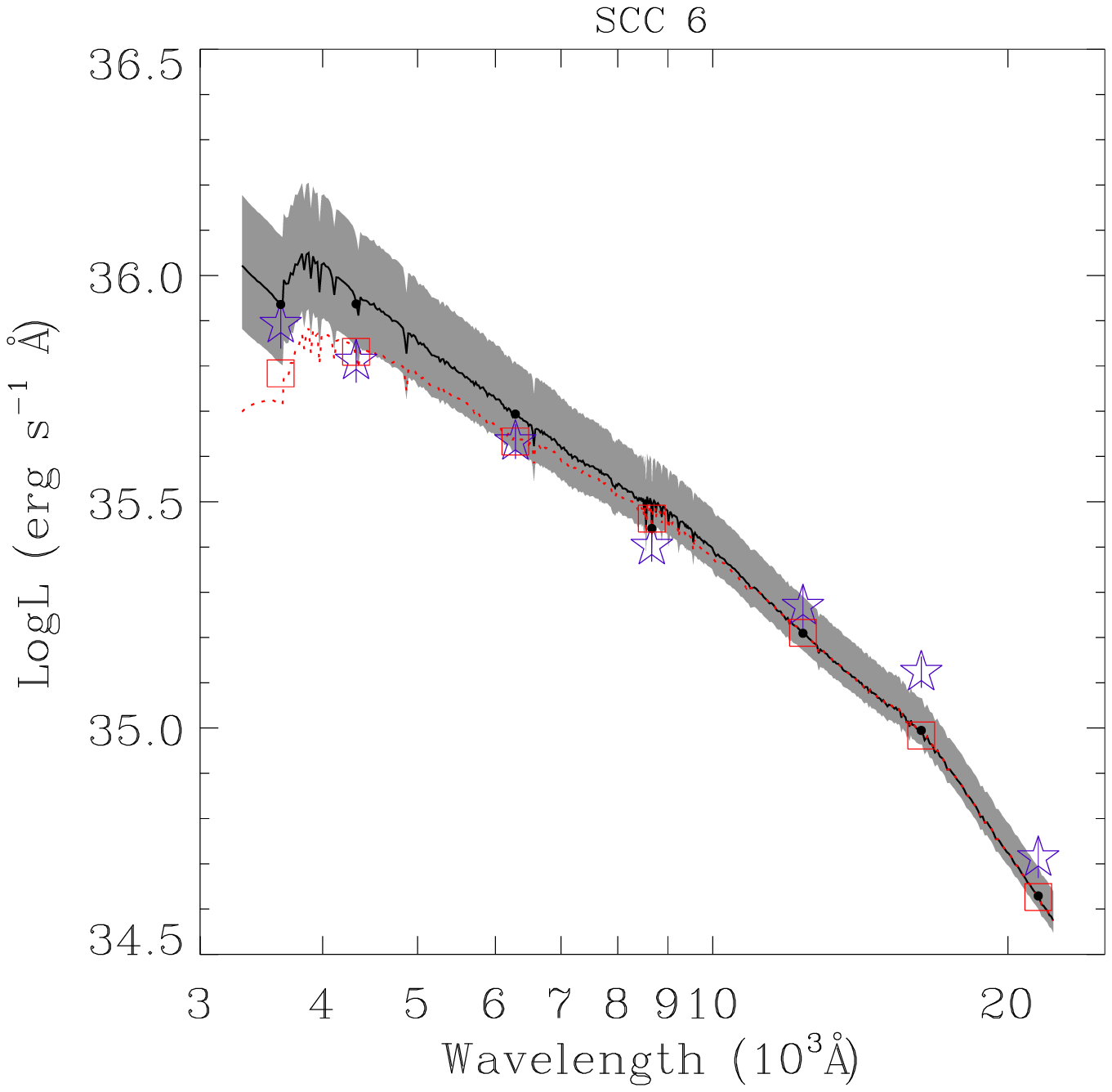}}} \\
\subfigure{
\resizebox{0.3\textwidth}{!}{\label{fig:sccfit36} \includegraphics{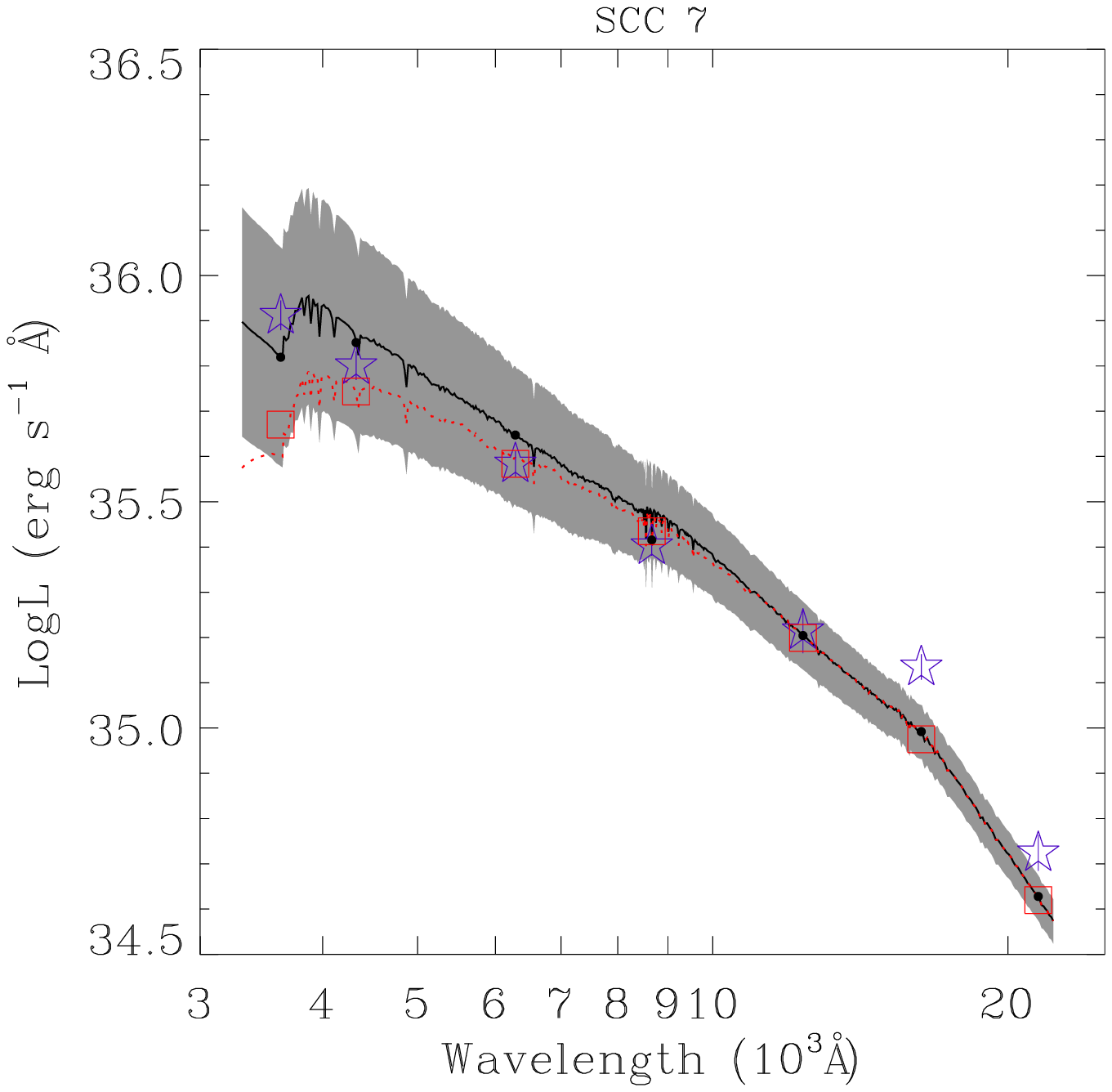}}} &
\subfigure{
\resizebox{0.3\textwidth}{!}{\label{fig:sccfit41} \includegraphics{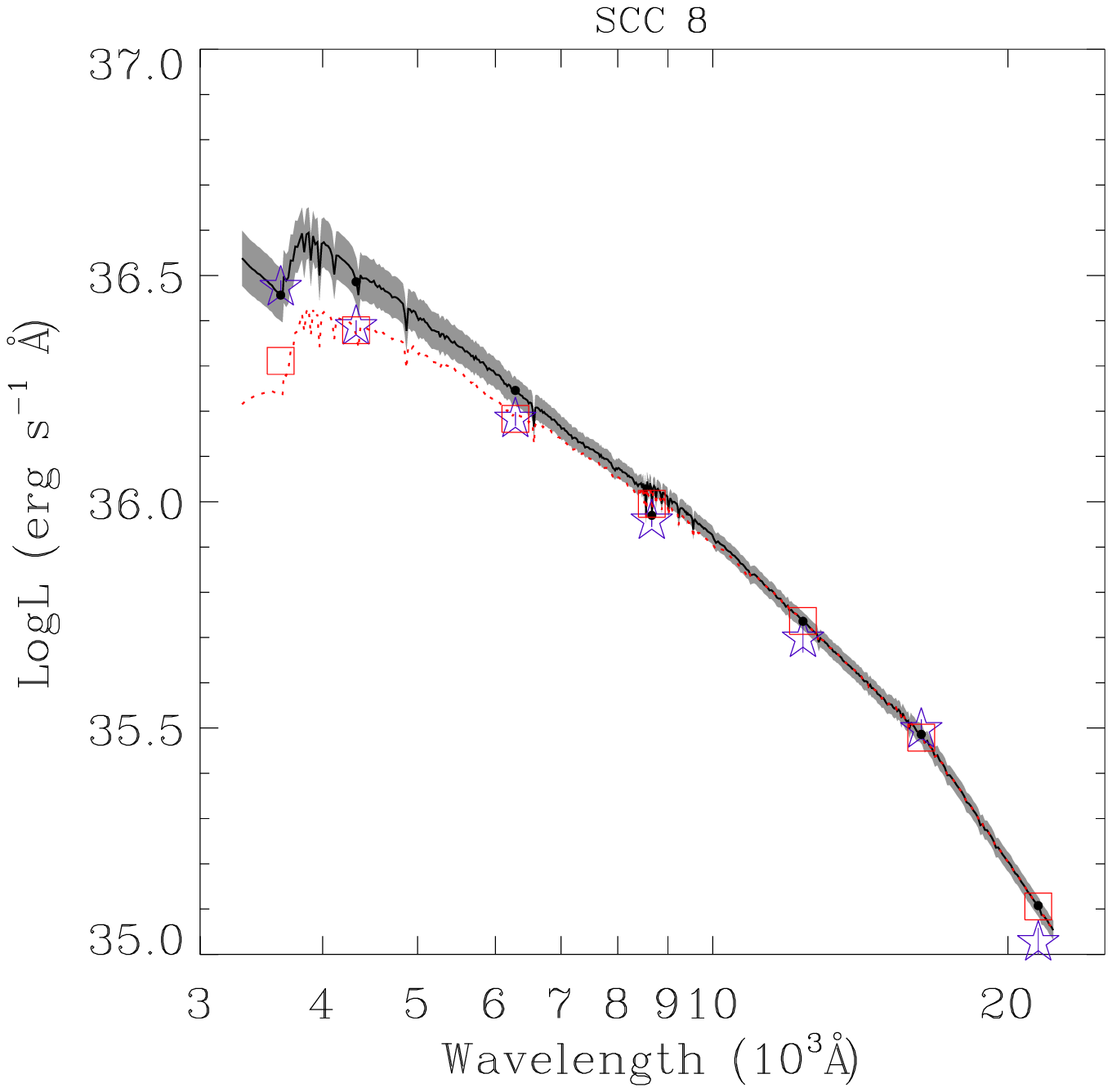}}} \\
\end{tabular}
\caption[]  {\small \label{fig:fitbb_scc_sed} Extinction corrected luminosity of the SCCs (purple stars) and the luminosity of the SSP model that best represents them before (solid-black line) and after (dotted-red line) being absorbed by the atmosphere. The red squares represent the modelled SSP integrated luminosity at each broad band after being affected by the atmospheric extinction.  %, and the solid squares are the luminosity at the filters central wavelength.

}
\end{figure*}

We have also characterised the stellar population dominating the SCCs described in Section \ref{sec:phot}.
The SED of the dominant stellar populations fitted to the SCCs are shown in Figure \ref{fig:fitbb_scc_sed}, 
and their properties are given in Table \ref{tab:art.sspfit}.
All of the SCCs are 10 Myr old, except for SCC \#1 and SCC \#3 which are 3 and 6 Myr old. 
All of the SCCs have masses greater than ${\rm 10^{4.5} \ M_{\sun}}$, which are similar to those of resolved star clusters in the low metallicity Haro 11 \citep{Ad10a} and Mrk 930 \citep{Ad11b}, but lower than those from the more metallic starburst galaxy M82 \citep[$\rm \overline{M_{SC}} \sim 10^{5.5} \ M_{\sun}$][]{Me05}.
%
%All of the SCCs have masses greater than ${\rm 10^{4.5} \ M_{\sun}}$, six of them are 10 Myr old and the other two clusters are only 3  and 5 Myr old. 
%
We briefly describe the properties of each SCC in the following paragraphs.

% 2SSP FIT TO THE ZONES INTEGRATED VALUES (ONLY USING DATA VALUES)
% PLOTS CREATED INSIDE P_PHOT11_MC_ELENA.PRO
%---------------------------------------------------------------------------------------------

SCC \#1 is associated with an age of 6 Myr, extinction  ${\rm A_V = 0.5 \ mag }$ and a
 mass of ${\approx 10^{5} \ M_{\sun}}$. 
SCC \#2 is older (10 Myr)  but with similar extinction and mass (${\rm A_V = 0.2 \ mag }$ 
 and ${\rm \approx 3\times10^{4} \ M_{\sun}}$). 
These two SCCs coincide with the brightest \mbox{H II} region in the emission line maps \mbox{H II} \#5 
 (Section \ref{subsec:phot_bbphot_hii}). 
Next to these complexes and towards the west  is  SCC \#3. 
This is the youngest (3 Myr old) and most extinguished (${\rm A_V \approx 1.5}$ mag) of 
 them all, and lies in between two \mbox{H II} regions (\#5 and \#4). 
The remaining five SCCs are 10 Myr old with masses ${\geq 10^{4.5}}$ ${\rm M_{\sun}}$. 
Three of them (SCC \#5, SCC \#6 and SCC \#8) are related to a slightly displaced \mbox{H II} region.
SCC \#8 is the second most massive cluster complex (${\approx 10^{5}}$ ${\rm M_{\sun}}$) and 
 its companion \mbox{H II} region is the second brightest in the emission line maps. 

For six of the eight SCCs the average change on the adopted age and extinction due to the
 model uncertainty is lower than 5\% and 14\% respectively.
The adopted age variation was the same independently of the uncertainty definition used.
The 5\% uncertainty in the models on the other hand, causes a 0.11 mag decrease  in the
 adopted extinction, which is twice the change induced by the uncertainty due to the time
 variation of the luminosity. 
For the other two SCCs the uncertainty of the model lead to a 5 Myr increment in both SCCs 
 adopted ages and a 40\% and 20\% change in the extinction value respectively.

%

%=============================================================================================
%       H II REGIONS AGE DETERMINATION
%=============================================================================================

% EDADES DE LAS REGIONES \mbox{H II}
\subsubsection{H II regions}
\label{subsec:hii_age}

% \mbox{H II} REGIONS AGE DATING 
%---------------------------------------------------------------------------------------------
%

The \mbox{H II} regions in this galaxy are being photoionised by star clusters that are 10 Myr old or younger, as shown in Section \ref{subsec:phot_nbphot_res} (Figure \ref{fig:mesh_histOIIIHa}) through the analysis of the cell's ${\rm \log [OIII]/H\alpha}$ value.
All of them present ${\rm \log [OIII]/H\alpha}$ that would be expected if ionisation of the gas was performed by photons, rather than shocks.
The ${\rm EW(H\alpha)}$  and U-B colour of a modelled SSP are used as tools to date the stellar populations that ionise \mbox{H II} regions \citep{Do81,Ch86,St96,Fe03,Ma08,Ha08b,Ma12}.
This procedure relies on: % the basis that:
\begin{enumerate}
\item The stellar extinction is either negligible or fixed and known.

\item The observed \mbox{H II} region encloses its ionising SCC.

\item The colours of the \mbox{H II} region are representative of the stellar ionising source.

\item And the effect of  an  underlying  stellar population is not included in the equivalent width of ${\rm H\alpha}$.

\end{enumerate}

In Section  \ref{subsec:phot_nbphot_res} we showed that extinction in the star-forming region of this galaxy is nearly uniform and can be considered negligible, 
therefore the first premise is fulfilled. 
In order to estimate the ages of the star clusters responsible for the ionisation 
 of the \mbox{H II} regions we followed the commonly used hypothesis that the other two 
 premises are verified as well \citep[ME00]{Al15}. 
This idea has been previously adopted by ME00, who dated  the \mbox{H II} regions by comparing their U-B colour and ${\rm EW(H\alpha)}$ with those predicted by SSP models from Starburst99 \citep{Le99}.

\begin{figure}
\centering
\resizebox{0.45\textwidth}{!}{\includegraphics{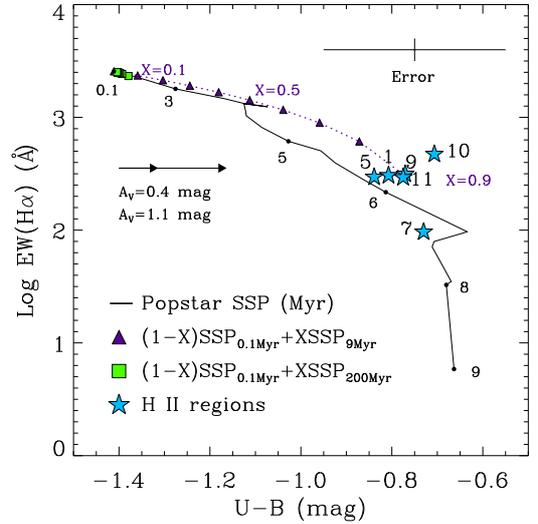}}  
\caption{ \small Optical U-B vs ${\rm \log EW(H\alpha)}$ diagram. The solid  line indicates the evolution of a PopStar single stellar population model and the numbers over the tracks give the age in Myr. The purple dotted line represents the resultant colour after combining the light of a young (0.1 Myr)  with an intermediate (10 Myr) stellar population. The purple triangles correspond to a combination of 90\% young and 10\% intermediate to 10\% young and 90\% intermediate populations. The green squares represent the mixture with a 200 Myr old population.
Blue stars indicate the position of the H II regions.
}
\label{fig:hii_ubewha}
\end{figure}

The ${\rm EW(H\alpha )}$ vs U-B diagram is shown in Figure \ref{fig:hii_ubewha}. The 6 HII regions with measured U-B colour are shown as blue stars. We can compare their position in the diagram with the trace of the evolution of a  PopStar SSP model  from 0.1 Myr to 9 Myr old. We can see that the contribution of an older stellar population in both the measured U-B colour and the ${\rm EW(H\alpha )}$ seems to be negligible.
The ${\rm EW(H\alpha )}$ and the U-B colours of the six H II regions are compatible with a ${\rm ~ 5 \ Myr}$ and  a ${\rm 7 \ Myr}$ old SSP model respectively.
We have compared our results with those obtained by ME00 who also find a 5 Myr old age for the ionising clusters from the ${\rm EW(H\alpha)}$.
But when the U-B colour of the H II regions is used, then the ionising cluster is matched with a $\approx 7$ Myr old model SSP, twice the age given by ME00 (${\rm \sim 3.5 \ Myr}$). 
This difference (which can be as high as 0.25 mag) might be explained by the fact that PopStar
 models predict a slightly redder (${\rm \Delta \approx 0.1 \ mag}$) U-B 
 colour than Starburst99.

% WOLF-RAYET STARS 
%---------------------------------------------------------------------------------------------
%
PopStar models consider that for a SSP with Z=0.004 the Wolf-Rayet phase starts at 3 Myr and ends at 5 Myr. 
If the star clusters that ionise the \mbox{H II} regions are 5 Myr old - as estimated from the ${\rm EW(H\alpha )}$ - then the presence of Wolf-Rayet stars cannot be ruled out.
This idea is supported by the U-B colour of the \mbox{H II} regions (${\rm U-B \approx 0.7 \ mag}$), since the characteristic U-B colour of Wolf-Rayet stars is 0.7 mag according to the sample of Wolf-Rayet stars collected by Paul Crowhter\footnote{http://pacrowther.staff.shef.ac.uk/WRcat/}.
The intense ${\rm H\alpha}$ emission from \mbox{H II} regions \#5, \#8, \#10 and \#11 (Table \ref{tab:tb_hiinb2_mix}) implies that they are being photoionised by more than ten massive O7 V stars (Table  \ref{tab:hiideriv}) and therefore, these regions are candidates to host Wolf-Rayet stars.
Another clue pointing to these \mbox{H II} regions as probably photoionised by Wolf-Rayet stars is their high excitation, deduced from their ${\rm \log \, [OIII] / H\alpha}$ values.
In Section \ref{subsec:phot_nbphot_res}, we showed that the ${\rm \log \, [OIII] / H\alpha}$  ratio is a good excitation indicator in the star-forming activity of this galaxy, and  
 the appearance of Wolf-Rayet stars in star clusters is likely to increase the nebular excitation of the surrounding \mbox{H II} region \citep{Bu05}. 
 This is in agreement with the results obtained by \citet{Me00}, and consistent with the identification of ${\rm He \, II \, \lambda 4686}$  in the spectrum of the galaxy by \citet{Ku85,Va92} and \citet{Sc99}.

%% file: art.sspfit2.tex
%------------------------------------------------------------------------------------------------
\begin{table}
\centering
\caption{\label{tab:art.sspfit} Properties  of the dominant stellar populations.}
{\small
\begin{threeparttable}
{
\renewcommand{\arraystretch}{1.5}
\begin{tabular}{ c c c c }
\hline
\hline
{\footnotesize ID} & {\footnotesize ${\rm \log Age}$ } & {\footnotesize ${\rm A_{V} }$} & {\footnotesize ${\rm \log Mass }$} \\
{ } & {\footnotesize (yrs)} & {\footnotesize (mag)} & {\footnotesize (${ M_{\sun} }$) } \\
\hline
North       & ${\rm 6.97^{+0.02}_{-0.03}}$ & ${\rm 1.38\pm0.02}$ & ${\rm 6.68^{+0.14}_{-0.16}}$ \\ 
West        & ${\rm 9.40^{+0.13}_{-0.15}}$ & ${\rm 0.00\pm0.01}$ & ${\rm 7.97^{+0.09}_{-0.25} }$ \\ 
South       & ${\rm 6.57^{+0.03}_{-0.03}}$ & ${\rm 1.50\pm0.00}$ & ${\rm 6.38^{+0.07}_{-0.10} }$ \\ 
East        & ${\rm 9.34^{+0.10}_{-0.08}}$ & ${\rm 0.00\pm0.01}$ & ${\rm 7.87^{+0.12}_{-0.27} }$ \\ 
Centre      & ${\rm 6.70^{+0.02}_{-0.02}}$ & ${\rm 1.50\pm0.00}$ & ${\rm 7.42^{+0.14}_{-0.06} }$ \\ 
Centre-LSBC & ${\rm 6.40^{+0.05}_{-0.06}}$ & ${\rm 0.76\pm0.02}$ & ${\rm 7.03^{+0.14}_{-0.06} }$ \\ 
SCC 1 & ${\rm 6.80^{+0.02}_{-0.03}}$ & ${\rm 0.34\pm0.06}$ & ${\rm 4.77^{+0.16}_{-0.18} }$ \\ 
SCC 2 & ${\rm 6.50^{+0.02}_{-0.03}}$ & ${\rm 0.94\pm0.17}$ & ${\rm 4.55^{+0.15}_{-0.06} }$ \\ 
SCC 3 & ${\rm 7.00^{+0.04}_{-0.01}}$ & ${\rm 0.50\pm0.06}$ & ${\rm 5.05^{+0.18}_{-0.03} }$ \\ 
SCC 4 & ${\rm 7.00^{+0.04}_{-0.02}}$ & ${\rm 0.52\pm0.05}$ & ${\rm 5.43^{+0.13}_{-0.06} }$ \\ 
SCC 5 & ${\rm 7.06^{+0.04}_{-0.04}}$ & ${\rm 0.04\pm0.39}$ & ${\rm 4.59^{+0.11}_{-0.06} }$ \\ 
SCC 6 & ${\rm 7.04^{+0.05}_{-0.05}}$ & ${\rm 0.62\pm0.20}$ & ${\rm 4.91^{+0.14}_{-0.06} }$ \\ 
SCC 7 & ${\rm 7.01^{+0.04}_{-0.02}}$ & ${\rm 0.72\pm0.36}$ & ${\rm 4.80^{+0.24}_{-0.03} }$ \\ 
SCC 8 & ${\rm 6.97^{+0.01}_{-0.02}}$ & ${\rm 0.52\pm0.08}$ & ${\rm 5.23^{+0.16}_{-0.04} }$ \\ 
\hline
\hline
\end{tabular}
}
\end{threeparttable}
} % END SMALL
\end{table}

%% file: tol02_5bv6c.tex
\section[]{DISCUSSION AND CONCLUSIONS}
\label{sec:disc}

%-FRACCION GAS DIFUSO, NUM DE ESTRELLAS OB EN EL GAS DIFUSO

%=============================================================================================
% ANALISIS DE LA DISTRIBUCIÓN DE LAS DIFERENTES POBLACIONES ESTELAES QUE SE ENCUENTRAN EN LA GALAXIA
%=============================================================================================
\subsection{Stellar populations across the galaxy}
\label{subsec:disc_sfh}

As shown in Sections \ref{sec:host} and \ref{fitbb_sedfit}, the galaxy is formed by a
 disk dominated by the light of a 1.5 Gyr old stellar population, which has a photometric
 mass of ${\rm 10^8 \, M_{\sun}}$, and is directly observed in the East and West zones of the
 galaxy.
A ${\rm 5 }$ Myr old star-forming event is located at its centre, formed by eight massive SCCs 
 (${\rm M_{SCC} > 10^4 \, M_{\sun}}$, see Table \ref{tab:art.sspfit}) and
 twelve \mbox{H II} regions (Figures \ref{fig:tol0957rgb} and \ref{fig:scchii_position}).
Since all the objects are younger than 10 Myr old, we can consider the idea of an instantaneous star-formation episode. 
 This result supports the assumption that the star-formation triggering mechanism of isolated \mbox{H II} 
 galaxies is stochastic, as proposed by \citet{Te10}.
So despite the fact that the star clusters appear to be aligned, it seems that the star formation has not propagated along this apparent alignment.

The mass of the SCCs is $> 10^{4}$ ${\rm M_{\sun}}$ and their total mass is
 ${\approx 10^{6}}$ ${\rm M_{\sun}}$, which represents only 8\% of the estimated mass in 
 the Centre zone (Table \ref{tab:art.sspfit}). 
SCCs are concentrated at the Centre zone of the galaxy, while the ${\rm H\alpha}$ emission and 
\mbox{H II} regions are distributed in the Centre, North and South zones (Figure \ref{fig:tol0957rgb}).
\mbox{H II} regions -with the exception of \#1- are displaced from the positions of the SCCs, 
so determining their ionising source is very difficult.
\mbox{H II} region \#1  is the brightest of them all and encloses two star cluster complexes, SCC \#1 and SCC \#2.
Since SCC \#1 is younger and more massive than its companion, it emits more ionising photons,
 which leads us to believe that SCC \#1 dominates the evolution of the \mbox{H II} region.
 Still, further spectroscopic information is required to accurately determine the age and  kinematics of the \mbox{H II} region.

%=============================================================================================
% PROBLEMA DEL AJUSTE EN LAS BANDAS NIR, EXCESO EN ELA BANDA H. SGRs??
%=============================================================================================
%\subsection{SCCs NIR colours}
\subsection{ SCCs red colours }
\label{subsec:scc_nir}

\begin{figure*}
\centering
\begin{tabular}{c c}
% TOL 0957-278
%\subfigure[Optical B]{
\subfigure{
\resizebox{0.45\textwidth}{!}{\label{fig:scchii_ubri} \includegraphics{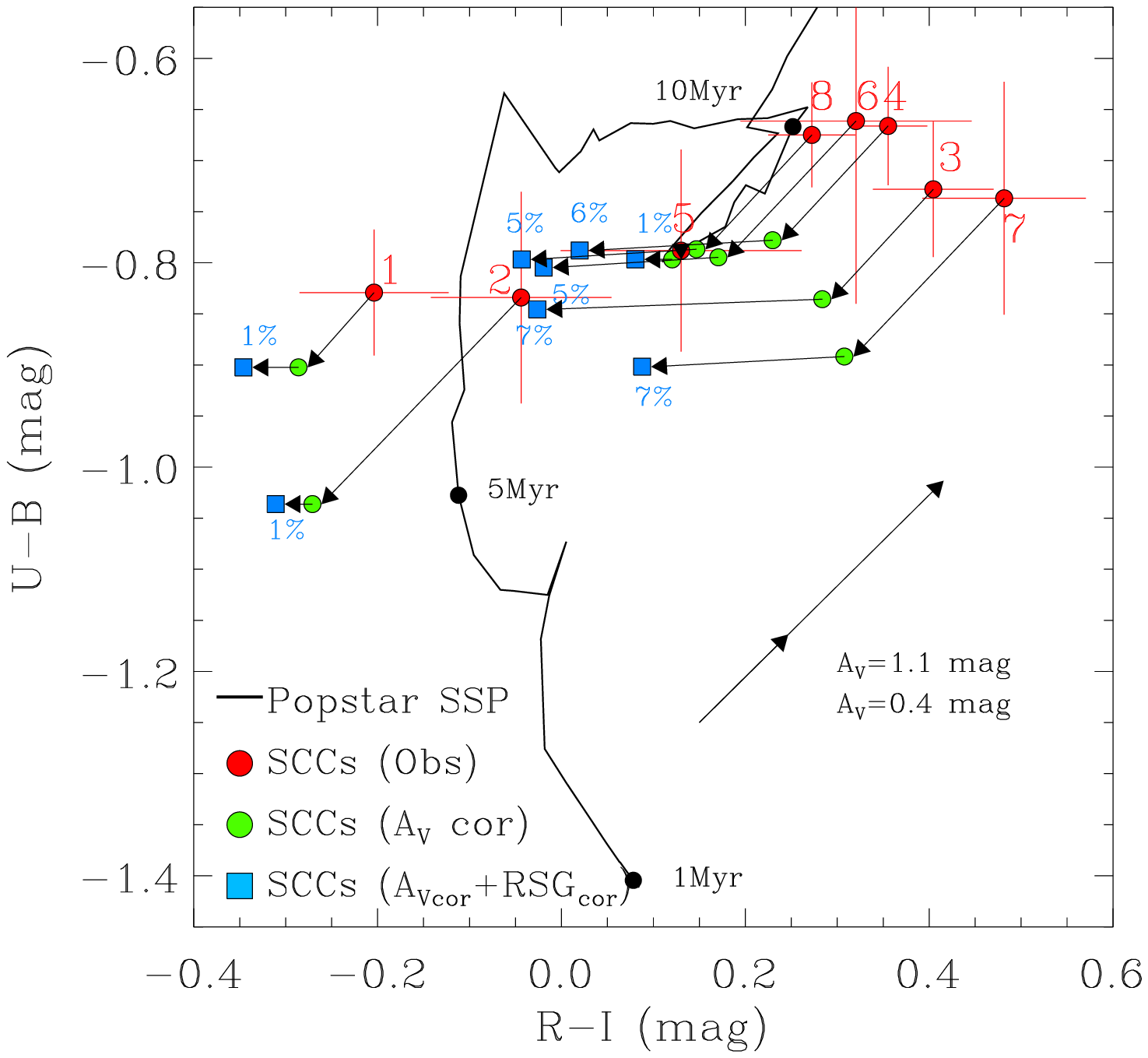}} } &
%\subfigure[NIR J]{
\subfigure{
\resizebox{0.45\textwidth}{!}{\label{fig:scchii_ijjh}\includegraphics{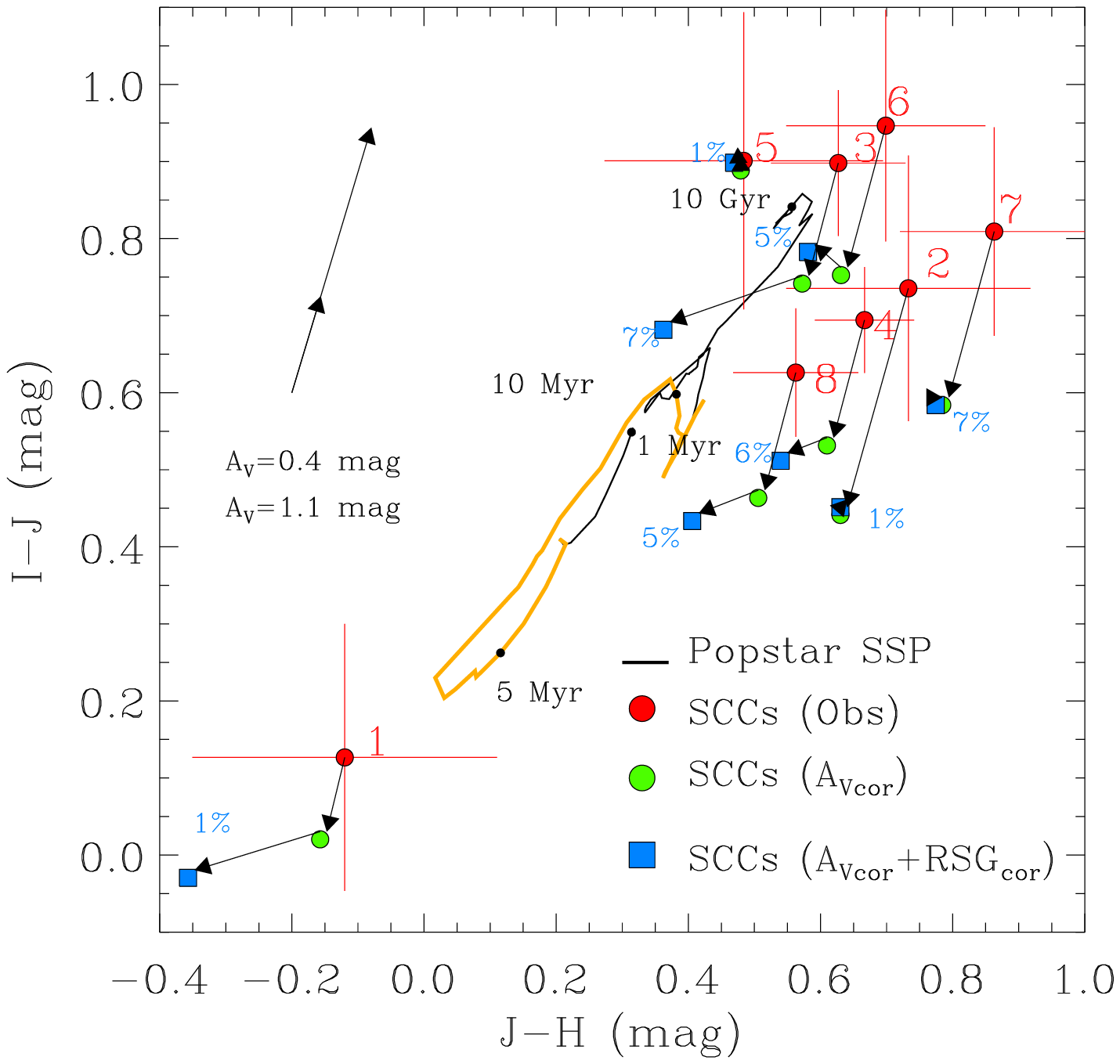}}} \\
\end{tabular}
\caption[] { Colour-colour diagrams showing the position of 
the observed SCCs  (red circles), the associated extinction vector  
 (see text), their extinction corrected colours 
(green circles) and after removing a fraction of RSG stars (blue squares; see
 section \ref{subsec:scc_nir_rsg}). Each SCC is tagged with its ID number in red and the fraction in R 
(as a percent) of the subtracted light (blue squares). Left: \mbox{U-B} vs \mbox{R-I} and right: \mbox{I-J} vs \mbox{J-H}; the RSG stars epoch is coloured 
in yellow. } 
\label{fig:scchii_colourcolour}
\end{figure*}

% Presentar el diagrama J-H vs H-K con la posición de mis cúmulos
%%---------------------------------------------------------------------------------------------
Figure \ref{fig:scchii_colourcolour} shows the position of the SCCs before (red circles) and after (green circles) correction by its associated SSP fit extinction value in the optical-NIR U-B vs R-I and NIR I-J vs J-H diagrams. The evolutionary track of an extinction-free PopStar SSP model is also overplotted.
% (solid line).
The SCCs U-B colour corresponds to a SSP younger than 10 Myr even before being corrected by extinction being the bluest SCCs (\#1 and \#2) the ones  coinciding with the brightest H II region (H II \#5). On the other hand, the SCCs R-I colour is redder than what the models predict, even after having been corrected by extinction. A similar excess in the R-I colour was observed previously by \citet{Di00} in a sample of nuclear star clusters. They suggested that this red excess could be attributed to Red Supergiant Stars not being properly taken into consideraton in SSP models. This issue will be addressed further in Section (\ref{subsec:scc_nir_rsg}).

H is the  band most affected by the emission lines; \mbox{J-H} turns bluer%\footnote{
%For example, when a SSP is 4 Myr old the emission lines would turn the  
% \mbox{J-H} colour bluer by ${\rm \Delta(J-H)=-0.25}$ mag but the \mbox{I-J} 
%redder by less than ${\rm \Delta(J-H)=+0.2}$ mag using GALEV models \citep{Ko09}. }
, thus moving the evolutionary track towards the left.
The emission lines contribution decreases with time, becoming negligible for a 10 Myr old SSP. 
% 
%On the other hand, figure \ref{fig:rsg_ijjh} shows that SCC \# 1 is located apart from the rest.
SCC   \#1 is an outlier at the bottom-left corner of the I-J vs J-H diagram. Its position suggests
 that its flux is affected by nebular emission lines, which is to be expected for the estimated age of  6 Myr.
The rest of the SCCs are found grouped at the top-right red area of the diagram 
(${\rm J-H> 0.4}$ mag and ${\rm I-J > 0.6}$ mag), closer to the evolutionary track 
at ages older than 10 Myr. Even after being corrected by  extinction,  
four of the SCCs (\#2, \#4, \#7 and \#8) are still apart from the evolutionary track, and 
the rest (with the exception of SCC \#1) are consistent with ages greater than 10  Myr, which contradicts 
the optical-NIR diagram.
%

% SCCs BROAD BAND HISTOGRAMS 
%---------------------------------------------------------------------------------------------

% Explicar dónde caen las galaxias BCD, por lo que se dice son comnflictivas
%---------------------------------------------------------------------------------------------
%
%
Reddening of SCCs NIR (I, J, H and K) bands has been observed previously in the literature \citep{Or99,Va05,Re08,Re10,Ad10a,Ga13}. 
The mismatch between young stellar populations observed NIR colours and the models was first
noticed by \citet{Or99} when studying the integrated colours of BCD galaxies. Subsequent 
studies on resolved stellar clusters in a variety of star-forming galaxies have come across 
this problem.
Ad hoc mechanisms have been proposed to explain it, such as 
 extended red emission \citep[ERE,][]{Re08,Ad10a}, nebular continuum and emission lines \citep{Re10}, 
 the use of large apertures to compute the photometry \citep{Ba14}, and that the
 number of RSG stars predicted by the models might be underestimated \citep{Or99,Va05,Ga13}, 
being this last one the most popular explanation.

% Explicar la posición de los cúmulos (zona BCD)
%---------------------------------------------------------------------------------------------
%

The 10 Myr estimated age for the SCCs (epoch at which the RSG population has 
 already begun its maximum contribution to the SCC light) supports the idea that the 
 mismatch between the colour of the SCCs and the models is related to the
 number of RSG predicted by the models. 
Therefore, motivated by \citet{Ga13}, who found that the J-H colour of a sample of stellar clusters was better reproduced by SSP models after increasing the number of RSG stars in them, 
we decided to analyse whether a dearth of RSG stars in PopStar models could be the cause of the mismatch between the models and the photometry of our SCCs.

% RED SUPERGIANT STARS AS THE REASON OF THE EXTREME RED \mbox{J-H} COLOR
%---------------------------------------------------------------------------------------------
\subsubsection{Dearth of RSG stars in the models?}
\label{subsec:scc_nir_rsg}

% Entonces... en el caso de que los modelos predigan mal el numero de RSG??
%---------------------------------------------------------------------------------------------
We can test whether the models underestimate the number of RSG stars by removing the 
 contribution of a number of RSG  from the luminosity of the SCCs and estimating the resulting NIR colours.
%
%Following this idea, we subtracted the luminosity of a number of typical  RSG from each SCC.
%
Considering that the metallicity of the galaxy might determine the spectral 
 type of its RSGs \citep{Ma03}, we chose a K5-7 RSG star of the low metallicity 
 Small Magellanic Cloud \citep{Ma03} as prototype for the RSG stars in Tol 02 and its 
 characteristics are shown in Table \ref{tab:tb_rsgstar2}.
We subtracted up to 20 times the luminosity of the prototype RSG star
 from the observed luminosity of each SCC. % in all of the bands. % extinction corrected 
The subtracted R band luminosity contribution represents a fraction of the SCC total 
 luminosity and is used as a measure to compare the colour changes suffered by different SCCs, 
 independently of their masses.
These colour changes are described below.

% TABLE WITH THE RSG STAR CHOSEN AS CHARACTERISTIC OF LOW METALLICITY GALAXIES
\input{tb_rsgstar2}

\begin{figure}
\centering
\resizebox{0.45\textwidth}{!}{\includegraphics{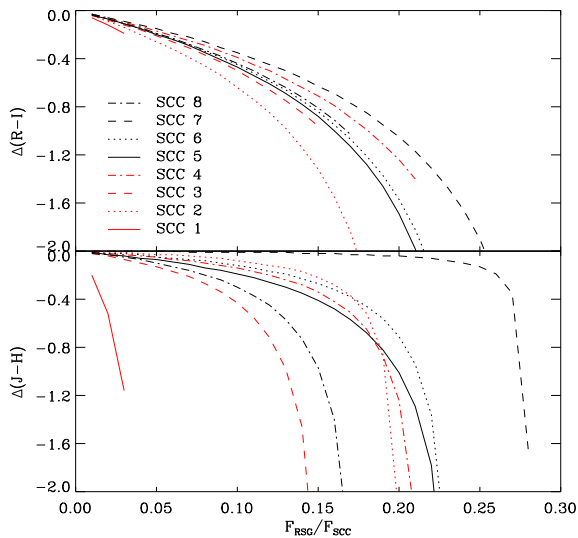}} 
\caption{ \small Change in the R-I (Top) and J-H (Bottom) colours of the SCCs due to the 
 subtraction of luminosity from RSG stars (${\rm \Delta = SCC_{Without RSG} - SCC_{Obs}}$). The colour change is plotted against the amount of subtracted luminosity in the R band, represented as a fraction of the SCC original luminosity (${\rm L_{RSG} / L_{Obs} }$).}
\label{fig:rsg_colorinc}
\end{figure}

% Color U-B, R-I explicar cómo se comportan los cúmulos al quitarles las estrellas RSG
%---------------------------------------------------------------------------------------------
The subtraction of RSG stars turns out to be  irrelevant for U-B.
 The change it produces in the colour is ${\rm \Delta ( U-B ) < 0.01 \ mag}$, which is 
 negligible compared to the increase of 0.7 magnitudes in U-B when the SSP ages from 2 to 6 Myr. 
On the contrary, when the subtracted R luminosity represents 20\% of the observed luminosity 
 (${\rm L_{RSG} / L_{SCC} = 0.2 }$), that turns R-I bluer by ${\rm \sim 1 \ mag}$,
 larger than the change of the R-I colour of an extinction free SCC during its lifetime (-0.1 \ mag < R-I < 0.7 \ mag).
This rapid change to the blue by artificially discounting RSG stars is shown in the top panel of Figure \ref{fig:rsg_colorinc}.
The \mbox{J-H} colour is also sensitive to the presence of RSG stars, as shown in the 
 bottom panel of Figure \ref{fig:rsg_colorinc}.
This figure shows that when the RSG stars R luminosity represents  ${\rm \sim 10\%}$ 
 of the SCCs R band luminosity, the change suffered by the J and H luminosities moves 
 0.1 mag to the blue the \mbox{J-H} colours of six clusters.

The change in I-J colour due to the subtraction of the RSG stars is more complex. Only four
SCCs become bluer when the RSG stars are subtracted from them,  while the remaining three 
(\#1, \#2 and \#6) become redder. SCCs \#1 and  \#2 
are located inside the brightest H II region, and the SCC \#6 aperture overlaps with the 
aperture of H II region \#7 in ${\rm \sim 75\%}$ of area. These three complexes can
be considered to be affected by the emission lines. Therefore the subtraction of the RSG 
stellar light enhances the effect of the contamination by the nebular emission lines and 
produces the curious effect of a redder colour of the SCC.

% ---------------------------------------------------------
The significance of the RSG stars to the SCC colours is shown in Figure \ref{fig:scchii_colourcolour}.
Each SCC is tagged with its ID number next to the original colour (red circle) and the
 fraction of the cluster subtracted R band light (as a percent) next to the SCC colour after a number of RSG stars has been subtracted (blue squares). % after being subtracted by a number of RSG stars (blue squares). }}
The pointing of the arrows in the figure shows that the subtraction of the light from the RSG stars moves the SCCs away from the models, 
 with the exception of SCC \#4 and \#8. When the flux of these two SCCs is subtracted from the light of RSG stars their resultant colours are in better agreement with the prediction of  the models.

It is clear from the figure that the mismatch between the SSP models and the photometry of the 
 SCCs can be caused by an underestimate of the number of RSG stars in the evolutionary
 tracks only for SCCs \#4 and \#8.

% CHANGE IN COLOUR DUE TO THE SUBTRACTION OF  FLUX FROM RED SUPERGIANT STARS
%---------------------------------------------------------------------------------------------

% J-H vs H-K CHANGING THE AMOUNT OF RSG STARS (art_rsgDucati.pro)

In summary: 

\begin{enumerate}
    \item The RSG stars do not contribute to the U-B colour.
    \item The increase in the number of RSG stars in a SCC reddens both  \mbox{J-H} and \mbox{R-I}, being the latter colour the most affected one.
\end{enumerate}

Therefore, at least for six of the eight SCCs  a ``simple'' change in the number of RSG stars of the models is not  capable of reproducing the observations. 
The problem seems to be more likely the modelling of the SED slope in the NIR  range 
  ${\rm 0.8 \mu}$m to 2.5 ${\rm \mu}$m, where the luminosity of a SSP model decreases steadily 
 as the wavelength increases. 
But if the SSP modelled luminosity  increased at  H, the 
\mbox{J-H} colour would redden, thus better reproducing the observation.
A NIR continuum spectrum with a luminosity increase that peaks towards the H band and 
then decreases towards the K band is observed in some metal rich star-forming spiral galaxies 
\citep{Ma13}. 
To assess whether this behaviour could be expected also in low metallicity smaller systems 
like H II galaxies (in particular  Tol 02) high spatial and spectral resolution NIR data on 
each of the SCCs would be required in order to confirm this hypothesis. 
%

%% file: tb_rsgstar2.tex
%--------------------------------------------------------------------------------------------------
%	FILTERS TABLE
%--------------------------------------------------------------------------------------------------
\begin{table}
\centering
\caption{\label{tab:tb_rsgstar2} Representative RSG star characteristics.}
{\small
%\begin{threeparttable}

\begin{tabular}{c c c c c c c c c c c c }
\hline
\hline
\multicolumn{12}{l}{ Star ID \citep{Ma03}:  059426  }\\
\multicolumn{12}{l}{ Star ID \citep{Pr83} : PMMR 158   }\\
\multicolumn{12}{l}{ Coords:   R.A. 01\fh 04\fm 30.28\fs   Dec. -72\degr 04\arcmin 36.1\arcsec  }\\
\multicolumn{12}{l}{ Radial velocity:  ${\rm 167.05 \pm 0.3 }$  }\\
\multicolumn{12}{l}{ Spectral Type and Luminosity class:  K5-7 I }\\
\multicolumn{12}{c}{  }\\
\multicolumn{3}{c}{U} & \multicolumn{3}{c}{B} & \multicolumn{3}{c}{R} & \multicolumn{3}{c}{I} \\
\multicolumn{3}{c}{${\rm 16.34 \pm 0.07 }$} & \multicolumn{3}{c}{${\rm 14.79 \pm 0.03}$} & \multicolumn{3}{c}{${\rm 12.09 \pm 0.00 }$} & \multicolumn{3}{c}{${\rm 10.67 \pm 0.36 }$} \\
\multicolumn{3}{c}{J} & \multicolumn{3}{c}{H} & \multicolumn{3}{c}{K} & \multicolumn{3}{c}{} \\
\multicolumn{3}{c}{${\rm 10.006 \pm 0.021}$} & \multicolumn{3}{c}{${\rm 9.174 \pm 0.026}$} & \multicolumn{3}{c}{${\rm 8.963 \pm 0.024}$} & \multicolumn{3}{c}{} \\
%\multicolumn{4}{c}{J} & \multicolumn{4}{c}{H} & \multicolumn{4}{c}{K} \\
%\multicolumn{4}{c}{${\rm 10.006 \pm 0.021}$} & \multicolumn{4}{c}{${\rm 9.174 \pm 0.026}$} & \multicolumn{4}{c}{${\rm 8.963 \pm 0.024}$} \\
\hline
\hline
\end{tabular}
%     {\scriptsize
%     \begin{description}  \addtolength{\itemsep}{-0.8\baselineskip} %tablenote
%       \item[] Notes: (1) Stars seem elongated, (2) h:m:s,(3) d:m:s, (4) Z, (5)  calibrated,
%       \item[]  (5) tasa de fotones capaces de ionizar el hidr\'ogeno, (6) densidad del gas de la regi\'on H II asociada al SCE y (7) radio fotom\'etrico en H$\alpha$.
       %\item[] (6) Densidad del gas de la regi\'on H II asociada al SCE y (7) radio fotom\'etrico en H$\alpha$.
       %\item[3] Densidad estelar del SCE.
       %\item[4] Masa del SCE \citep{Me05}.
%     \end{description} %tablenotes
%     } %scritpstize
%  \end{threeparttable}
} %small
\end{table}
%--------------------------------------------------------------------------------------------------

%% file: tol02_6v6c.tex
\subsection[]{CONCLUSIONS}
\label{sec:con10}

% PHOTOMETRIC RESULTS
%--------------------------------------------------------------------------------------------
A detailed study of the star-forming regions and star cluster complexes in the 
 \mbox{H II} galaxy Tol 02 was performed using a broad (U, B, R, I, J, H, K, ${\rm H \alpha}$ 
 and deep ${\rm [OIII]\lambda 5007}$) set of images of the galaxy.
The ${\rm [OIII]\lambda 5007}$ map is presented for the first time for this galaxy.
These images, combined with SSP PopStar models, have allowed us to conclude that:
\begin{itemize}
\item The galaxy can be represented by a disk whose light is dominated by an 
 \mbox{extinction-free} 1.5 Gyr old stellar population with a total mass of  
 ${\approx 4 \times 10^{8}}$ ${\rm M_{\sun}}$, and a ${\rm \sim 5}$ Myr old central starburst 
 whose light spreads towards the North and South of the galaxy, dominating over the red light 
 of the disk component.
\item The star-forming region of this galaxy is formed by twelve bright (${\rm L_{H\alpha} > 10^{38} \ erg \ s^{-1}}$) H II regions, the most luminous  of which is a giant H II region (${\rm L_{H\alpha} \leq 10^{39} \ erg \ s^{-1}}$).
 The U-B colour, ${\rm EW(H\alpha)}$ and ${\rm \log \, [OIII]/ H \alpha}$ values of the four brightest H II regions suggest that these H II regions might be photoionised by clusters containing  Wolf-Rayet stars. This is in agreement with the results obtained by \citet{Me00}, and consistent with the identification of ${\rm He \, II \, \lambda 4686}$  in the spectrum of the galaxy by \citet{Ku85,Va92} and \citet{Sc99}.

\item Eight massive (${\rm M > 10^4 \, M_{\sun}}$) SCC located in the brightest \mbox{H II} regions 
 were identified, however the physical connection between each H II region and the SCCs is 
 not clear.
The crowding of the H II regions and SCCs makes it impossible to determine to which  \mbox{H II} region a given SCC relates.
Fitting SEDs of model SSP using PopStar is consistent with six of the SCC being 10 Myr old,
 so expected to contain RSG stars, while the other two are 3 and 5 Myr old and therefore are Wolf-Rayet 
 cluster candidates.
\item The young ages ascribed to the SCCs and H II regions (${\rm \le 10}$ Myr) suggest an instantaneous and stochastic star formation mode that depends on short scale inhomogeneities of the interstellar medium.

An ad-hoc experiment subtracting a combination of typical RSG stars indicates 
that the R-I excess observed in the SCCs is well explained if the models predict  an incorrect number of RSG 
stars, as previously suggested by \citet{Di00}.

\item The I-J and J-H colours of SCCs \#4 and \#8 (most massive clusters identified) 
can be explained if the models are underestimating the number of RSG stars. However, 
the NIR colours of the remaining SCCs can not be explained in the same way.

\end{itemize}

This article presents the first results of a project aimed at characterising the 
 spatially resolved dominant stellar populations in six low metallicity H II galaxies,
 which helps to understand whether these galaxies  with intensive \mbox{star-formation} 
  are prone to form massive  star clusters and if their star formation process is 
 stochastic or sequentially induced, therefore providing a better understanding of the 
 \mbox{star-formation} process undergone in low metallicity systems.

%% file: tol02_7v6c.tex
\section*{Acknowledgements}

The authors wish to thank an anonymous referee whose recommendations greatly improved the clarity of the paper.
AT-C is thankful to Patricio Lagos, Erique P\'{e}rez, Lino Rodr\'{i}guez-Merino, Abraham Luna,   Mercedes Moll\'{a}, Polychronis Papaderos, Nate Bastian and Divakara Mayya for fruitful discussions, and to her sister Nora for a careful reading of the manuscript.  
We also would like to thank the Time Allocation Committee and the very helpful supporting staff at SOAR. 

This research has made use of the NASA/IPAC Extragalactic Database (NED) which is operated by the Jet Propulsion Laboratory, California Institute of Technology, under contract with the National Aeronautics and Space Administration.

This work was supported by CONACyT (M\'{e}xico) student fellowship grant 46360 and research project numbers CB-2011/167281. Partial financial support from projects: AYA2013-47742-C4-3-P from the Ministerio de Econom\'{i}a y Competitividad (MINECO, Spain) and
SELGIFS: FP7-PEOPLE-2013-IRSES-612701 the Research Executive Agency (REA, EU) is acknowledged.
Ana Torres-Campos is grateful to the hospitality of the Departamento de F\'{i}sica Te\'{o}rica in Madrid and the Observatorio Nacional in Rio, during work visits when part of this work was completed.

%% file: tol02_8.tex
\section{APENDIX A. FLUX CALIBRATION}
\label{ap:a}

%---------------------------------------------------------------------------------------------
\subsection{Broad band flux calibration}
%---------------------------------------------------------------------------------------------
The magnitude of the standard star is related to the star image counts rate:

\begin{equation}
        m_{\lambda \star} = -2.5\log C_{\star} - k_{\star} X_{\star} + ZP_{\lambda} \label{eqn:sstar_mag}
\end{equation}

\noindent where $m_{\lambda \star}$ is the standard star magnitude, $C_{\star}$ is the number of counts per second received by the detector from the standard star, $k_{\star}$ is the extinction coefficient at wavelength $\lambda$, $X_{\star}$ is the airmass at which the standard star was observed and $ZP_{\lambda}$ is the zeropoint, a constant that relates the standard star image counts per second to the published magnitude value. Equation \ref{eqn:sstar_mag} is solved for the $N_{\star}$ standard stars observed and a median value  for $ZP_{\lambda}$ is obtained as:

\begin{equation}
  \bar{ZP}_{\lambda} = \frac{1}{N_{\star}}   \sum_{i=1}^{N_{\star}} Z_{\lambda \, i}  \quad , \\
\end{equation}

The observed magnitude of an object with total counts per second $C_{ob}$ inside an aperture with N pixels is:

\begin{equation}
        m_{\lambda ob} = -2.5\log \left[ C_{ob} - \left(N_{pix}C_{sky}\right) \right] - k_{\star} X_{\star} + \bar{ZP}_{\lambda} \label{eqn:counts2mag_star}
\end{equation}

\noindent where $C_{sky}$ is the sky counts per second, and the object magnitude uncertainty is given by:

\begin{equation}
        \sigma_{m_{\lambda ob}} = \frac{1.085}{C_{ob}-(N_{pix}C_{sky})} \sigma_{C_{ob}}
\end{equation}

\noindent where $C_{ob}$ is the counts per second of the object and its uncertainty is given by:

\begin{equation}
        \sigma_{C_{ob}} = \sqrt{\frac{C_{ob}}{Gain}+N_{pix}*\sigma_{sky}^2}
\end{equation}

The magnitude to flux conversion is computed using the relation:

\begin{equation}
%        Flux_{\lambda \, ob} = 10^{(8.9-zp_{\lambda}-m_{\lambda \, ob})/2.5}
        F_{\lambda \, ob} = F_{0 \, \lambda} 10^{-m_{\lambda \, ob}/2.5}
\end{equation}

\noindent where $F_{0 \, \lambda}$ is the flux of Vega \citep{Be98}. 

%---------------------------------------------------------------------------------------------
\subsection{Narrow band flux calibration}
%---------------------------------------------------------------------------------------------
For the narrow band flux calibration we followed \citet{Ja87}, where the standard star flux is given by equation \citep{Ja87}:
 %Integrated flux equation (Lagos et al, 2007 and Jacoby et al 1987)
\begin{equation}
	F_{NB \, \star} = \frac{\int F_{\lambda} T_{\lambda} \; \mathrm{d}\lambda}{\int T_{\lambda} \; \mathrm{d}\lambda}  \quad , %\,\, ,
\end{equation}

\noindent in units of [${\rm erg \, cm^{-2} \, s^{-1}}$]. Where $F_{\lambda}$ is the monochromatic flux above the atmosphere for the standard star in units of [${\rm erg \, cm^{-2} \, s^{-1} \AA^{-1}}$] and $T_{\lambda}$ is the narrow band filter transmission curve.

The CCD system sensitivity (in units of [erg cm$^{-2}$ s$^{-1}$ counts$^{-1}$]) was computed using the standard stars images:

% STANDARD STAR CCD DETECTORSYSTEM SENSITIVITY (Lagos et al, 2007 and Jacoby et al 1987)
\begin{eqnarray}
        \bar{S}_{NB} = \frac{1}{n_{\star}}   \sum_{i=1}^{n_{\star}} \frac{F_{NB \, \star}}{C_{\star} 10^{0.4 X_{\star} k_{\star}}}  \quad , \\
	  \sigma_{\bar{S}_{NB}} = \frac{1}{n_{\star}} \sqrt{ \sum_{i=1}^{n_{\star}}  \left( \frac{\sigma_{C_{\star}}}{C_{\star}} S_{NB} \right)^2 }   \quad ,
\end{eqnarray}

\noindent $C_{\star}$ is the count (photon) rate\footnote{The number of counts of the standard stars is determined using the \textit{daophot} and \textit{phot} tasks from the \textit{noao.digiphot.daophot} package.}, $X_{\star}$ is the airmass and $k_{\star}$ is the extinction coefficient\footnote{The extinction coefficient curve is the Cerro Tololo International Observatory (CTIO) one used in IRAF's task \textit{extinction} \citep{St83,Ba84}.} for the narrow band filter image of the standard star, $\sigma_{C_{\star}}$ is the standard star counts rate uncertainty ($\sigma_{C_{\star}} = \sqrt{C_{\star }+2\left(\sigma_{sky}\right)^2}$)  and $\sigma_{sky}$ is the uncertainty of the sky counts per second.

The targe flux is related to the sensitivity of the system by:

\begin{equation}
	F_{NB \, g} = \bar{S}_{NB} \left( C_g 10^{0.4 X_g k_g} \right)  \quad , \label{eqn:sys_sens_gf}
\end{equation}

\noindent where $F_{NB \, g}$ is the target flux observed in the narrow band, $C_g$ is the count rate from the target at any pixel of interest, $X_g$ is the airmass at which the target was observed and $k_g$ is the extinction coefficient for the narrow band filter. The emission line flux emitted by the target (in this case a galaxy) $F_{g_{\lambda}}$ is essentially a delta function over the filter bandpass centre at the wavelength  of the emission line, L. Then, 

\begin{equation}
	F_{NB \, g} = \int F_{g_{\lambda}} T_{\lambda}  \mathrm{d}\lambda = F_{g_L} T_{L} \;  \quad , \label{eqn:sys_gf_delta}
\end{equation}

\noindent where $T_{L}$ denotes the filter transmission at the emission line wavelength. So, if we substitute equation \ref{eqn:sys_sens_gf} in equation \ref{eqn:sys_gf_delta} we obtain the observed flux of the galaxy:

\begin{equation}
	F_{g_L} = \frac{\bar{S}_{NB}}{T_{L}} C_g 10^{0.4 X_g k} = C_g f_{pc}\quad , %\,\, ,
\end{equation}

%We can re-write this equation as:
 %flux of the galaxy (Jacoby et al 1987)
%\begin{equation}
%	F_{g_L} = C_g f_{pc} \quad , %\,\, ,
%\end{equation}

\noindent where $f_{pc}$ is the flux per count in the galaxy image, in units of [${\rm erg \, cm^{-2} \, s^{-1} \, count^{-1} }$]. As this equation states, to flux calibrate the flat corrected galaxy image we just have to multiply it by the flux per count coefficient ($f_{pc}$). The counts to flux transformation equations are:

% FLUX PER COUNT FACTOR AND UNCERTAINTY
\begin{equation}
	 f_{pc} = \frac{\bar{S}_{NB}}{T_{L}} 10^{0.4 X_g k}  \quad and \quad  \sigma_{f_{pc}} = \frac{\sigma_{\bar{S}_{NB}}}{T_{L}} 10^{0.4 X_g k}  \quad
\end{equation}

% FLUX PER COUNT FACTOR AND UNCERTAINTY
%\begin{eqnarray}
%	 f_{pc} = \frac{\bar{S}_{NB}}{T_{L}} 10^{0.4 X_g k}  \quad , \\
%	 \sigma_{f_{pc}} = \frac{\sigma_{\bar{S}_{NB}}}{T_{L}} 10^{0.4 X_g k}  \quad ,
%\end{eqnarray}

The counts per second inside an object aperture with N pixels is given by:

% FLUX PER COUNT FACTOR AND UNCERTAINTY
\begin{eqnarray}
	 C_{ob} = \left( \sum_{i=1}^{N} C_{i} \right) - \left( N*sky \right)  \quad , \\
	 \sigma_{C_{ob}} = \sqrt{\frac{C_{ob}}{N_{f}}+\left( N*\sigma_{sky}^2 \right)}  \quad ,
\end{eqnarray}

\noindent where $C_{i}$ is the counts per second in pixel i, sky is the mean sky value, 
$\sigma_{sky}$ is the standard deviation of the sky measurement and $N_{f}$ the number of 
median combined galaxy images. Finally flux of the object ($F_{ob}$) and its uncertainty ($\sigma_{F_{ob}}$) in units of erg ${\rm s^{-1}}$ ${\rm cm^{-2}}$ is calculated as:
% FLUX PER COUNT FACTOR AND UNCERTAINTY
\begin{eqnarray}
	F_{ob} = C_{ob} * f_{pc}  \quad , \\
	\sigma_{F_{ob}} = \sqrt{ \left( f_{pc}*\sigma_{C_{ob}} \right) ^2 + \left( C_{ob}*\sigma_{f_{pc}} \right) ^2} \quad ,
\end{eqnarray}

%---------------------------------------------------------------------------------------------
%       NARROW BAND CONTINUUM SUBSTRACTION
%---------------------------------------------------------------------------------------------
%\subsection{Narrow band images continuum subtraction}